\documentclass[referee,a4paper,12pt,traditabstract]{swsc} 
\usepackage{longtable,lscape}
\usepackage{booktabs}
\usepackage{supertabular}
\usepackage{hhline}
\usepackage{multirow}
\usepackage{rotating}
\usepackage{pdflscape}
\usepackage{tabularx}
\usepackage{longtable}
\usepackage{adjustbox,lipsum}
\usepackage[margin=1in]{geometry}
\usepackage{amsmath}
\usepackage{graphicx}
\usepackage{txfonts}
\usepackage{subfigure}
\usepackage{epstopdf}
\usepackage{lineno}
\usepackage[authoryear,round]{natbib}
\usepackage[backref]{hyperref}
\usepackage{url}
\usepackage{graphicx}
\usepackage{txfonts}
\usepackage{url}
\usepackage{amssymb}                    
\usepackage{color}
\usepackage[normalem]{ulem}

\usepackage{url}                         
\usepackage{multirow}
\usepackage{tabularx}
\usepackage{longtable}
\usepackage{array} 
\usepackage{longtable}
\usepackage{booktabs}
\usepackage{supertabular}
\usepackage{natbib}
\bibpunct{(}{)}{;}{a}{}{,}

\hypersetup{colorlinks=true,citecolor=cyan,urlcolor=cyan,linkcolor=blue}

\begin{document}

\title{The Probabilistic Solar Particle Event foRecasting (PROSPER) Model}

   
   \titlerunning{The PROSPER model}

   \authorrunning{Papaioannou et al.}

   \author{Athanasios Papaioannou
          \inst{1},
          Rami Vainio
          \inst{2},
           Osku Raukunen
          \inst{2,3},
           Piers Jiggens
          \inst{4},
           Angels Aran
          \inst{5,6,7},
           Mark Dierckxsens
          \inst{8},
           Sotirios A. Mallios
          \inst{1},
          Miikka Paassilta
          \inst{2},
          \and
          Anastasios Anastasiadis
          \inst{1}
          }
   \institute{Institute for Astronomy, Astrophysics, Space Applications and Remote Sensing (IAASARS), National Observatory of Athens,
              I. Metaxa \& Vas. Pavlou St.
GR-15236, Penteli, Greece\\
              \email{\href{mailto:atpapaio@astro.noa.gr}{atpapaio@astro.noa.gr}}
         \and
        Department of Physics and Astronomy, University of Turku, 20014 Turku, Finland
         \and
         Aboa Space Research Oy, Tierankatu 4B, 20520 Turku, Finland
         \and
         European Space Research and Technology Centre (ESTEC), Space Environment and Effects Section, Keplerlaan 1, 2200AG Noordwijk, The Netherlands 
         \and
             Dep. F\'{i}sica Qu\`{a}ntica i Astrof\'{i}sica (FQA), Universitat de Barcelona (UB),  c. Martí i Franquès, 1, 08028 Barcelona, Spain
             \and
             Institut de Ci\`{e}ncies del Cosmos (ICCUB), Universitat de Barcelona (UB), c. Martí i Franquès, 1, 08028 Barcelona, Spain
             \and
             Institut d'Estudis Espacials de Catalunya (IEEC), c. Gran Capit\`{a}, 2-4, 08034 Barcelona, Spain
                          \and
                  Royal Belgian Institute for Space Aeronomy (BIRA-IASB), Avenue Circulaire 3, 1180 Uccle, Belgium
             }


 
 \abstract
   {The Probabilistic Solar Particle Event foRecasting (PROSPER) model predicts the probability of occurrence and the expected peak flux of Solar Energetic Particle (SEP) events. Predictions are derived for a set of integral proton energies (i.e. E$>$10, $>$30 and $>$100 MeV) from characteristics of solar flares (longitude, magnitude), coronal mass ejections (width, speed) and combinations of both. Herein the PROSPER model methodology for deriving the SEP event forecasts is described and the validation of the model, based on archived data, is presented for a set of case studies. The PROSPER model has been incorporated into the new operational Advanced Solar Particle Event Casting System (ASPECS) tool to provide nowcasting (short term forecasting) of SEP events as part of ESA's future SEP Advanced Warning System (SAWS). ASPECS also provides the capability to interrogate PROSPER for historical cases via a run on demand functionality.}


   \maketitle

\section{Introduction}
\label{sec:intro}
Solar energetic particle (SEP) events constitute a significant component of the near Earth radiation environment and consist of protons, electrons and heavier ions \citep{vainio2009dynamics}. Such events originate from particle acceleration in solar flares (SFs) and/or shocks associated with coronal mass ejections (CMEs) \citep[e.g.][]{reames2015sources,2019RSPTA.37780095V}. Once energetic particles are accelerated and injected into open magnetic field lines, they are consequently routed through the interplanetary magnetic field (IMF) \citep{belov2005proton, cane2006introduction}. Subsequently, when an observer (i.e. spacecraft) is magnetically connected to the source of the particles, enhancements by several orders of magnitude above the pre-event background are observed \textit{in-situ}, with SEP events often observed in a broad range of solar longitudes \citep{rouillard2012longitudinal, lario2016longitudinal}. SEPs can last from a few hours to several days and their relative composition varies by many orders of magnitude from event to event \citep{2013SSRv..175...53R,2016LRSP...13....3D}. The classical paradigm divides SEP events into two categories, see e.g. \cite{reames1999particle}: those particles that are accelerated at SFs \citep{aschwanden2002particle} are known as impulsive events, and other particle populations that are accelerated by near-Sun CME-driven shocks are termed as gradual \citep{reames1999particle,reames2002magnetic,kahler2001correlation,cane2006introduction}. However, this ``two class" picture does not match the diversity and wealth of the observed SEP event properties which indicate a more complex nature \citep{cane2010study, 2016JSWSC...6A..42P, 2019RSPTA.37780095V}.  

SEP events have a direct space weather impact on electronics and humans \citep{baker2004specifying}. For example, the survivability of a spacecraft is directly affected by the total energy deposited by the passage of energetic particles. This may result in the degradation and ultimate failure of its electronic components due to ionisation or displacement damage mechanisms \citep{daly1996space,feynman2000space}. Effects on electronics also include single event effects (SEEs) \citep{pellish2010impact}, these appear when particles deposit sufficient energy or charge in a sensitive region of a component, with non-permanent (soft) errors such as bit flips, as well as, permanent (hard) errors such as latchups (i.e. SELs) and burnouts (SEBs) taking place \citep{sexton2003destructive, harboe201340}. Furthermore, SEP events are a major threat to human spaceflight outside the protective shield of the Earth’s magnetosphere \citep{bizzarri2017journey, townsend2021effects} and pose a severe danger for aircrews and passengers on polar flights \citep{tobiska2015advances,mishev2015computation, miroshnichenko2018retrospective}. The ionizing particle radiation can lead to damage to human cells and DNA alternations \citep{azzam2012ionizing}. There are two basic parameters that have a decisive role in the radiation effects in humans: the strength of the exposure and the specific organ/tissue that was encountered. For example, a short-term exposure to high doses of radiation depositing energy in the eye can lead to acute radiation effects such as cataracts. Long-term exposure to low doses of radiation deposing energy in bone marrow may progressively lead to leukaemia or other types of cancer. The former effects are categorized as deterministic while the latter as stochastic \citep{cucinotta2003radiation}. The high-energy tail of the SEP spectrum, in which ions are accelerated to relativistic energies, is dominant in spacecraft orbits well within the magnetosphere. In addition, high-energy particles upon interaction with matter, from e.g. the shielding material of a spacecraft, produce secondaries that can enhance the radiation effect of SEP events and are of major concern in heavily shielded environments such as human spaceflight \citep{2018JSWSC...8A...4R}. Nonetheless, these high-energy particles can also reach the Earth's atmosphere and generate secondaries through nuclear reactions. Consequently, these secondaries are recorded as a significant sudden increase at ground level, as detected by e.g. neutron monitors (NMs). Such events are termed as Ground-level enhancements (GLEs) \citep{butikofer2009solar,asvestari2017analysis,mishev2018first}. This results in enhanced ionisation together with modifications of the local chemistry of the high polar atmosphere of the Earth \citep{usoskin2011ionization,mironova2014possible}.   

SEPs, along with trapped radiation in planetary environments, are responsible for cumulative (dose) effects on spacecraft electronics and materials under low and moderate amounts of shielding. These are modelled statistically to produce specification (climatological) models to predict degradation over the duration of a mission. Such models make use of fits to SEP flux data in the form of lognormal, truncated power law, and exponential cut-off power law distributions to derive fluxes as a function of mission duration and confidence (\citet{jiggens2018updated, 2018JSWSC...8A...4R} and references therein). However, these models cannot make short-term forecasts of the SEP environment to provide warnings for operational missions, human spaceflight, launch operators and aircraft operators.

A small fraction of all solar flares and CMEs lead to SEP events and as a result the prognosis of SEPs is not a trivial task, since this is a highly imbalanced problem \citep[see the relevant discussion in][]{lavasa2021assessing}. The scientific questions that one needs to address in such studies \citep[see e.g.][]{anastasiadis2019solar} are summarized as: \textit{If we know the characteristics of the parent solar events, could the probability of SEP occurrence be reliably inferred} and \textit{how do the characteristics of SEP events (e.g. peak flux) map to the characteristics of their parent solar events} ? Both questions are being largely addressed through the implementation of databases and the establishment of empirical and/or semi-empirical statistical relations \citep[see e.g.][]{gopalswamy2003large,gopalswamy2004intensity,garcia2004forecasting1,
garcia2004forecasting,belov2005proton,laurenza2009technique,laurenza2018short,nunez2011predicting,trottet2014statistical,dierckxsens2014relationship,anastasiadis2017predicting,papaioannou2018nowcasting,kahler2018forecasting,richardson2018prediction}. The underlying idea is to identify a proper proxy (or combinations of proxies) that can be used for the unfolding of patterns and relationships among the parameters of SEP events and their parent solar events, using observational evidence at hand.In turn, such empirical relations point to the underlying physical processes of the SEP generation \citep{balch2008updated}. For example, \cite{laurenza2009technique} provides prognosis of the SEP occurrence based on solar flare location, size (i.e. soft X-ray (SXR) fluence) and evidence of particle escape (i.e. radio fluence at $\sim$1 MHz); \cite{papaioannou2018nowcasting} provides short-term forecasts of the SEP occurrence and the corresponding peak flux utilizing CME characteristics (e.g. width and speed); \cite{kahler2018forecasting} has used the SXR peak flux ratio, following \cite{garcia2004forecasting1}, to establish the probability of SEP occurrence and \cite{richardson2018prediction} used the CME speed and the direction relative to the observer's site (i.e. magnetic connection) to predict the peak intensity of protons. In addition, the work from \cite{posner2007up}, has proven the concept of short-term forecasting of the appearance and intensity of solar ion events by means of relativistic electrons, making use of the higher speed of these electrons propagating from the Sun to 1 AU. Hence, it appears that such empirical or semi-empirical relations, can be used for the forecasting of solar radiation storms. Given the complexity and the incomplete knowledge of the underlying physical mechanisms at work, recent studies attempt to make use of higher order statistical relations \citep[see e.g.][]{papaioannou2018nowcasting2} and machine learning approaches \citep[see][]{lavasa2021assessing} in order to infer the probability of SEP occurrence. Such studies make use of the complete parameter space at hand, utilizing both solar flare and CME characteristics and have shown promising results.      
In this work, the Probabilistic Solar Particle Event Forecasting (PROSPER) Model, is presented. This model is integrated into the Advanced Solar Particle Event Casting System (ASPECS) operational tool that provides predictions of the probability of occurrence of SEP events, the expected peak flux and the resulting SEP time profile (\url{http://phobos-srv.space.noa.gr/}). PROSPER applies a novel, data-driven methodology to predict SEP events probabilistically for a set of integral energies, namely E$>$10, $>$30 and $>$100 MeV\footnote{PROSPER was further extended to E$>$60 MeV but the results are not shown through the ASPECS operational system}. As it is going to be presented in detail, PROSPER takes advantage of the Bayes theorem \citep{1763RSPT...53..370B}, taking into account all observational evidence at hand, without any bias. Bayesian approaches have been increasingly applied to the field of solar physics \citep{arreguirecent} and to solar flare forecasting, in particular \citep[see e.g.][and reference there in]{2004ApJ...609.1134W, 2005SpWea...3.7003W}. The model comes with three modes of operation depending on the available inputs : (a) CME characteristics (width, speed); (b) SF characteristics (longitude, magnitude) and (c) combinations of both CME and SF characteristics. The methodology is outlined and discussed in Section \ref{sec:model}. The application of the methodology to all modes of operation is detailed in Section \ref{sec:results}. Validation results for a set of case studies are presented in Section \ref{sec:validation}. Finally, the results are summarised and discussed in Section \ref{sec:discuss}.

\section{Analysis}
\label{sec:model}
\subsection{Data}
For the development of the PROSPER model, a catalogue of SEP events that includes 314 SEP events from 1984–2013 was utilized \citep[e.g.][]{2016JSWSC...6A..42P}. This SEP event catalogue is based on Geostationary Operational Enviromental Satellite (GOES)/Energetic Particle Sensor (EPS) data (for details on the data set see \cite{sandberg2014cross}) and it includes key information on the proton peak flux and the total fluence of the identified SEP events in the three integral energy channels (namely, E $>$ 10; $>$30; $>$100 MeV)\footnote{PROSPER will further be extended to E$>$300 MeV as part of the ASPECS tool}. It further includes the associated solar sources of the SEP events in terms of solar flares and CME characteristics. In particular, it includes solar SXR flux measurements, provided by GOES, recorded in the same period including 35306 C, M and X class flare events (\url{ftp://ftp.ngdc.noaa.gov/STP/space-weather/solar-data/solar-features/
solar-flares/x-rays/goes/}). From the initial sample of SXR flares 14818 events for which the location was not available were excluded, leading to a sample of 20429 C, M and X class
flares. At the same time, the CME identifications (i.e. plane-of-sky speeds and angular widths - AW) are made by the Large Angle and Spectroscopic Coronagraph (LASCO) \citep{brueckner1995large} onboard the Solar and Heliospheric Observatory (SoHO) in the period from 1997–2013. The initial CME sample consists of 22143 events, which was reduced to 3693 events when association criteria between CMEs and solar flares were applied (see details in \cite{2016JSWSC...6A..42P})\footnote{by construction, CMEs associated to far side solar flares are not included in our sample}. The CME identifications, utilized in this study, are included in the Coordinated Data Analysis Web (CDAW) online CME Catalog \citep[\url{https://cdaw.gsfc.nasa.gov/CME_list/};][]{2009EM&P..104..295G}. 

\subsection{Mathematical formulation of PROSPER} \label{sec:math}
In this section the general methodological approach for the implementation of PROSPER is presented. PROSPER provides both the Probability of SEP Occurrence $P(\rm SEP)$ (subsection \ref{sec:math1}) and the expected Peak Proton Flux (subsection \ref{sub_result2}). The first part, starts with the construction of Cumulative Distribution Functions (CDFs) by the data, continues with the implementation of Probability Density Functions (PDFs) and concludes with the application of the Bayes formula, from which the $P(\rm SEP)$ is directly provided. The second part, presents the establishment of the peak proton flux with the construction of the relevant CDFs.     

\subsubsection{Probability of SEP occurrence} \label{sec:math1}

 This section presents the procedure for the establishment of $P(\rm SEP)$ within PROSPER mode of operation (a), i.e. utilising CME characteristics alone. PROSPER is applied to E$>$10, $>$30, ($>$60) and $>$100 MeV and for three modes of operation i.e., (a), (b) and (c) in Table \ref{tab:table1} which provides an overview of the modes of operation for PROSPER, the inputs used, as well as, the bins applied based on the characteristics of their parent solar events. In particular, CMEs are grouped with respect to their angular width \citep[see details in][]{papaioannou2018nowcasting}, while solar flares by their longitude. Solar events originating in the west of the Sun, as observed by an observer on Earth, are more likely to produce large SEPs detectable at Earth \citep[see e.g.][]{2017SoPh..292..173S}, especially those that lie within W20$^{o}$-W80$^{o}$ \citep[see][and references therein]{2012ApJ...756L..29C}. From our sample roughly half of the SEP events ($\sim$46\%) are associated with a solar flare originating west of W20$^{o}$, hence our choice of positional requirement in the algorithms. Such events are termed as: ``well connected" whereas those east of W20$^{o}$ as ``poorly connected". 

\begin{table}[h!]
\caption{Details on PROSPER's modes of operation, inputs and initial binning}
\begin{tabular}{lccc}
\hline
\bf{Mode of operation} & \bf{Inputs} &  \bf{Bins} & \bf{Resulting function}\\
                       &             &            &     \bf{per bin}       \\ \hline
 (a) CME &  speed      &  continuous speed       & \\
           &              &   AW = 360$^{o}$ (Halo - H) & $f(V_{CME})$ [3]\\ 
         &     \& AW          &     120$^{o}$ $\le$ AW $<$ 360$^{o}$ (Partial Halo - PH) &\\
         &              &      AW $<$ 120$^{o}$ (Non-Halo - NH) & \\\hline
          (b) flare & Soft X-ray & continuous magnitude & \\
 & flare magnitude  & lon. $\ge$20$^{o}$ (well connected - WC) & $f(F_{SXR})$ [2] \\ 
           &         \& longitude & lon. $<$20$^{o}$ (poorly connected - PC)  & \\  \hline
 (c) flare \& CME & all of the above & [H or PH or NH] \& WC &  $f(V_{CME}, F_{SXR})$ [6]\\ 
                  &                  & [H or PH or NH] \& PC  &                       \\  \hline
\textit{Acronyms \& bins}&  &  & \\
                  \textit{AW = angular width} & \textit{H = Halo} & \textit{PH= Partial Halo} & \textit{NH = Non-Halo} \\
                   \textit{lon.= longitude} & \textit{WC = well connected} & \textit{PC = poorly connected} \\
                   \textit{with respect to the} & (\textit{lon. $\geq$20$^{o}$}) & (\textit{lon. $<$20$^{o}$}) \\
                   \textit{Sun-Earth line} & & \\
\end{tabular}%
\label{tab:table1}
\end{table}

Column 1 denotes the mode of operation, column 2 declares the inputs used per mode, column 3 details the initial bins applied to the data based on the inputs and column 4 presents the resulting number of functions per bin. The numbers in the brackets of column 4 denote the number of resulting functions per mode and integral energy. These functions are the P($\rm SEP$) ones which are going to be detailed in the following steps for each of the selections. Hence for each integral energy (e.g. E$>$10 MeV) there are 3 $f(V_{CME})$, 2 $f(F_{SXR})$ and 6 $f(V_{CME}, F_{SXR})$ functions in accordance to column 3. PROSPER is applied independently to each of the integral energies considered.

\paragraph{\textbf{Implementation of Cumulative Distribution Functions}}

First, we bin the data in accordance to Table \ref{tab:table1}. The empirical cumulative distribution function (CDF) for each sub-sample was constructed by sorting the events in the ascending order of the continuous parameter (CME speed and/or solar flare magnitude; see also Table \ref{tab:table1}), and using the standard fractional ordinates as probabilities. As parent solar events (i.e. parameterized by flares and/or CMEs) can be either associated with SEP events or not, it is possible to construct two CDFs: one for each sample, i.e. one for all parent solar events (i.e. parameter, $\Lambda$) of the specific binning: $P(\Lambda<\lambda)$\footnote{$\lambda$ refers to the continuous parameter of Table \ref{tab:table1} per case} and one for all parent solar events that were associated with an SEP event in that bin, $P(\Lambda<\lambda\,|\,\text{SEP})$.\\

Next, the resulting (filtered) data were fit with a log-normal distribution \citep[see e.g.][]{2018JSWSC...8A...4R}, which was found to be the most robust of the flux distributions applied to SEP fluxes for specification models, in this work. This is given by the following equation:

\begin{equation} \label{eq:1}
F(\lambda) = \frac{1}{2} \left[1+\mathrm{erf}\left(\frac{\log_{10}(\lambda)-\mu}{\sigma \sqrt{2}}\right)\right],    
\end{equation}

\noindent where $F(\lambda)$ is the probability of $\Lambda$ being lower than $\lambda$, $\mathrm{erf}(x)$ is the error function, and $\mu$ and $\sigma$ are the mean and standard deviation of $\log_{10}(\lambda)$, respectively. 

\paragraph{\textbf{Implementation of the Probability Density Functions}}

These CDFs were consequently used in order to construct probabilities for which the CME speed/flare magnitude, $\lambda$ lies within a certain range defined by an upper and a lower limit, i.e. $\in[\lambda_{1},\lambda_{2}]$. The resulting independent probability of \textit{observing} $\alpha$, i.e. the probability of $\Lambda$ to fall within a certain range of values is given by the subtraction of two CDFs constructed by all data points in a sample one for the lower limit, i.e. $\Lambda<\lambda_{1}$ and one for the upper one, i.e. $\Lambda<\lambda_{2}$ leading to $P\left(\Lambda<\lambda_{2}\right)$ and $P\left(\Lambda<\lambda_{1}\right)$. The subtraction of which results in: 

\begin{equation} \label{eq:2}
 P(\alpha) =P\left(\Lambda \in[\lambda_{1},\lambda_{2}]\right)=P\left(\Lambda<\lambda_{2}\right)-P\left(\Lambda<\lambda_{1}\right).  
\end{equation}

The independent probability of \textit{observing} $\beta$, i.e. the probability of SEP events in a sample, is based on the number of SEP associated events ($N_{SEP}$) and the total number of events ($N_{total}$) in the sample and is defined as:

\begin{equation} \label{eq:2a}
 P(\beta) = P(SEP) = N_{SEP} / N_{total}.
\end{equation}

 The conditional probability of \textit{observing} $\alpha$ under the condition of \textit{observing} $\beta$, $P(\alpha\,|\,\beta)$, i.e. the probability of the $\Lambda$ to fall within a certain range of $\lambda$, under the condition that these values refer to SEP events. This is also given as the subtraction of two CDFs. In particular, the first CDF refers to the lower limit, i.e. $\Lambda<\lambda_{1}\,|\,\text{SEP}$ and the second to the upper one, i.e. $\Lambda<\lambda_{2}\,|\,\text{SEP}$ leading to $P\left(\Lambda<\lambda_{1}\,|\,\text{SEP}\right)$ and $P\left(\Lambda<\lambda_{2}\,|\,\text{SEP}\right)$. As a result the conditional probability is defined as: 
 
 \begin{equation} \label{eq:3}
  P(\alpha\,|\,\beta) =P\left(\Lambda\in[\lambda_{1},\lambda_{2}]\,|\,\text{SEP}\right)=P\left(\Lambda<\lambda_{2}\,|\,\text{SEP}\right)-P\left(\Lambda<\lambda_{1}\,|\,\text{SEP}\right).
  \end{equation}

Using the Bayes theorem \citep{1763RSPT...53..370B, joyce1999foundations} one may define the inverse conditional probability:

\begin{equation}\label{eq:4}
P(\beta\,|\,\alpha) = P\left(\text{SEP}\,|\,\Lambda, \in[\lambda_{1},\lambda_{2}]\right)
\end{equation}

\noindent which is a PDF and provides the probability of SEP occurrence when $\Lambda$ $\in[\lambda_{1},\lambda_{2}]$. The Bayes formula dictates that: 

\begin{equation} \label{eq:5}
    P(\beta\,|\,\alpha)=\frac{P(\alpha\,|\,\beta)P(\beta)}{P(\alpha)}
\end{equation}

Merging Equations \ref{eq:2}--\ref{eq:5}, the following formula is obtained:

\begin{equation} \label{eq:6}
    P\left(\text{SEP}\,|\,\Lambda \in[\lambda_{1},\lambda_{2}]\right) = \frac{\left[P\left(\Lambda<\lambda_{2}\,|\,\text{SEP}\right)-P\left(\Lambda<\lambda_{1}\,|\,\text{SEP}\right)\right]P(\text{SEP)}}{P\left(\Lambda<\lambda_{2}\right)-P\left(\Lambda<\lambda_{1}\right)}.
\end{equation}

Hence, Equation \ref{eq:6} shows the probability to have an SEP event under the condition that $\Lambda \in[\lambda_{1},\lambda_{2}]$. For example, in the case of CMEs, substituting $\Lambda$ with $V_{CME}$, $\lambda_{1}$ with $V_{1}$ and $\lambda_{2}$ with $V_{2}$, Equation \ref{eq:6} becomes:

\begin{equation} \label{eq:7}
    P\left(\text{SEP}\,|\,V_{\mathrm {CME}}\in\left[V_{1},V_{2}\right]\right) = \frac{\left[P\left(V_{{\rm CME}}<V_{2}\,|\,\text{SEP}\right)-P\left(V_{{\rm CME}}<V_{1}\,|\,\text{SEP}\right)\right]P(\text{SEP})}{P\left(V_{{\rm CME}}<V_{2}\right)-P\left(V_{{\rm CME}}<V_{1}\right)}.
\end{equation}

Moreover, setting $V_{1}=V$ and $V_{2}=V+{\rm dV}$ and dividing both nominator and denominator by $dV$ while letting ${\rm d}V\to0$ results to:

\begin{equation}\label{eq:8}
    P\left(\text{SEP}\,|\,V_{{\rm CME}}=V\right) =\frac{P'\left(V_{{\rm CME}}<V\,|\,{\rm SEP}\right)}{P'\left(V_{{\rm CME}}<V\right)}P({\rm SEP})\\
  =\frac{f_{{\rm SEP}}(V)}{f(V)}P({\rm SEP}).
\end{equation}

Both Equations \ref{eq:7} and \ref{eq:8} provide the probability of occurrence of an SEP event given a CME of speed, $V$. \\

\begin{figure}[ht]
\centering
\includegraphics[width=0.69\textwidth]{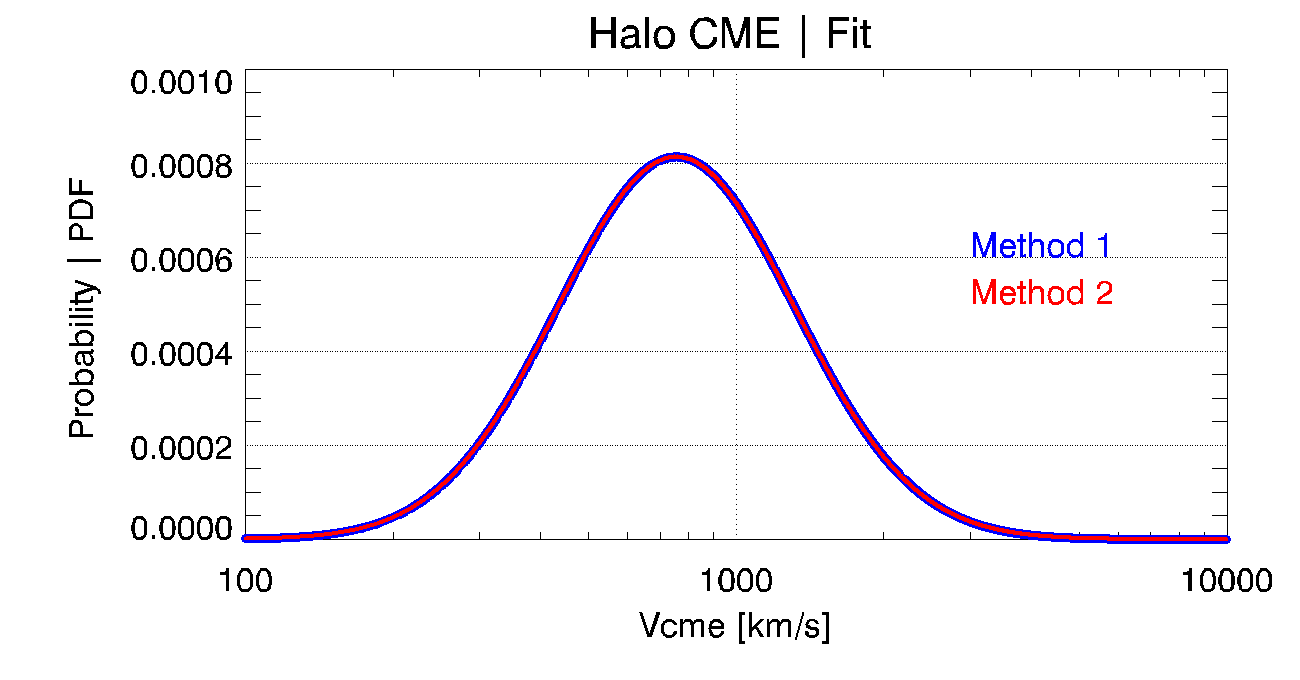}
\caption{An illustration of the two methods (\textit{Method 1}; (blue color) and \textit{Method 2}; (red color)), as a superposition investigated in this work. Both fits in the plots correspond to the Probability Distribution Functions (PDFs) for the case of Halo CMEs, with respect to the CME speed $V_{CME}$. See text for details}
\label{fig:method}
\end{figure}

In Equation \ref{eq:8}, $f(V)=P'(V_{{\rm CME}}<V)$ is the probability density function (PDF) of the CME speeds for the binned sub-sample of choice (see Table \ref{tab:table1}) and $f_{{\rm SEP}}(V)$ is the PDF for those CMEs of the bin that are associated with SEP events. In order to construct the PDFs (i.e. $f(V)$ \& $f_{{\rm SEP}}(V)$) from the relevant CDFs, there are two ways: (a) apply numerical differentiation (labeled as \textit{Method 1}) and (b) fit the data points with an empirical CDF and directly take the derivatives of the relevant distributions, as a function of the CME speed ($V_{CME}, [km/s]$) (labeled as \textit{Method 2}). As it can be seen in Figure \ref{fig:method}, for the case of Halo CMEs and E$>$10 MeV SEP events, the output is equivalent. Therefore, from the empirical CDFs, which are based on data, and the applied log-normal fits, one can directly obtain the relevant PDFs and from the application of the Bayes formula (Equation \ref{eq:8}) obtain an analytical expression that answers to the question: \textit{What is the probability that an SEP event [P(SEP)] will occur if one knows the CME speed ($V_{CME}$) (and width) of the driving CME} ? 

Equation \ref{eq:8} also stands for the case of solar flares. In this case, the application of the Bayes formula based on solar flare longitude and flux (F) results in:

\begin{equation}\label{eq:8a}
    P\left(\text{SEP}\,|\,F_{{\rm SXR}}=F\right) =\frac{P'\left(F_{{\rm SXR}}<F\,|\,{\rm SEP}\right)}{P'\left(F_{{\rm SXR}}<F\right)}P({\rm SEP})\\
  =\frac{f_{{\rm SEP}}(F)}{f(F)}P({\rm SEP}).
\end{equation}

At this point, it is noted that Equation \ref{eq:5} acquires very large values and tends to infinity as $P(\alpha)\rightarrow 0$. For this reason, using some elementary mathematical properties of sets and probabilities, Equation \ref{eq:5} can be written in a form which is more stable and appropriate for numerical calculations, as:

\begin{equation} \label{eq:9}
    P(\beta\,|\,\alpha)=\frac{P(\alpha\,|\,\beta)P(\beta)}{P(\alpha|\beta)P(\beta)+P(\alpha\,|\,\beta^{c})P(\beta^{c})},
\end{equation}

\noindent with $\beta^{c}$ being the complementary event of $\beta$ \citep{maritz2018empirical}. Although the denominators of Eq. \ref{eq:5} and Eq. \ref{eq:9} are equivalent, the former $\rightarrow 0$ faster than the latter one. As a result Equation \ref{eq:9} has an optimal numerical behaviour and allows one to obtain outputs at a wider range compared to Equation \ref{eq:5} before the denominator of Equation \ref{eq:5} $\rightarrow 0$.
Therefore, the formulas used in PROSPER calculations are the more complex expressions of Equations \ref{eq:6}--\ref{eq:8a} based on substitution of Equations \ref{eq:2}--\ref{eq:4} in Equation \ref{eq:9}.

Additionally, in the case of a solar flare and an associated CME the Bayes formula used in PROSPER gets the general form:

\begin{equation} \label{eq:7a}
    P(\beta_{i,j}\,|\,V,F)=\frac{f_{\beta_{i}}(V)f_{\beta_{j}}(F)P(\beta_{i,j})}{[f_{\beta_{i}}(V)f_{\beta_{j}}(F)P(\beta_{i,j})+f_{\beta_{i}^{c}}(V)f_{\beta_{j}^{c}}(F)P(\beta_{i,j}^{c})]}.
\end{equation}

where $i$ refers to CME $AW$, i.e. Halo, Partial Halo, Non Halo and $j$ to the solar flare longitude (``well" and/or ``poorly" connected); see details per selection in Table \ref{tab:table1}). $ P(\beta_{i,j}$ and $P(\beta_{i,j}^{c})$ are calculated directly by the measurements. 

\subsubsection{Peak flux estimation} \label{sub_result2}

PROSPER provides a peak flux estimation for a given (user-defined) confidence based on the $P(\text{SEP})$ described here above. First the data are filtered, based on the $P(\rm SEP)$ that was derived in subsection \ref{sec:math1}. In particular, once $P(\rm SEP)$ is obtained, peak flux data are filtered based on the integral energy that the SEP event is expected to reach. That said, if i.e. an SEP event is expected to reach E$>$100 MeV only the events from the list with significant peak fluxes reaching E$>$100 MeV are employed across all integral energies. The same stands for filtering the data samples based on other integral energies based on the obtained $P(\rm SEP)$ (i.e. if an event is expected to reach E$>$30 MeV). At any case, the presence of a high energy population is an important filter for lower energy fluxes since it ensures spectral coherence. Next, additional binning is applied using the available characteristics of the parent solar events (i.e. SF magnitude \& location, CME speed \& width). Once the data samples are retrieved we utilize the tabulated peak proton flux (PPF) for the SEP events in the list employed in \citet{2016JSWSC...6A..42P} in order to derive the relevant distributions.

The filtered peak fluxes were fit with the exponential cut-off power law that was found to be the most robust of the flux distributions applied for specification models (consistent with findings from both \cite{jiggens2018updated} and \cite{2018JSWSC...8A...4R}). This is given by the following equation:

\begin{equation} \label{eq:11}
    P\left(\rm F_{p}\ge PPF_{0} \,|\, SEP \right) = 1- \frac{PPF_{0}^{-\gamma}\exp\left(\frac{x_{low}}{x_{lim}}\right)}{x_{low}^{-\gamma}\exp\left(\frac{PPF_{0}}{x_{lim}}\right)},
\end{equation}

\noindent with $\gamma$ being the power-law exponent, $x_{low}$ a parameter related to the lower limit of the $F_{P}$ distribution, and $x_{lim}$ a parameter related to the upper limit of the $F_{P}$ distribution.

Equation \ref{eq:11} answers the question: \textit{What is the expected probability that the peak flux will exceed a certain threshold, ($PPF_{0}$) for a given integral energy, provided that it is certain an SEP event will occur}? From Equation \ref{eq:11}, one can find the expected peak flux that corresponds to a specific probability threshold, solving Equation \ref{eq:11} for $PPF_{0}$ for a given value of $ P\left(\rm F_{P}\ge PPF_{0} \,|\, SEP \right) = P_{thres}$. Following this, the $P(\rm SEP\,|\, \Lambda = \lambda_{0})$ is used to modulate the prediction of the expected peak flux. In essence, a weighted average is employed, driven by the obtained $P(\rm SEP\,|\, \Lambda = \lambda_{0})$. There are two components in this averaging: the first one is the $PPF_{0}$ for a specific user defined threshold multiplied by $P(\rm SEP\,|\, \Lambda = \lambda_{0})$ and the second one is the background value of the specific integral energy multiplied by the 1 - $P(\rm SEP\,|\, \Lambda = \lambda_{0})$. The averaging leads to $PPF_{thres}$ and the obtained $P(\rm SEP\,|\, \Lambda = \lambda_{0})$ is used as a weight. To this end PROSPER provides $F_{P\rm thres}$ obtained from Equation \ref{eq:12}:

\begin{equation} \label{eq:12}
   F_{P\rm thres}=PPF_{0}  \cdot P(\rm SEP \,|\, \Lambda = \lambda_{0}) + \textit{background flux} \cdot \left[1-P(\rm SEP\,|\, \Lambda = \lambda_{0})\right]
\end{equation}

In the limiting cases of the binary extremes of $P(\rm SEP | \Lambda = \lambda_{0})$, the following expression is obtained:

\begin{equation} \label{eq:14}
  F_{P\rm thres} = \begin{cases}
PPF_{0} &\mbox{, } P(\rm SEP \,|\, \Lambda = \lambda_{0}) = 1 \\
\textit{background flux} &\mbox{, } P(\rm SEP \,|\, \Lambda = \lambda_{0})= 0\\
\end{cases}
\end{equation}

\begin{figure}[ht]
\centering
\begin{minipage}[b]{0.48\linewidth}
\includegraphics[width=8cm,height=4.5cm]{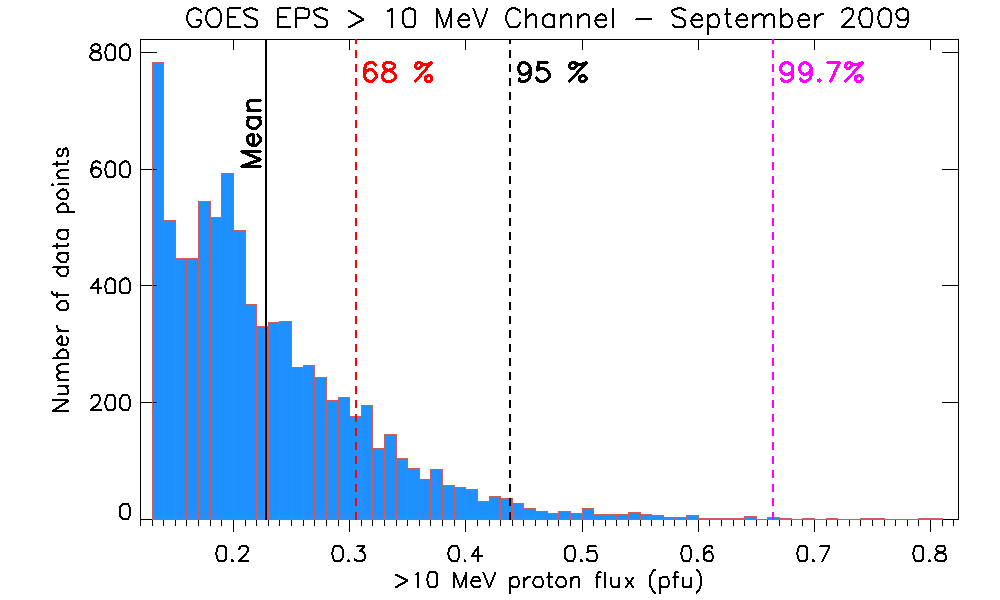}
\end{minipage}
\begin{minipage}[b]{0.45\linewidth}
\includegraphics[width=8cm,height=4.5cm]{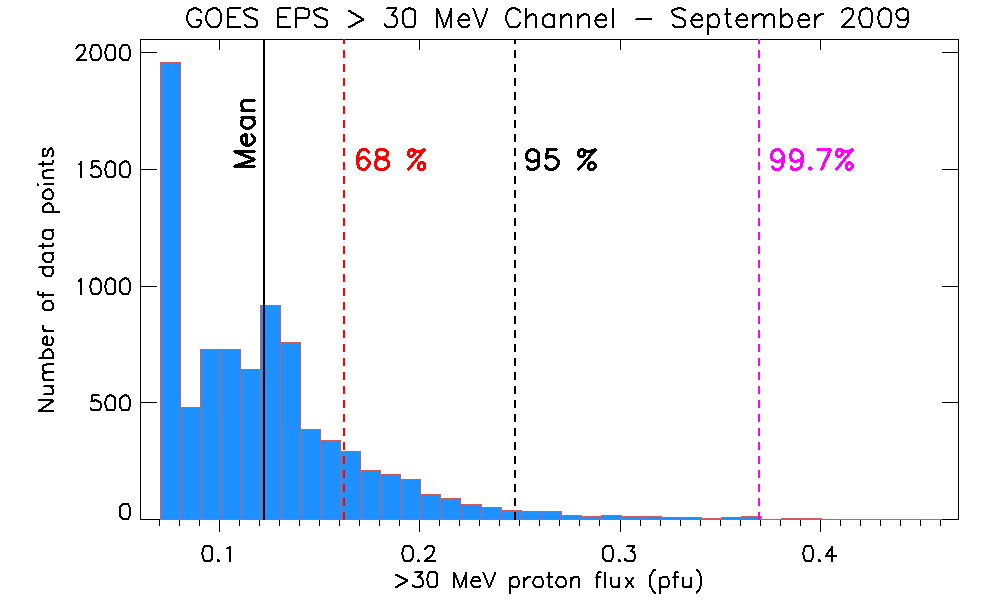}
\end{minipage}
\begin{minipage}[b]{0.48\linewidth}
\includegraphics[width=8cm,height=4.5cm]{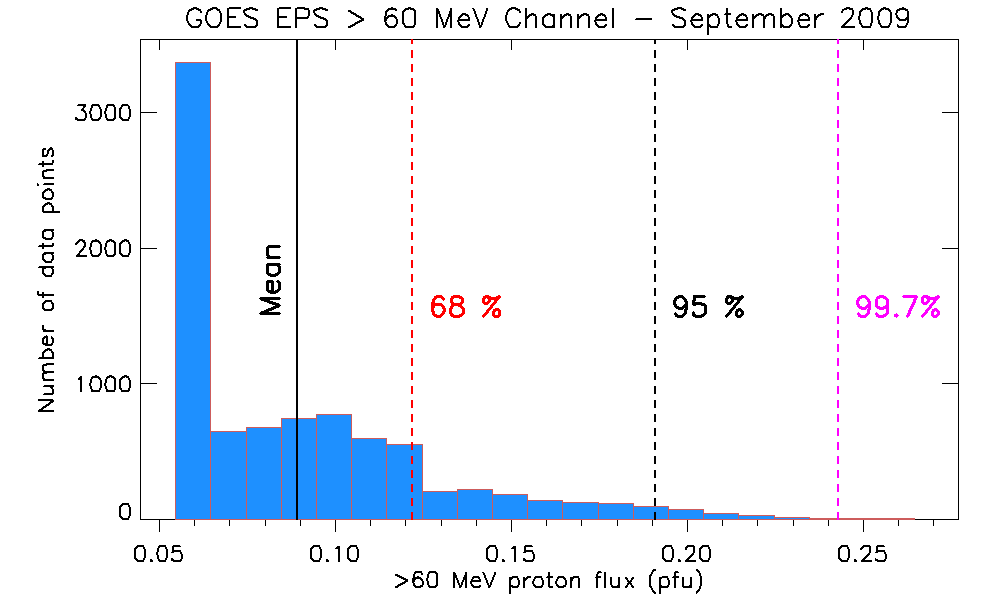}
\end{minipage}
\begin{minipage}[b]{0.45\linewidth}
\includegraphics[width=8cm,height=4.5cm]{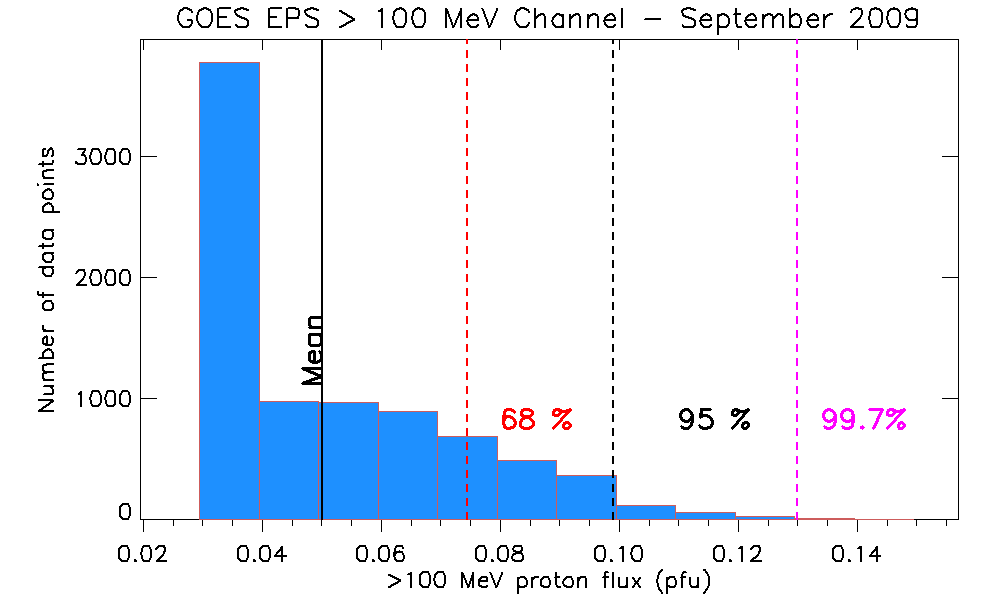}
\end{minipage}
\quad
\caption{The distributions of the GOES/EPS data per integral energy of interest for September 2009. The relevant percentiles of 68\%-95\% \& 99.7\% of the distributions are depicted as vertical dashed lines. The mean values of these distributions (presented as solid vertical lines) are used as the background level per channel in Equation \ref{eq:12}.}
\label{fig:bcg}
\end{figure}

The \textit{background flux} was calculated on the basis of the recorded intensity of each integral energy of interest (i.e. E$>$10-; $>$30-; and $>$100 MeV)\footnote{data taken from \url{https://satdat.ngdc.noaa.gov/sem/goes/data/avg/}} and corresponds to the mean value obtained in September 2009 (e.g. during the solar minimum), which is: 0.23, 0.122, 0.089 and 0.050 pfu, respectively (see Figure \ref{fig:bcg}). Evidently, depending on the CL, $thres$, of choice (by default in the ASPECS tool, PROSPER provides $thres$=50\%, and 90\%, which were defined by the user community), and thus the model provides $PPF_{\rm 50}$ and $PPF_{\rm 90}$. 

\section{Results}
\label{sec:results}
\subsection{Coronal Mass Ejections} \label{sub_res_CME}

In the case of CMEs, PROSPER's concept was applied across three CME AW bins (see Table \ref{tab:table1}), for all integral energies of our database (E $>$ 10; $>$30; ($>$60) and $>$100 MeV) and for both the distribution of all CMEs speeds ($V_{CME}$) in each sub-sample, as well as, the distribution of the speeds of CMEs related to SEPs. Figure \ref{fig:CDFs_fit} depicts the obtained empirical CDFs from the data (i.e. $P(V_{{\rm CME}}<V)$). From top to bottom, on the left hand side, the column refers to \textit{all} Halo, \textit{all} Partial Halo and \textit{all} Non Halo CMEs of our sample. Additionally, similar plots for $P(V_{{\rm CME}}<V\,|\,{\rm SEP})$ for an integral energy of E$>$10 MeV is provided on the right hand side of the plot, for each case, following the same labeling. Each sub-plot of Figure \ref{fig:CDFs_fit} presents the data points (in black color) for each distribution. Additionally, the log-normal fit (from Equation \ref{eq:1}) to these points is presented as a red-line, with the fraction of the data (in \%) represented by the fit. Finally, the mean absolute error (MAE)\footnote{MAE=$\sum_{i=1}^{n} \,|\, y_{i} -x_{i} \,|\, /n $, with $n$ being the number of pairs, $y$ and $x$ the fitted and the actual value, respectively} was calculated at each case and is imprinted on every plot. Although the related results are not shown in Figure \ref{fig:CDFs_fit}, it is noteworthy that when shifting to higher energies, i.e. from E$>$10 MeV to E$>$100 MeV, a reduction in the number of SEP events that reach higher energies is apparent. In turn, fits for these energies rely on fewer points. 

\begin{figure}
   \centering
   \includegraphics[width=12cm]{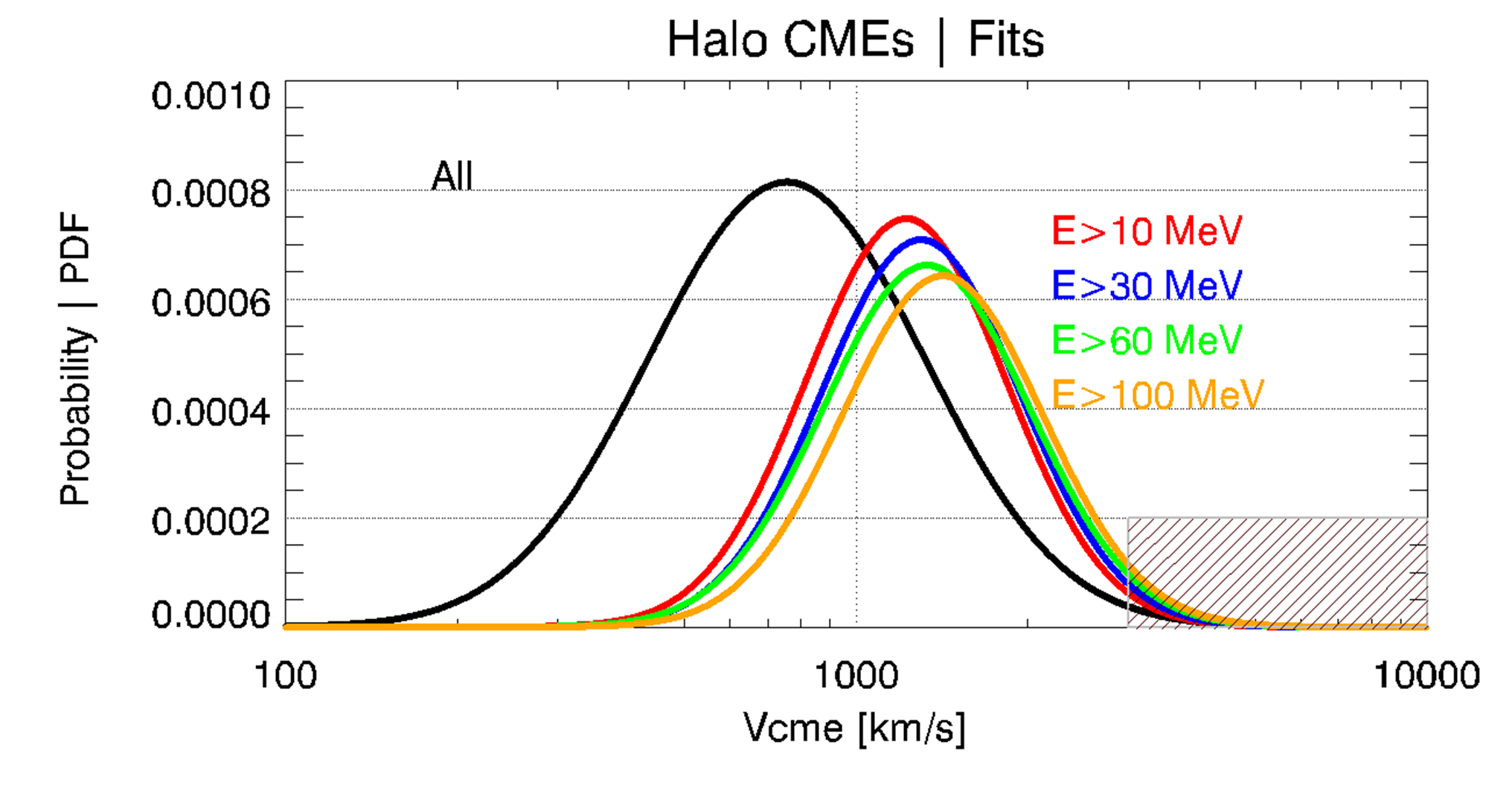}
   \includegraphics[width=12cm]{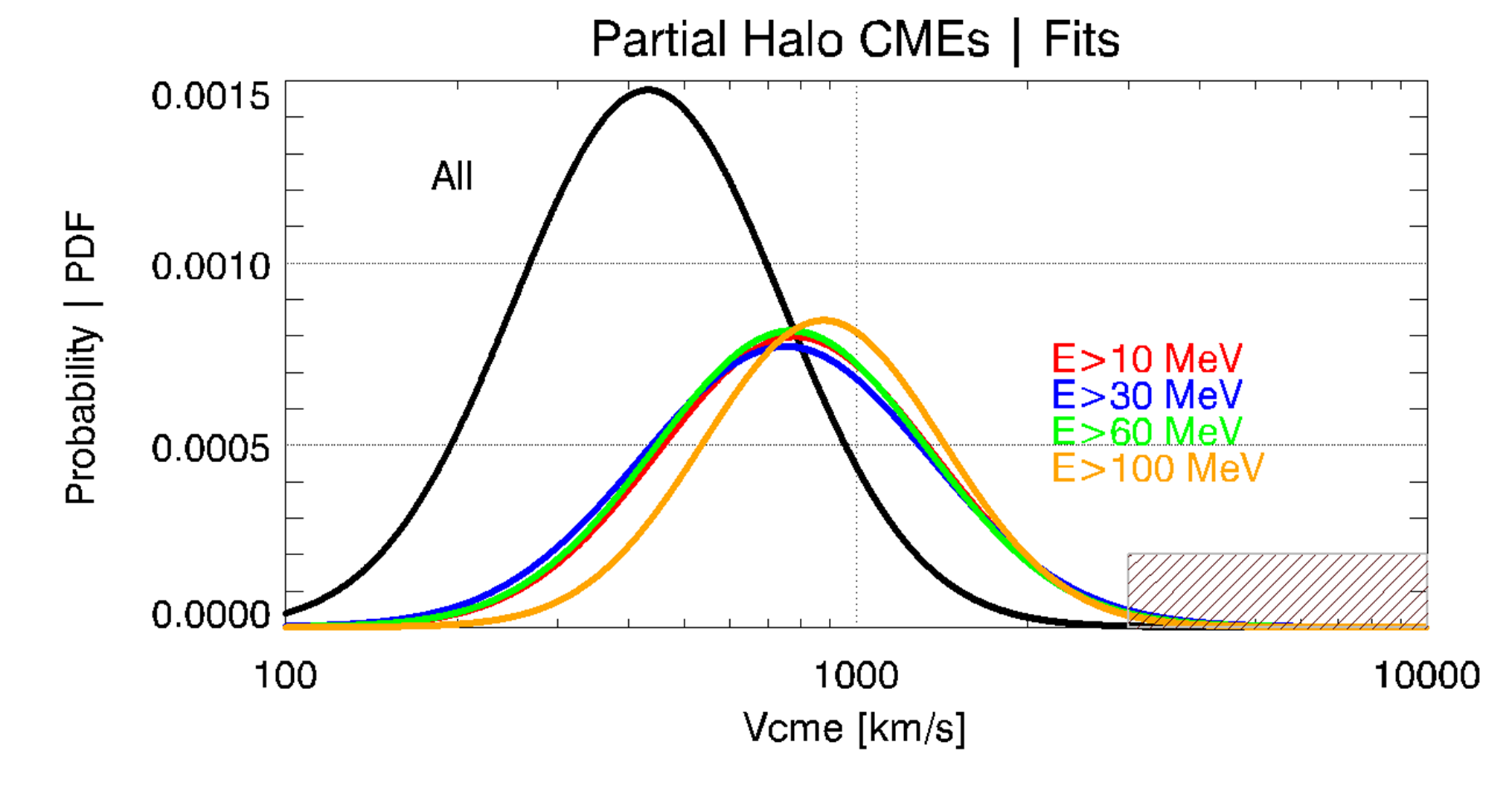}
   \includegraphics[width=12cm]{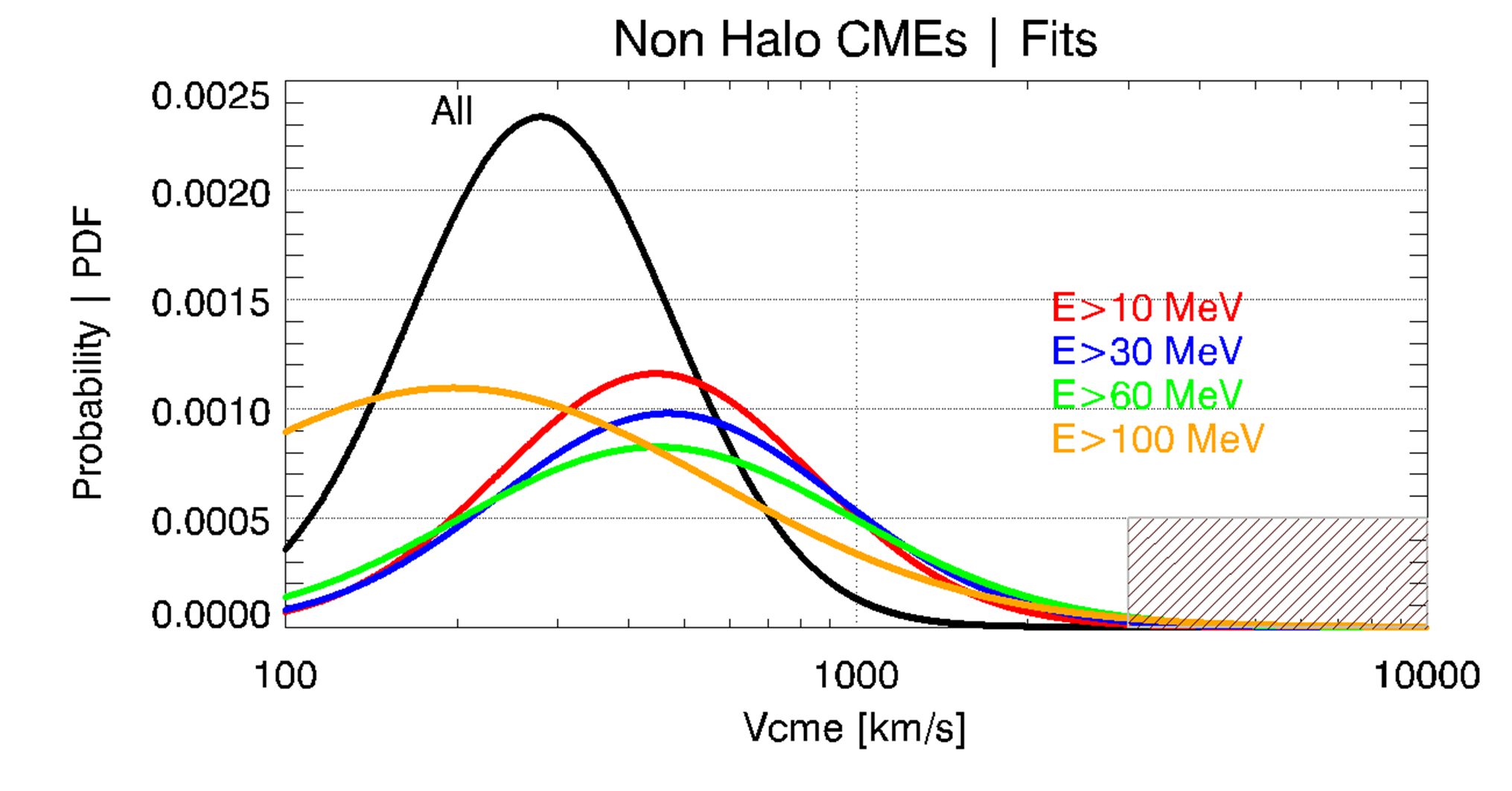}
      \caption{The Probability Distribution Functions (PDFs) for the case of Halo (top panel), Partial Halo (middle panel) and Non Halo (bottom panel) CMEs,  as derived by the CDFs, applying \textit{Method 2}. The black color represents all CMEs in the sample, whereas  CMEs associated with SEP events for different integral energies are colour coded. The gray border hatched rectangle area provides the limit for the $V_{CME}$.}
      \label{fig:pdf}
   \end{figure}

A combined representation of these CDFs, but for all integral energies of interest (e.g. E$>$10; $>$30; ($>$60) and $>$100 MeV) is given in Figure \ref{fig:cdfs_2}\footnote{Although E$>$60 MeV is not implemented in the ASPECS tool, we present the obtained fits for $P( \rm SEP)$ in this and consequent Figures for consistency}. Each panel provides the fitted CDFs for the $P(V_{{\rm CME}}<V)$ (black line) and  $P(V_{{\rm CME}}<V\,|\,{\rm SEP})$ distributions - the latter for each integral energy, color coded as: E$>$10 MeV - red line;  E$>$30 MeV - blue line; E$>$60 MeV - green line and E$>$100 MeV - orange line. From top to bottom, panels refer to Halo, Partial Halo and Non Halo CMEs, respectively. The PDFs from the derivatives of the distributions $f(V)$ and $f_{{\rm SEP}}(V)$ are consequently presented in Figure \ref{fig:pdf}, following the same reasoning (and labeling) as in Figure \ref{fig:cdfs_2}. It should be noted that the X-axis (i.e. $V_{CME}$) in all panels extend to a simulated value of 10000 km/s. However, the actual data point to an upper limit $V_{CME}$ of $\sim$3000 km/s (see Figure \ref{fig:CDFs_fit}). That said, the derived Bayes $P(\rm SEP)$ is obtained up to that $V_{CME}$. However, we present the whole evolution of the fit and we overlay a gray border hatched rectangle area after that limit. 

The final step in this formulation is to employ the Bayes formula (Equation \ref{eq:8}) and retrieve the fits that reply to the question: \textit{what is the probability of SEP occurrence from a CME with known width (AW) and speed ($V_{CME}$)}? These fits are presented in Figure \ref{fig:bayes-fits}. Each panel corresponds to one integral energy of interest, and each line within every panel refers to Halo (blue line); Partial Halo (red line) and Non Halo (green line) CMEs. Hence, for each integral energy a total of 3 $f(V_{CME})$ functions are obtained (see Table \ref{tab:table1}). The abscissa of Figure \ref{fig:bayes-fits} gives the $V_{CME}$ and the ordinate provides the probability of SEP occurrence. Hence, for a CME with known AW, one may select the proper fit and from the known $V_{CME}$ directly obtain the expected probability, $P(\rm SEP)$. It can be noted that up to the limit of $V_{CME}$ $\sim$3000 km/s for any given CME speed, the largest $P(\rm SEP)$ is obtained for Halo (blue line), followed by Partial Halo (red line) and Non Halo (green line). Moreover, both the red and the green curve would have a similar behaviour as the blue curve, if the X-axis would have been expanded to $>$10000 km/s. In addition, when crossing the $V_{CME}$ $\sim$ 3000 km/s limit it can be seen that Partial Halo (red line) and Non Halo (green line) CMEs lead to a higher $P(\rm SEP)$ compared to Halo CMEs (blue line). This is the output of the application of Equation \ref{eq:8} and it points to the possibility to get a higher $P(\rm SEP)$ for Partial and Non Halo CMEs provided that these CMEs are faster than any CME ever observed. Nonetheless, the application of an upper limit for $V_{CME}$ that is inferred by the actual sample used, results into higher probabilities for Halo CMEs up to that limit. That said, all observed CMEs with $V_{CME} >$ 3000 km/s are treated as having a $V_{CME}$ = 3000 km/s.     

As noted in subsection \ref{sub_result2}, based on user consultations, a focused probability range from 50-90\% has been selected, while taking into account, as a filter, the probability to derive an SEP event at a respective energy. If an SEP event is expected to reach e.g. E$>$100 MeV the data for all energies is binned based on this energy filtering and consequently the probabilities are derived based on CME speed and width. The obtained results for the case of CME inputs are depicted in Figure \ref{fig:PPF}. There are three different filters based on the predicted integral energy that the particles will reach and different bins based on the CME characteristics. In particular, panels (a) and (b) of Fig.~\ref{fig:PPF} refer to SEP events that are expected to reach an integral energy of E$>$10 MeV. The red (blue) lines depict the fast (slow) bins and each of the panels refer to Halo (panel (a)) and Not Halo (panel (b)) CMEs. In addition, panels (c)-(f) of Fig.~\ref{fig:PPF} refer to SEP events that are expected to reach an integral energy of E$>$30 MeV. The blue (orange) lines in each of the panels present the E$>$10 (E$>$30) MeV fits, whereas each of the panels stands for a different selection of AW and CME speed -- panel (c): Halo and slow; panel (d) Halo and fast; panel (e) not Halo and slow and panel (f) not Halo and fast CMEs. Finally, panel (g) of Fig.~\ref{fig:PPF} represents the fits for a filtering at E$>$100 MeV and each line corresponds to an integral energy of interest color coded as: red (E$>$10 MeV); green (E$>$30 MeV) and blue (E$>$100 MeV). All bins and selections are detailed in Table \ref{tab:cme_peak}.

\begin{table}[h!]
\caption{Details on PROSPER's peak flux fits for the CME bins, derived by Equation \ref{eq:11}. The filtering on the integral energy is first applied and then there are two bins. One for the width of the CME (i.e. AW = 360$^{o}$Halo or AW $<$360$^{o}$Not Halo) and one for the CME speed (i.e. $V_{CME}$ being $<$ or $\ge$ 1250 km/s). These fits are presented in Figure \ref{fig:PPF}.}
\begin{tabular}{l l|ll}
\hline
\hline
\multicolumn{4}{l}{\bf{if an event reaches E$>$10 MeV}} \\
\hline
\hline
 AW = 360$^{o}$ (Halo - H)   &             &      AW $<$ 360$^{o}$ (Not Halo)     &            \\ \hline
 $V_{CME}$ $<$ 1250 km/s     &  $V_{CME}$ $\ge$ 1250 km/s &   $V_{CME}$ $<$ 1250 km/s     &  $V_{CME}$ $\ge$ 1250 km/s  \\ \hline
 $\gamma$  = 0.36 &       $\gamma$  = 0.26   &   $\gamma$  = 0.53 &       $\gamma$  = 0.32 \\
 $x_{low}$ = 3.68E+00     &    $x_{low}$ = 6.38E+00   &  $x_{low}$ = 2.50E+00    &   $x_{low}$ = 4.00E+00 \\ 
 $x_{lim}$ = 5.37E+01     &    $x_{lim}$ = 1.18E+04   &   $x_{lim}$ = 1.49E+03   &   $x_{lim}$ = 8.95E+03\\ \hline
 \hline
 \multicolumn{4}{l}{\bf{if an event reaches E$>$30 MeV}} \\
\hline
\hline
 AW = 360$^{o}$ (Halo - H)   &             &  AW = 360$^{o}$ (Halo - H)       &            \\ \hline
 $V_{CME}$ $<$ 1250 km/s     &  $V_{CME}$ $\ge$ 1250 km/s &$V_{CME}$ $<$ 1250 km/s & $V_{CME}$ $\ge$ 1250 km/s \\ \hline
 \bf{E$>$10 MeV} &    &   \bf{E$>$30 MeV} & \\ \hline
 $\gamma$ = 0.30 &       $\gamma$  = 0.26   &  $\gamma$ = 1.09 &       $\gamma$  = 0.30 \\
 $x_{low}$ = 3.68E+00     &    $x_{low}$ = 6.38E+00   &  $x_{low}$ = 1.23E+00     &    $x_{low}$ = 1.24E+00 \\ 
 $x_{lim}$ = 5.27E+01     &     $x_{lim}$ = 1.19E+04   &  $x_{lim}$ = 1.54E+02    &    $x_{lim}$ = 2.27E+03 \\ \hline
 AW $<$ 360$^{o}$ (Not Halo)   &             &    AW $<$ 360$^{o}$ (Not Halo)        &            \\ \hline
 $V_{CME}$ $<$ 1250 km/s     &  $V_{CME}$ $\ge$ 1250 km/s &  $V_{CME}$ $<$ 1250 km/s     &  $V_{CME}$ $\ge$ 1250 km/s\\ \hline
 $\gamma$  = 0.45 &       $\gamma$  = 0.30   &  $\gamma$  = 0.58 &       $\gamma$  = 0.32 \\
 $x_{low}$ = 3.94E+00    &   $x_{low}$ = 5.72E+00    &  $x_{low}$ = 9.06E-01    &   $x_{low}$ = 6.71E-01 \\ 
 $x_{lim}$ = 9.96E+02   &   $x_{lim}$ = 8.55E+03   &  $x_{lim}$ = 3.23E+03   &   $x_{lim}$ = 3.51E+03 \\ \hline
 \hline
 \multicolumn{4}{l}{\bf{if an event reaches E$>$100 MeV}} \\
\hline
\hline
 \multicolumn{4}{l}{No bin}           \\ \hline
 \multicolumn{4}{l}{All data points} \\ \hline
  \multicolumn{1}{l|}{\bf{E$>$10 MeV}} & \bf{E$>$30 MeV} & \bf{E$>$100 MeV} \\ \hline
  \multicolumn{1}{l|}{$\gamma$  = 0.26} &  $\gamma$  = 0.28  &   $\gamma$  = 0.41 \\
  \multicolumn{1}{l|}{$x_{low}$ = 9.50E+00}    &   $x_{low}$ = 2.30E+00    &  $x_{low}$ = 3.02E-01\\ 
  \multicolumn{1}{l|}{$x_{lim}$ = 1.19E+04}     &   $x_{lim}$ = 2.04E+03   &  $x_{lim}$ = 2.58E+02\\ \hline
 \end{tabular}%
\label{tab:cme_peak}
\end{table}

\subsection{Solar Flares} \label{sub_res_sf}
A similar probabilistic approach was further applied for solar flare inputs  replacing V$_{\textrm{CME}}$ with SXR peak flux and imposing two longitudinal source location ranges: well connected ($longitude \ge 20^{o}$), poorly connected ($longitude < 20^{o}$) in place of CME angular width. Applying the same methodology we obtained the following fits presented in Figure \ref{fig:pf}. These fits are similar to Figure \ref{fig:bayes-fits} but for the SF case. Each panel corresponds to one integral energy of interest, and each line within every panel refers to well connected (red line) and poorly connected (blue line) SFs. As a result, for each integral energy a total of 2 $f(F_{SXR})$ functions are obtained (see Table \ref{tab:table1}). The X-axis of sub-panels in Figure \ref{fig:pf} gives the $F_{SXR}$ and the Y-axis provides the probability of SEP occurrence. Hence, for a solar flare with known longitude, one may select the proper fit and from the magnitude of the solar flare directly obtain the expected probability, $P(\rm SEP)$. In this case, data point to an upper limit $F_{SXR}$ of X28 and thus the derived Bayes $P(\rm SEP)$ is obtained up to that limit. Again, the total evolution of the simulated fit is presented and we overlay a gray hatched area after that limit.

In the case of the peak flux, the focused range of 50 \& 90\% is maintained across all modes of operation. Figure \ref{fig:PPF_sf} provides the outputs per filtered integral energy and bins based on solar flare characteristics. Specifically, Fig.~\ref{fig:PPF_sf}(a) refers to a filtering of E$>$10 MeV with each of the presented fits color coded as: red ($F_{SXR} <$ M3.0); green (M3.0 $\le F_{SXR} <$ X1.0) and blue ($F_{SXR} \ge$ X1.0). Moreover, Fig.~\ref{fig:PPF_sf}, panels (b), (d) and (f) refer to a filtering at E$>$30 MeV. Red (green) fits present the integral energy of E$>$10 (E$>$30) MeV, while each of the three panels point to individual solar flare bins. Finally, panels (c), (e) and (g) of Fig.~\ref{fig:PPF_sf} are obtained with a filtering at E$>$100 MeV. In this case, each of the lines represent an integral energy: red (E$>$10 MeV); green (E$>$30 MeV) and blue (E$>$100 MeV). Again, each of these three panels have to do with a selection based on solar flare magnitude. All of the fits depicted in Figure \ref{fig:PPF_sf} are detailed in Table \ref{tab:sf_peak}.      

\begin{table}[h!]
\caption{Details on PROSPER's peak flux fits for the SF bins, derived by Equation \ref{eq:11}. The filtering on the integral energy is first applied and then there are three bins based on the magnitude of the solar flare. These fits are presented in Figure \ref{fig:PPF_sf}.}
\begin{tabular}{l|l|l}
\hline
\hline
\multicolumn{3}{l}{\bf{if an event reaches E$>$10 MeV}}  \\
\hline
\hline
 Flare flux $<$ M3.0   &      M3.0 $\leq$ Flare flux $<$ X1.0          &     Flare flux $\ge$ X1.0   \\ \hline
 $\gamma$  = 0.56 &      $\gamma$  = 0.40   & $\gamma$  = 0.27 \\
 $x_{low}$ = 2.63E+00     &    $x_{low}$ = 2.97E+00   &$x_{low}$ = 7.69E+00    \\ 
 $x_{lim}$ = 1.46E+03     &    $x_{lim}$ = 1.17E+04   & $x_{lim}$ = 1.39E+04   \\ \hline
 \hline
 \multicolumn{3}{l}{\bf{if an event reaches E$>$30 MeV}}    \\
\hline
\hline
  Flare flux $<$ M3.0   &      M3.0 $\leq$ Flare flux $<$ X1.0          &     Flare flux $\ge$ X1.0   \\ \hline
 \multicolumn{3}{l}{\bf{E$>$10 MeV}} \\ \hline
 $\gamma$  = 0.52 &      $\gamma$  = 0.36   & $\gamma$  = 0.28 \\
 $x_{low}$ = 5.66E+00     &    $x_{low}$ = 4.81E+00   &$x_{low}$ = 1.29E+01     \\ 
 $x_{lim}$ = 1.36E+03     &    $x_{lim}$ = 1.01E+04   & $x_{lim}$ = 1.43E+04   \\ \hline
 \multicolumn{3}{l}{\bf{E$>$30 MeV}}\\ \hline
 $\gamma$  = 0.76 &      $\gamma$  = 0.53   & $\gamma$  = 0.28 \\
 $x_{low}$ = 1.13E+00     &    $x_{low}$ = 9.29E-01   &$x_{low}$ = 1.59E+00     \\ 
 $x_{lim}$ = 3.27E+02     &    $x_{lim}$ = 5.72E+03   & $x_{lim}$ = 2.29E+03   \\ \hline
 \hline
 \multicolumn{3}{l}{\bf{if an event reaches E$>$100 MeV}}   \\ \hline
 \hline
  Flare flux $<$ M3.0   &      M3.0 $\leq$ Flare flux $<$ X1.0          &     Flare flux $\ge$ X1.0   \\ \hline
  \multicolumn{3}{l}{\bf{E$>$10 MeV}}\\ \hline
 $\gamma$  = 0.47 &      $\gamma$  = 0.27   & $\gamma$  = 0.21 \\
 $x_{low}$ = 7.74E+00     &    $x_{low}$ = 1.06E+01   &$x_{low}$ = 2.66E+01     \\ 
 $x_{lim}$ = 2.05E+04     &    $x_{lim}$ = 9.42E+03   & $x_{lim}$ = 1.60E+04   \\ \hline
 \multicolumn{3}{l}{\bf{E$>$30 MeV}}\\ \hline
 $\gamma$  = 0.57 &      $\gamma$  = 0.36   & $\gamma$  = 0.20 \\
 $x_{low}$ = 1.91E+00     &    $x_{low}$ = 2.74E+00   &$x_{low}$ = 5.33E+00     \\ 
 $x_{lim}$ = 5.56E+02     &    $x_{lim}$ = 2.52E+03   & $x_{lim}$ = 2.16E+03   \\ \hline
 \multicolumn{3}{l}{\bf{E$>$100 MeV}}\\ \hline
 $\gamma$  = 0.72 &      $\gamma$  = 0.53   & $\gamma$  = 0.23 \\
 $x_{low}$ = 3.67E-01     &    $x_{low}$ = 3.33E-01   &$x_{low}$ = 7.12E-01     \\ 
 $x_{lim}$ = 5.03E+00     &    $x_{lim}$ = 1.88E+02   & $x_{lim}$ = 1.65E+02   \\ \hline
 \end{tabular}%
\label{tab:sf_peak}
\end{table}

\subsection{Solar Flares \& CMEs} \label{sub_res_sfCME}
For the case of a combination of a solar flare (with known magnitude and longitude) and a CME (with known AW and speed) being identified Equation \ref{eq:7a} was employed but enforcing more conditions, depending on the obtained characteristics (e.g. flare longitudinal bin and magnitude, CME width and velocity). In this case, all combinations were considered (e.g. Halo, Partial halo and Non halo, together with well and poorly connected) for all respective energies. Figure \ref{fig:Pfcme} depicts derived probability of
SEP occurrence using both flare and CME inputs for the case of Halo and well connected events for an integral energy of E$>$10 MeV. Each line corresponds to a specific solar flare class. In particular, C1.0 (black), M1.0 (blue), X1.0 (red) and X10.0 (magenta) lines are shown. Based in the above, the X-axis of Figure \ref{fig:Pfcme} is extended up to $V_{CME}$ = 3000 km/s. From this plot one can derive the probability to get an SEP event, in case the longitude of the parent solar flare falls within the bin (lon $\ge$20$^{o}$ - well connected) and the corresponding CME is a Halo one (AW = 360$^{o}$). Then from the specific magnitude of the flare, the proper curve (fit) is selected and the speed of the CME (X-axis on the plot) leads to the probability of SEP occurrence (Y-axis on the plot). Similar fits have been constructed for all combinations of: (lon. x AW x integral proton energy). As a result, Equation \ref{eq:7a}  was applied in total in 18 cases (not shown). 

Figure \ref{fig:PPF_sf_cme} shows the obtained fits per filtered energy and bins based on all obtained inputs. Similar to the previous cases the fits are presented in detail in Table \ref{tab:sf_cme_peak}. As a rule, the dashed lines in the case of E$>$10 MeV (Fig.~\ref{fig:PPF_sf_cme})(a) and E$>$100 MeV (Fig.~\ref{fig:PPF_sf_cme})(b), (d), (f) present the bins with the largest CME speed per case for a given solar flare flux bin, while the continuous lines depict the cases with the slower CMEs for the same solar flare bins (see details of the bins in each panel of Fig.~\ref{fig:PPF_sf_cme} and Table \ref{tab:sf_cme_peak}). Panel (a) of Figure \ref{fig:PPF_sf_cme} shows E$>$10 MeV fits for six different selections of solar flares \& CMEs (see detail in Table \ref{tab:sf_cme_peak}). Each pair of lines (i.e. solid and dashed) colored with the same color refer to a similar solar flare bin. In particular, these are: $F_{SXR} <$ M3.0 (red); M3.0 $\le F_{SXR} <$X1.0 (blue) and $F_{SXR} \ge$ X1.0 (magenta) (see panel (a)). Moreover, panels (b), (d) and (f) of Fig.~\ref{fig:PPF_sf_cme} represent the case of applying a filtering at E$>$100 MeV. In all these panels each line refers to an integral energy color coded as: red (E$>$10 MeV); blue (E$>$30 MeV) and magenta (E$>$100 MeV). However, there was no event associated with a solar flare of magnitude $<$X3.0 for any CME speed, reaching an integral energy of E$>$100 MeV. Therefore, the relevant panel (f) in Figure \ref{fig:PPF_sf_cme} shows only the fits for events associated with $\geq$X3.0 for each of the three CME bin (i.e. only dashed lines) (see Table \ref{tab:sf_cme_peak}). In addition, no reliable fit for SEP events associated with solar flares of magnitude $<$M6.5 and CME speed $V_{CME}$ $<$ 1350 km/s, at E$>$10 MeV, when a filter on reaching an integral energy of E$>$100 MeV is applied, was found (see Fig.~\ref{fig:PPF_sf_cme}(b) and Table \ref{tab:sf_cme_peak}). Hence, there is no solid red line displayed in this panel. Finally, panels (c)-(e), (g)-(i) and (h)-(j) present the cases with a filtering at E$>$30 MeV. Each panel provides two fits color coded per integral energy as: E$>$10 MeV (red) and E$>$30 MeV (blue). Each of the mentioned pairs of panels refer to the same solar flare bin (i.e. $F_{SXR} <$ M3.0; M3.0 $\le F_{SXR} <$X1.0 and $F_{SXR} \ge$X1.0) while differ in the CME speed as detailed in Table \ref{tab:sf_cme_peak}. As it can be seen,in two cases that an event reaches E$>$30 MeV these seems to be a cross-over between the obtained fit at E$>$10 and E$>$30 MeV (i.e. panels (c) and (g)). However, these crossings are below a threshold of 40\%. Evidently, if a CL $<$50\% is chosen then in this particular case (i.e. E$>$30 MeV) a spectral incoherence arises. This means that in a future work an update of the selection in PROSPER for the solar flare \& CME case that will lead to spectral coherence at all cases for lower CL should be pursued.  

\begin{landscape}
\begin{table}[h!]
\caption{Details on PROSPER's peak flux fits for the combined solar flares \& CME bins, derived by Equation \ref{eq:11}. The filtering on the integral energy is first applied and then bins on the solar flare magnitude and CME speed are applied. These fits are presented in Figure \ref{fig:PPF_sf_cme}.}
\begin{tabular}{l|l|l | l | l | l}
\hline
\hline
 \multicolumn{3}{l |}{\bf{if an event reaches E$>$10 MeV}} &  \multicolumn{3}{l}{\bf{if an event reaches E$>$100 MeV}} \\
\hline
\hline
 Flare flux $<$ M3.0  &  M3.0 $<$ Flare flux $\leq$ X1.0          &    Flare flux $>$ X1.0      & Flare flux $<$ M6.0  &  M6.0 $<$ Flare flux $\leq$ X3.0   &    Flare flux $>$ X3.0 \\ \hline
 $V_{CME}$ $<$ 1250 km/s     & $V_{CME}$ $<$ 1400 km/s   & $V_{CME}$ $<$ 1600 km/s &   \multicolumn{3}{l}{\bf{E$>$10 MeV}}\\ \hline
 $\gamma$  = 0.45 &     $\gamma$  = 0.69   & $\gamma$  = 0.22 & $V_{CME}$ $<$ 1350 km/s     & $V_{CME}$ $<$ 1350 km/s   & $V_{CME}$ $<$ 1600 km/s \\\cline{4-6}
 $x_{low}$ = 2.98E+00     &  $x_{low}$ = 3.30E+00  &$x_{low}$ = 4.63E+00  &  $\gamma$= & $\gamma$ = 0.48& $\gamma$ = \\
 $x_{lim}$ = 3.85E+01     &  $x_{lim}$ = 6.49E+02   & $x_{lim}$ = 4.27E+02   &  $x_{low}$ =      &  $x_{low}$ = 8.70E+00  &$x_{low}$ = \\ \cline{1-3}
  $V_{CME}$ $\ge$ 1250 km/s      &  $V_{CME}$ $\ge$ 1400 km/s &  $V_{CME}$ $\ge$ 1600 km/s &  $x_{lim}$ =      &  $x_{lim}$ = 8.49E+02  & $x_{lim}$ =  \\ \hline  
 $\gamma$  = 0.35 &$\gamma$  = 0.23 & $\gamma$  = 0.21 &  $V_{CME}$ $\ge$ 1350 km/s      &  $V_{CME}$ $\ge$ 1350 km/s &  $V_{CME}$ $\ge$ 1600 km/s  \\ \cline{4-6}  
  $x_{low}$ = 4.23E+00 &$x_{low}$ = 4.87E+00 &$x_{low}$ = 2.82E+01 & $\gamma$=0.36 & $\gamma$ = 0.18& $\gamma$ =0.18  \\
 $x_{lim}$ = 1.58E+03   &$x_{lim}$ = 8.19E+03 &$x_{lim}$ = 1.35E+04 &   $x_{low}$ = 2.47E+01 &$x_{low}$ = 1.85E+01 &$x_{low}$ = 6.16E+01 \\\cline{1-3}
\cmidrule{1-3}
  \multicolumn{3}{l|}{\bf{if an event reaches E$>$30 MeV}} &  $x_{lim}$ = 2.38E+04  &$x_{lim}$ = 1.17E+04 &$x_{lim}$ = 1.35E+04 \\
   \hline
  \cmidrule{1-3} 
  Flare flux $<$ M3.0  &  M3.0 $<$ Flare flux $\leq$ X1.0          &    Flare flux $>$ X1.0     &  \multicolumn{3}{l}{\bf{E$>$30 MeV}}        \\ \hline
    \multicolumn{3}{l|}{\bf{E$>$10 MeV}} & $V_{CME}$ $<$ 1350 km/s     & $V_{CME}$ $<$ 1350 km/s   & $V_{CME}$ $<$ 1600 km/s \\ \hline
  $V_{CME}$ $<$ 1250 km/s     & $V_{CME}$ $<$ 1350 km/s   & $V_{CME}$ $<$ 1650 km/s & $\gamma$=0.41 & $\gamma$ = 0.47& $\gamma$ = \\ \cline{1-3}
    $\gamma$  = 0.17 &     $\gamma$  = 0.64   & $\gamma$  = 0.26 &  $x_{low}$ = 2.96E+00 &$x_{low}$ = 2.38E+00 &$x_{low}$ = \\
 $x_{low}$ = 6.14E+00     &  $x_{low}$ = 4.05E+00  &$x_{low}$ = 7.80E+00  & $x_{lim}$ = 9.71E+00    &  $x_{lim}$ = 1.01E+02   & $x_{lim}$ =  \\ \cline{4-6}
 $x_{lim}$ = 2.98E+01     &  $x_{lim}$ = 7.45E+02   & $x_{lim}$ = 4.59E+02  &  $V_{CME}$ $\ge$ 1350 km/s      &  $V_{CME}$ $\ge$ 1350 km/s &  $V_{CME}$ $\ge$ 1600 km/s\\ \hline
 $V_{CME}$ $\ge$ 1250 km/s      &  $V_{CME}$ $\ge$ 1350 km/s &  $V_{CME}$ $\ge$ 1650 km/s & $\gamma$=0.24 & $\gamma$ = 0.26 & $\gamma$ = 0.16 \\ \cline{1-3}
 $\gamma$  = 0.32 &$\gamma$  = 0.22 & $\gamma$  = 0.21 &  $x_{low}$ = 2.23E+00 &$x_{low}$ = 4.97E+00 &$x_{low}$ = 1.64E+01\\
  $x_{low}$ = 6.06E+00 &$x_{low}$ = 5.27E+00 &$x_{low}$ = 2.82E+01 & $x_{lim}$ = 4.47E+02   &$x_{lim}$ = 3.14E+03 &$x_{lim}$ = 1.76E+03 \\ \cline{4-6}
 $x_{lim}$ = 1.50E+03   &$x_{lim}$ = 8.04E+03 &$x_{lim}$ = 1.35E+04  & \multicolumn{3}{l}{\bf{E$>$100 MeV}}\\ \hline
  \multicolumn{3}{l|}{\bf{E$>$30 MeV}} & $V_{CME}$ $<$ 1350 km/s     & $V_{CME}$ $<$ 1350 km/s   & $V_{CME}$ $<$ 1600 km/s \\ \hline
  $V_{CME}$ $<$ 1250 km/s     & $V_{CME}$ $<$ 1350 km/s   & $V_{CME}$ $<$ 1650 km/s  & $\gamma$=1.88 & $\gamma$ = 0.77 & $\gamma$ = \\ \cline{1-3}
    $\gamma$  = 0.29 &     $\gamma$  = 0.05   & $\gamma$  = 0.40   &  $x_{low}$ = 4.04E-01 &$x_{low}$ = 4.60E-01 &$x_{low}$ = \\
 $x_{low}$ = 1.26E+01     &  $x_{low}$ = 5.86E+00  &$x_{low}$ = 1.33E+00 & $x_{lim}$ = 4.66E+02     &  $x_{lim}$ = 4.67E+00  & $x_{lim}$ =   \\ \cline{4-6} 
 $x_{lim}$ = 1.48E+00     &  $x_{lim}$ = 4.77E+00  & $x_{lim}$ = 2.82E+02   &  $V_{CME}$ $\ge$ 1350 km/s      &  $V_{CME}$ $\ge$ 1350 km/s &  $V_{CME}$ $\ge$ 1600 km/s\\ \hline
 $V_{CME}$ $\ge$ 1250 km/s      &  $V_{CME}$ $\ge$ 1350 km/s &  $V_{CME}$ $\ge$ 1650 km/s  & $\gamma$=0.34 & $\gamma$ = 0.32, & $\gamma$ = 0.26\\ \cline{1-3}
 $\gamma$  = 0.63 &$\gamma$  = 0.33 & $\gamma$  = 0.18 &  $x_{low}$ = 3.75E-01 &$x_{low}$ = 2.33E-01 &$x_{low}$ =1.18E+00 \\
  $x_{low}$ = 1.19E+00 &$x_{low}$ = 1.08E+00 &$x_{low}$ = 2.14E+00  & $x_{lim}$ = 6.82E+00     &  $x_{lim}$ = 1.11E+02  & $x_{lim}$ = 2.92E+02 \\ 
 $x_{lim}$ = 1.23E+03   &$x_{lim}$ = 2.66E+03 &$x_{lim}$ = 2.00E+03 \\ \hline
\end{tabular}%
\label{tab:sf_cme_peak}
\end{table}
\end{landscape}

\section{Validation}
\label{sec:validation}
Table \ref{tab:valid} provides all inputs of the NASA CCMC SEP scoreboard campaign events\footnote{https://ccmc.gsfc.nasa.gov/assessment/topics/SEP/campaign2020.php} that have been investigated in this part of the validation of PROSPER, presented in this work. The purpose of this challenge is to facilitate collaboration of as many as possible SEP modelers and to provide a standardized validation procedure, with specific inputs being used by all parties. Such a challenge is based on curated events and greatly facilitates the comparison of outputs from different prediction methods. Nonetheless, larger samples of events are needed for a more complete evaluation of these methods, including PROSPER. The inputs (i.e. solar flare and/or CME characteristics) have been utilized in order to achieve the PROSPER's $P(\rm SEP)$ per mode of operation. PROSPER peak flux outputs are compared to the derived peak fluxes by the SEP challenge at two integral energies, i.e. E$>$10 and E$>$100 MeV. These values are presented in the last two columns of Table \ref{tab:valid}. The SEP event on 06 January 2014 was not associated with a solar flare (see Table \ref{tab:valid}), although, a back-sided origin of the driving CME and a (probably occulted) C2.1 solar flare\footnote{https://cdaw.gsfc.nasa.gov/CME\_list/sepe/} have been reported \citep{thakur2014ground}. Moreover, the sequence of SEP events at 04 \& 06 September 2017 are considered as one event leading to a single peak flux for the E$>$10 MeV. An SEP event started on 04 September 2017 with a significant enhancement recorded at E$>$10 MeV. However, on 06 September 2017 the strongest solar flare of solar cycle 24 was marked (i.e. X9.3) \citep{jiggens2019situ} and an SEP event was clearly visible in higher energies (up to E$>$100 MeV), while characterized also as a sub-GLE \citep{2017JSWSC...7A..28M}. Nonetheless, the E$>$10 MeV flux continued to be sustained after its initiation on 04 September 2017 and throughout both SEP events lasting several days. Finally, four SEP events did not exceed a threshold of 1 pfu at E$>$100 MeV and thus are presented in Table \ref{tab:valid} as ``N/A" entries. It is important to note that in this part of our analysis, the parameters of Table \ref{tab:valid} are taken ``as is" from the campaign event website and all comparisons and results are driven by the parent solar event parameters and the observed peak proton fluxes at E$>$10 \& E$>$100 MeV.

\begin{table}[h!]
\caption{Details of the SEP events' solar event parameters (i.e. solar flare and CME characteristics) and observed peak proton flux at E$>$10-; \& E$>$100 MeV, used in the validation, provided under the SHINE/ISWAT/ESWW SEP Model Validation Challenge.}
\begin{tabular}{r|lll|lll|ll}
\hline
\bf{SEP event} &
  \multicolumn{3}{l|}{\bf{Flare information}} &
  \multicolumn{3}{l|}{\bf{CME information}} &
  \bf{E\textgreater{}10 MeV} &
  \bf{E\textgreater{}100 MeV} \\
  \hline
Date &
  Onset &
  Mag. &
  Lon. &
  Start&
  Speed &
  Width &
  Peak Flux & Peak Flux
   \\
   & Time & (W/m$^{2}$) & ($^{\circ}$) & Time & (km/s) & ($^{\circ}$)  & \multicolumn{2}{c}{(pfu)} \\
               & (HH:MM) &  &  & (HH:MM)  & &  &  & \\
   \hline
2012-03-07    & 0:02  & X5.4 & E15 &0:24  & 2684 & 360 &6529.8 & 69.272 \\
2012-05-17     & 1:25  & M5.1 & W89 &1:48  & 1582 & 360 &255.44 & 20.445 \\
2012-07-12    & 15:37 & X1.4 & W02 &16:48 & 885  & 360 &96.08  & N/A    \\
2013-04-11   & 6:55  & M6.5 & E12 &7:24  & 861  & 360 &113.55 & 2.0297 \\
2014-01-06  & -                   &  -    & -   &$\,|\,$ 8:00  &  1402 & 360 &42.17  & 4.079  \\
2014-01-07  & 18:04 & X1.2 & W11 &18:24 & 1830 & 360 &1026.1 & 4.2687 \\
2017-07-14    & 1:07  & M2.4 & W33 &1:25  & 1200 & 360 &22.374 & N/A    \\
2017-09-04 & 20:28 &   M5.5 &  W16 &20:36 &  1418 &  360 & \multirow{2}{*}{844.38} &  \multirow{2}{*}{N/A} \\
2017-09-06 & 11:53 & X9.3 & W34 &12:24 & 1571 & 360 &      &        \\
2017-09-10 & 15:35 & X8.2 & W88 &16:00 & 3163 & 360 &1493.5 & 68.128\\
\hline
\end{tabular}%
\label{tab:valid}
\end{table}


\subsection{Probability of detection}
\begin{figure}[h!]
\centering
\includegraphics[width=11cm]{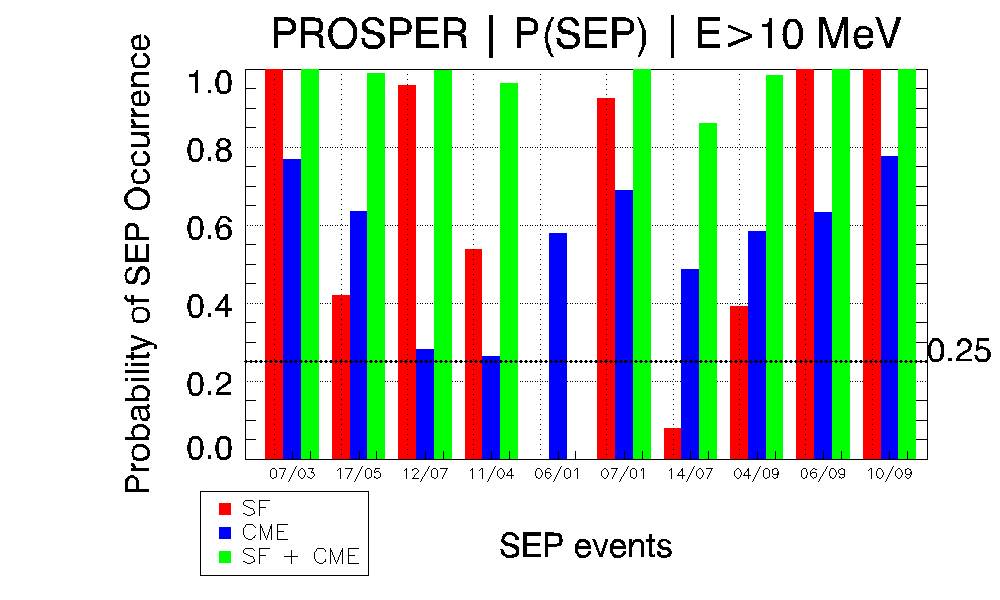}
\includegraphics[width=11cm]{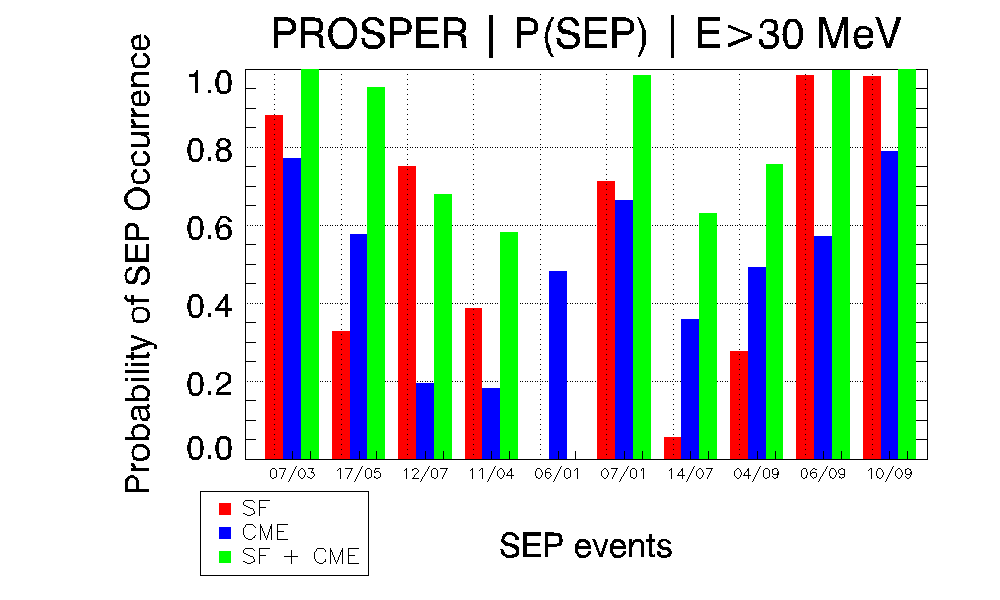}
\includegraphics[width=11cm]{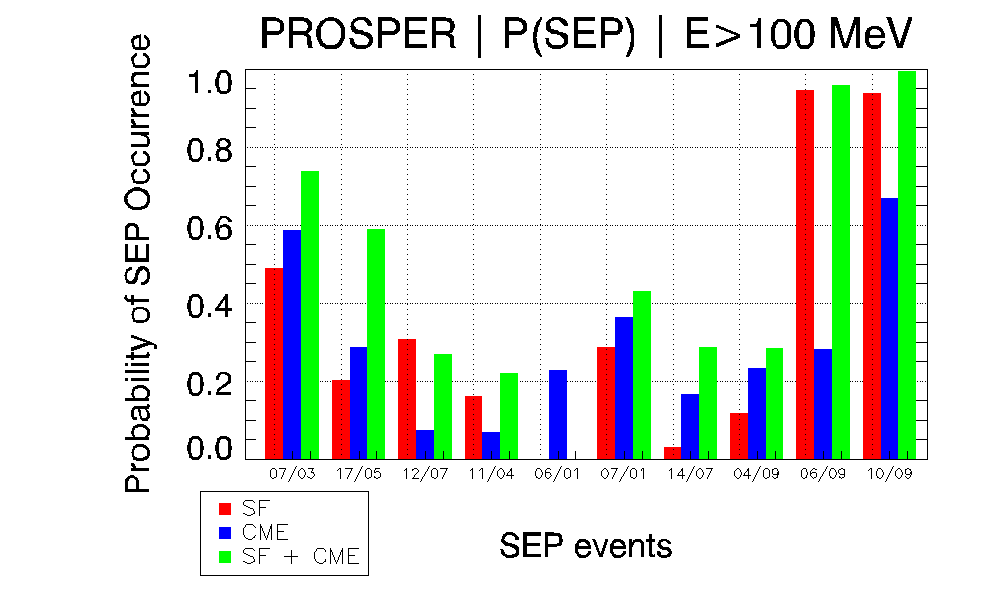}
\caption{Comparison of the predicted probabilities of SEP detection per PROSPER's mode of operation for all cases included in Table \ref{tab:valid}, for each integral energy of interest (i.e. E$>$10-; E$>$30-; and E$>$100 MeV). In each panel, the red histogram represents the outputs of PROSPER based on SF data, the blue histogram represents the outputs of the model based on CME data, and the green histogram represents the outputs of the model based on both SF \& CME inputs. In the upper panel the horizontal line represent the threshold above which most SEP prediction modules would issue a notification for a forthcoming SEP event – see details in Table 1 of \cite{anastasiadis2017predicting}.}
\label{fig:psepval}
\end{figure}

Using the associated parent solar events as input parameters (see Table \ref{tab:valid}), derived the probability of SEP detection per event and for all three PROSPER's modes of operation (see Table \ref{tab:table1}). Figure \ref{fig:psepval} presents the obtained outputs. Results are presented in red (flare mode of operation), blue (CME mode of operation) and green (solar flare \& CME mode of operation). Each panel corresponds to each integral energy (i.e. E$>$10-; $>$30-; and $>$100 MeV). Within the panel of E$>$10 MeV  (upper panel) the threshold ($pt$) above which the probabilistic forecast is transformed into a categorical one is over-plotted with a dotted horizontal black line. This $pt$ was directly obtained by Table 1 of \cite{anastasiadis2017predicting} and corresponds to the threshold above which an SEP event would have been identified. Such a $pt$ is obtained at the perfect skill of a forecasting algorithm and is usually applied for a direct discrete analysis of SEP yes/no. However, it must be noted that PROSPER gives a probabilistic forecast. From the inspection of the E$>$10 MeV panel in Figure \ref{fig:psepval}, it seems that the PROSPER's mode of operation based on solar flare input (red bars) would spot almost all of the events: $\sim$90\% (i.e. 8/9 events for which flare input was available). The SEP event that would not have been spotted by the PROPSER's flare mode of operation is the one that occurred on 14 July 2017. This SEP event was associated with a ``well-connected" (W33$^{o}$) solar flare but with relatively modest-magnitude (M2.4) that, although led to a probability of SEP occurrence of $\sim$20\%, was lower than the $pt$ and thus would not have been spotted by PROSPER. Moreover, the $P(\rm SEP)$ seems to be quite successful (100\%) for the CME module (i.e. 10/10 events for which CME input was available were identified as SEP events (i.e. crossing the $pt$). Adding to this, the 14 July 2017 event that was ``missed" by the flare mode of operation was associated with a halo CME with a speed of 1200 km/s -- which enhanced the probability of SEP occurrence to $\sim$ 58\% in the CME mode and thus pushed it above the $pt$ threshold, resulting into an identification of the event by PROSPER. Finally, the green bars represent PROSPER's outputs for the solar flare \& CME mode of operation. As it can be seen, the hit rate is 100\% (i.e. 9/9 events for which both flare and CME inputs were available).

These outputs (i.e. $P(\rm SEP)$) stem from the fits presented in Section \ref{sec:results} here above. In particular, 5/9 ($\sim$66\%) of the parent solar flares included in Table \ref{tab:valid} are ``poorly connected" (i.e. longitude $<$20$^{o}$) and 4/9 ($\sim$44\%) are ``well connected" (i.e. longitude $\geq$ 20$^{o}$). Hence, the fits used in these cases are presented in Figure \ref{fig:pf}. Figure \ref{fig:psepval} shows that at an integral energy E$>$10 MeV the largest differences in the achieved $P(\rm SEP)$ are due to the magnitude of the associated solar flare. For example, the first two events on the list are the 07 March 2012 (X5.4/E15 -``poorly connected") and 17 May 2012 (M5.1/W89 - ``well connected") with the obtained probability being higher for the former of the two (see Figure \ref{fig:psepval}). When shifting to higher energies (E$>$100 MeV) the differences in the obtained $P(\rm SEP)$ values (red bars) become larger because of the additional separation of the fits due to the longitudinal difference. Moreover, the two events on 06 \& 10 September 2017 result to very high probabilities of SEP occurrence, which seem to be preserved in each integral energy of interest. The specific SEP events are associated with very strong ($>$X8.0) and ``well connected" solar flares, which seem to be dominant. All driving CMEs are Halo and fast, thereby all of the $P(\rm SEP)$ values (blue bars) are obtained from the blue lines (fits) of Figure \ref{fig:bayes-fits}, and the differences are due to the $V_{CME}$ inputs, with larger energies requiring a higher value of $V_{CME}$ in order to lead to a larger $P(\rm SEP)$. That said, for the same $V_{CME}$ the obtained $P(\rm SEP)$ is lower when shifting to higher energies in all tested cases. Finally, for the flare \& CME mode of operation (green bars) one can notice that for the cases of 12 July 2012 and 11 April 2013 that are both ``poorly connected" ones, associated with comparable CMEs (see Table \ref{tab:valid}) the difference between the achieved $P(\rm SEP)$ is driven by the magnitude of the associated solar flare and the former case leads to higher $P(\rm SEP)$ at all integral energies. For the SEP event on 04 September 2017 which is also a ``poorly connected" one the $V_{CME}$ is almost a factor of $\sim$1.7 larger than the previous two cases, although the magnitude of the associated solar flare is M5.5 (i.e. less strong than the previous two cases). However, the obtained $P(\rm SEP)$ is higher than both cases at all integral energies, pointing to the role of the $V_{CME}$ in the establishment of that probability. Finally, the SEP event on 14 July 2017 which is a ``well connected" one and associated with a CME of $V_{CME}$ = 1200 km/s leads to a higher $P(\rm SEP)$ compared to the case of the ``poorly connected" SEP event at 11 April 2013 at all integral energies. In the latter case the associated flare is stronger (M6.5 \textit{versus} M2.4) but the CME speed is slower (861 \textit{versus} 1200 km/s). Thus the better magnetic connection and the higher $V_{CME}$ seem to lead to a higher $P(\rm SEP)$.  

\subsection{Peak proton flux}
\begin{figure}[h!]
\centering
\includegraphics[width=0.32\linewidth]{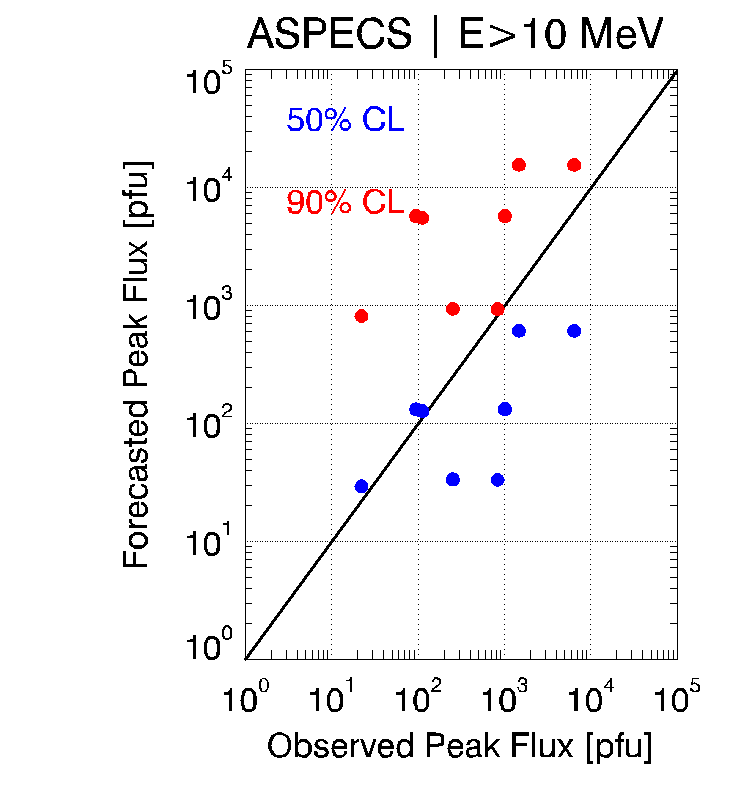}
\includegraphics[width=0.32\linewidth]{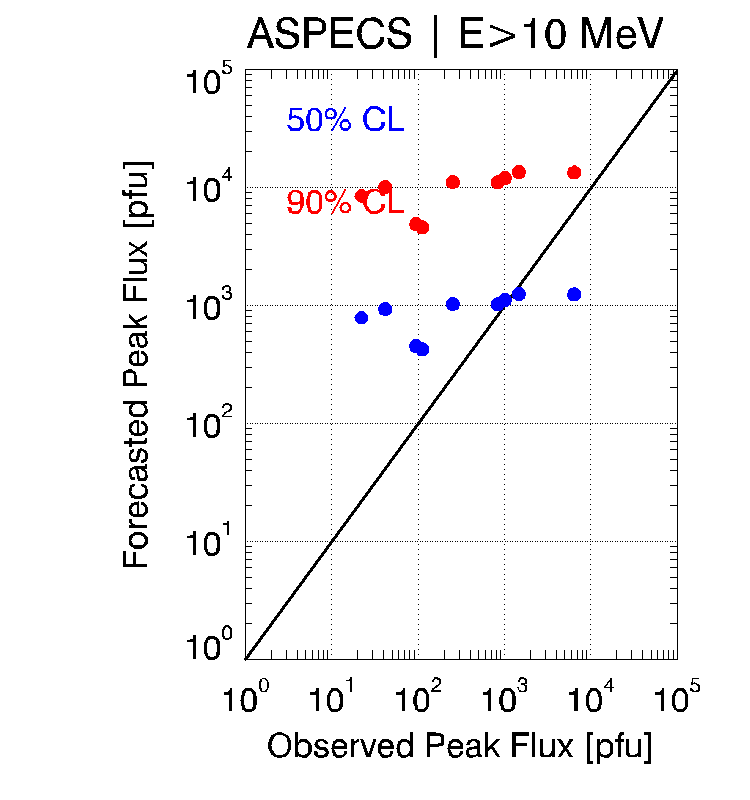}
\includegraphics[width=0.32\linewidth]{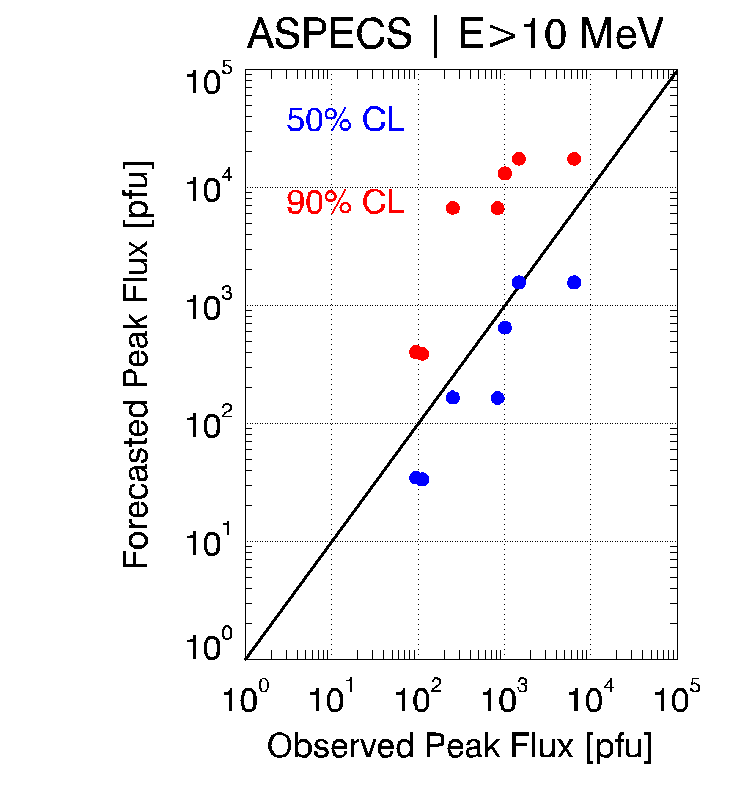}
\includegraphics[width=0.32\linewidth]{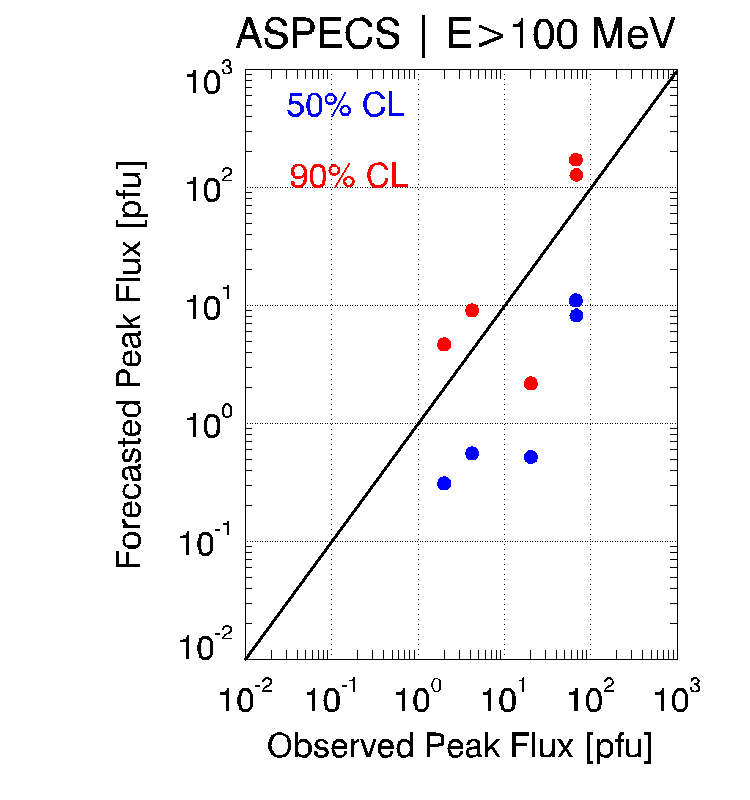}
\includegraphics[width=0.32\linewidth]{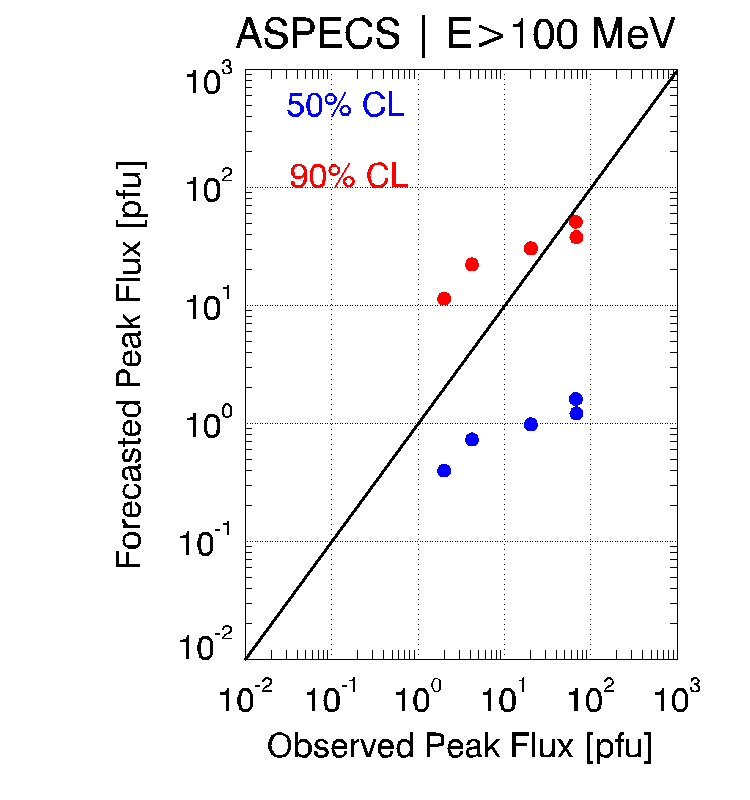}
\includegraphics[width=0.32\linewidth]{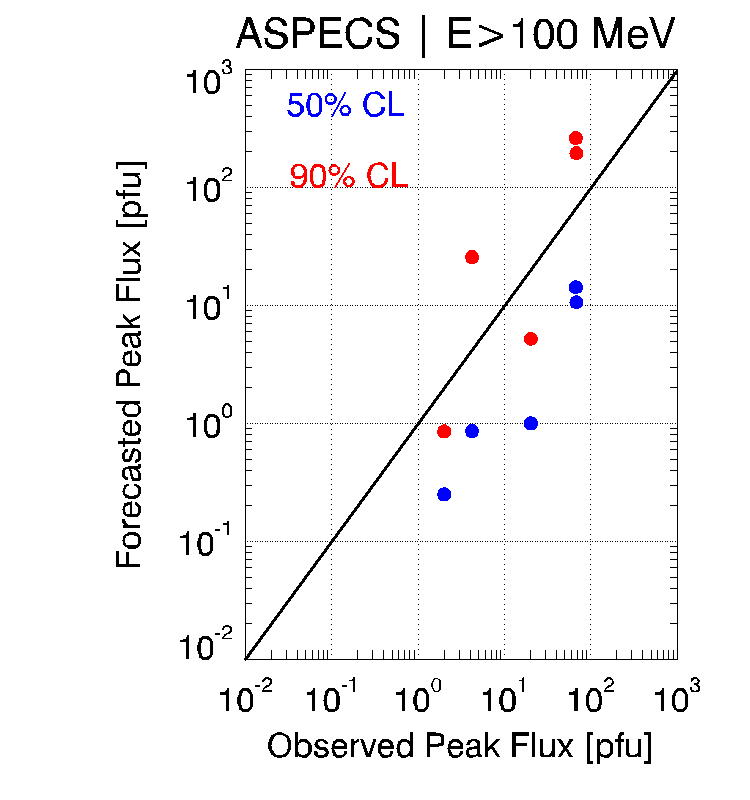}
\caption{Comparison of predicted with observed peak flux for the SEP events of Table \ref{tab:valid}, based
on solar flare (column at the left hand side), CME (column in the middle) and solar flare \& CME (column at the right hand side) input data. The top row refers to an integral energy of E$>$10 MeV and the bottom to E$>$100 MeV.}
\label{fig:pfval}
\end{figure}

Figure \ref{fig:pfval} provides scatter plots of the predicted peak flux versus the observed peak flux at E$>$10 MeV (top row) and E$>$100 MeV (bottom row) for each of the three PROSPER's modes of operation. In particular the first column corresponds to the PROSPER's outputs based on SF input, the middle column refers to the outputs of PROSPER utilizing CME input alone and the third column represents outputs from PROSPER's third mode of operation that makes use of both SF and CME inputs. Blue circles denote the obtained outputs at a 50\% CL (lower limit) and red circles present the same outputs for a 90\% CL (upper limit). 

The predicted peak flux at E $>$ 10 MeV as a function of the observed peak flux at the same energy, utilizing only SF information (upper panel on the left hand side of Figure \ref{fig:pfval}), shows a strong correlation especially in the lower limit (i.e. 50\% CL) with the correlation coefficient ($cc$) being 0.76 (lower limit) and 0.72 (upper limit). Moreover, more than half of the predicted values seem to be very close to the dichotomous line which indicates a perfect prediction. However, the upper limit predictions (90\% CL) show an over-forecasting being above the perfect dichotomous prediction line, with these predictions being within one order of magnitude from the observed peak fluxes, but not always. Nonetheless, the lower and the upper limits seem to capture the actual peak flux quite well at E$>$10 MeV. When moving to the CME mode of operation (column in the middle) for E$>$10 MeV, a weaker correlation is observed with the $cc$ being 0.33 (lower limit) and 0.32 (upper limit) respectively. The tendency for over predicting is also visible in this case and holds for the majority of the events both at the lower and the upper limit. At the same time, the CME mode of operation, seems not to provide granularity since all of the points are predicted using as input halo and fast CME and with the expectation of an SEP event reaching E$>$100 MeV. Hence the differences are small between events, since all outputs are derived by the fits of the bottom panel of Figure \ref{fig:Pfcme}. As a result, those small differences are corroborated with the small differences between the established $\rm P(SEP)$ (see Equation \ref{eq:12}). 
Moving to the results of the combined SF \& CME mode of operation (column at the right hand side), a stronger correlation is achieved ($cc$ is 0.77 - lower limit \& 0.71 - upper limit). Furthermore, the predictions at both the lower and the upper limit seem to be very close to the dichotomous line, with a small tendency for over-forecasting in the upper limit predictions. Nonetheless, the predictions at the lower limit for the strongest SEP events (i.e. those achieving the largest peak flux) at E$>$10 MeV seem to fall at the perfect dichotomous line. 

In the case of E$>$100 MeV, the peak flux seems to be captured well by all three modes of operation, with the obtained $cc$ being on average $\geq$ 0.90. Especially with the upper limit in the case of the solar flare input (panel at the left hand side at the bottom row) and the flare \& CME (panel on the right hand side at the bottom row) modules the agreement is quite reasonable with most of the events predicted at the upper limit (90\%) being either on the dichotomous line or very close to it (i.e. within less then 1 order of magnitude).   

The results presented in Figure \ref{fig:pfval}, underline the inherent difficulty of the SEP characteristics prognosis. However, the usage of the upper (90\%) and lower (50\%) limit in PROSPER shows promise and seems to capture the expected peak flux of the SEP events, reasonably well. 

\section{Discussion \& Conclusions}
\label{sec:discuss}

\begin{figure}[ht]
\centering
\begin{minipage}[b]{0.48\linewidth}
\includegraphics[width=8cm,height=4.5cm]{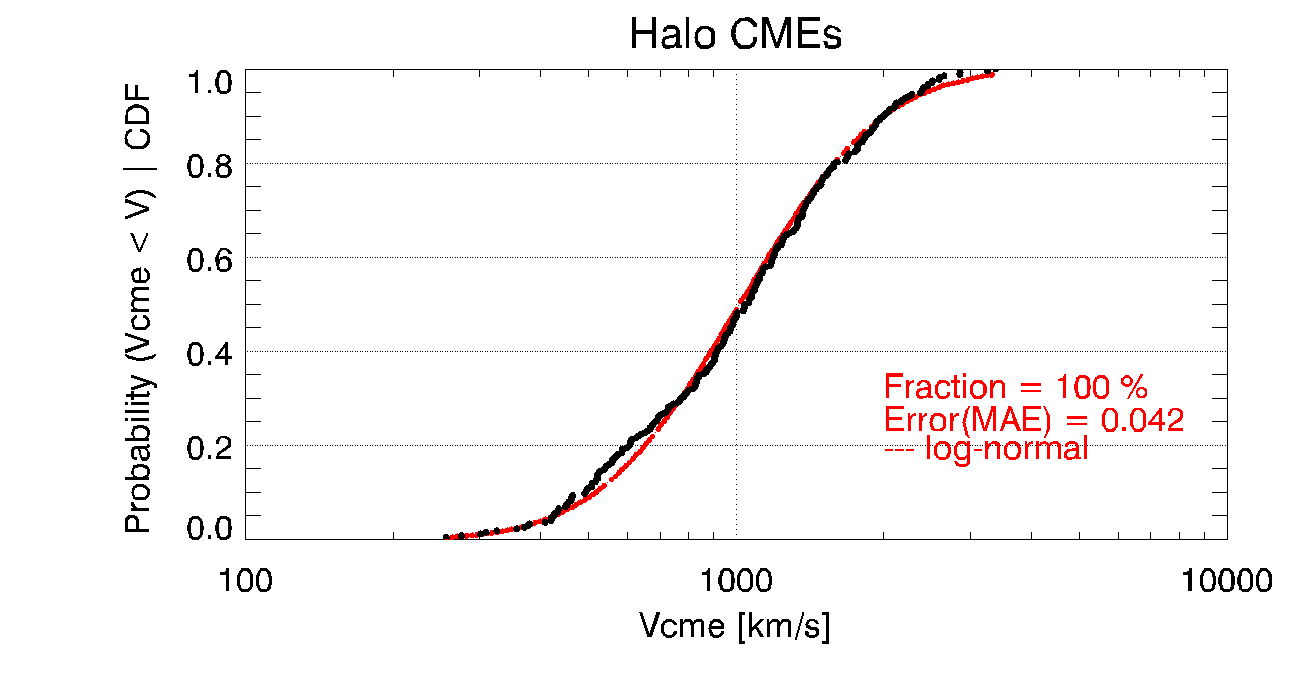}
\end{minipage}
\begin{minipage}[b]{0.45\linewidth}
\includegraphics[width=8cm,height=4.5cm]{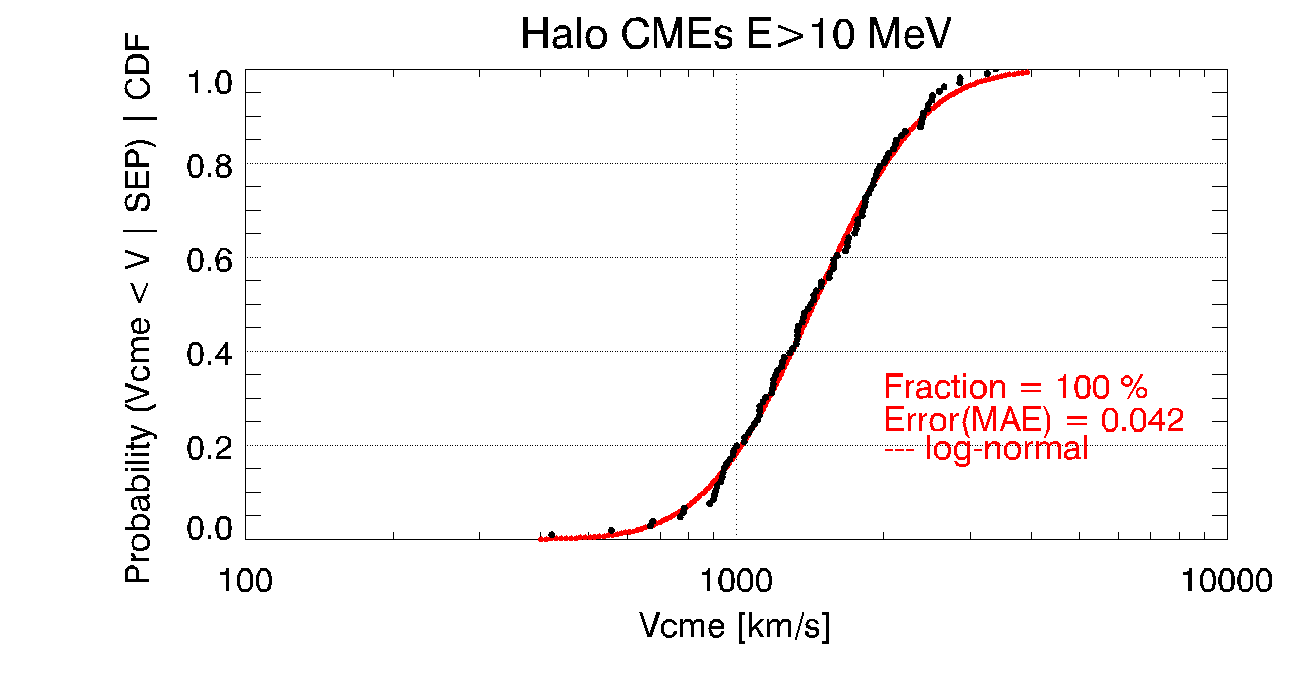}
\end{minipage}
\begin{minipage}[b]{0.48\linewidth}
\includegraphics[width=8cm,height=4.5cm]{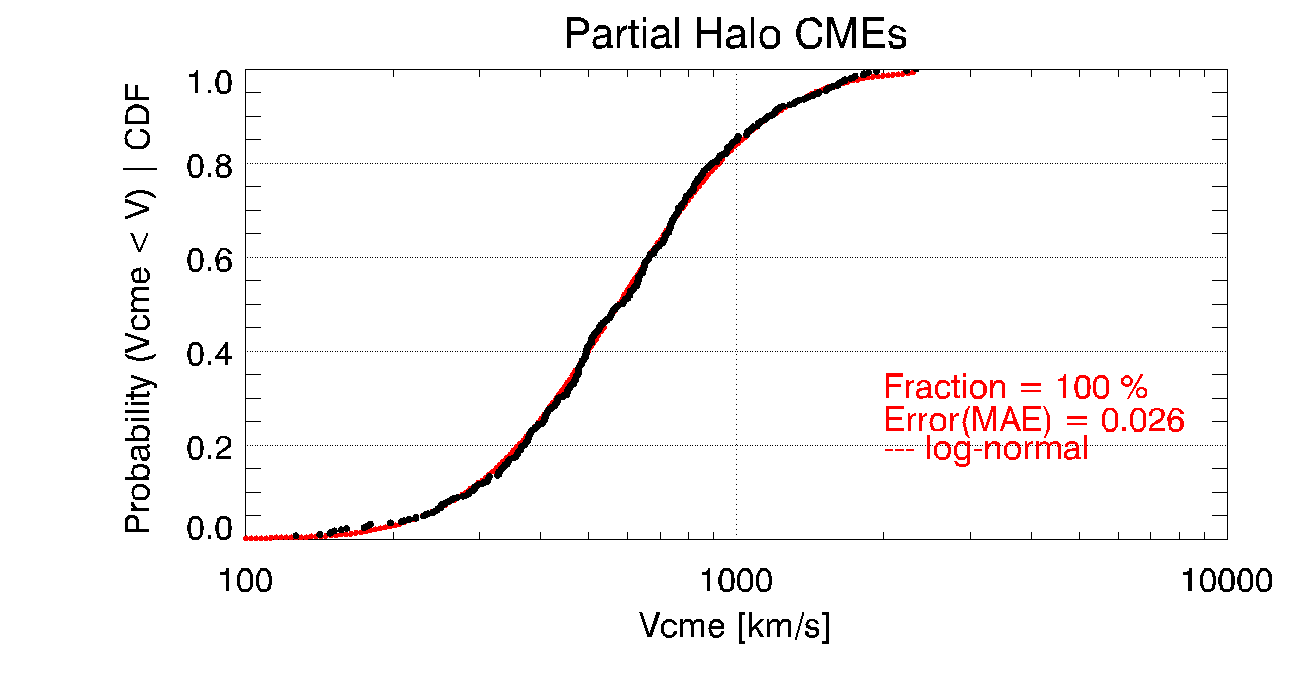}
\end{minipage}
\begin{minipage}[b]{0.45\linewidth}
\includegraphics[width=8cm,height=4.5cm]{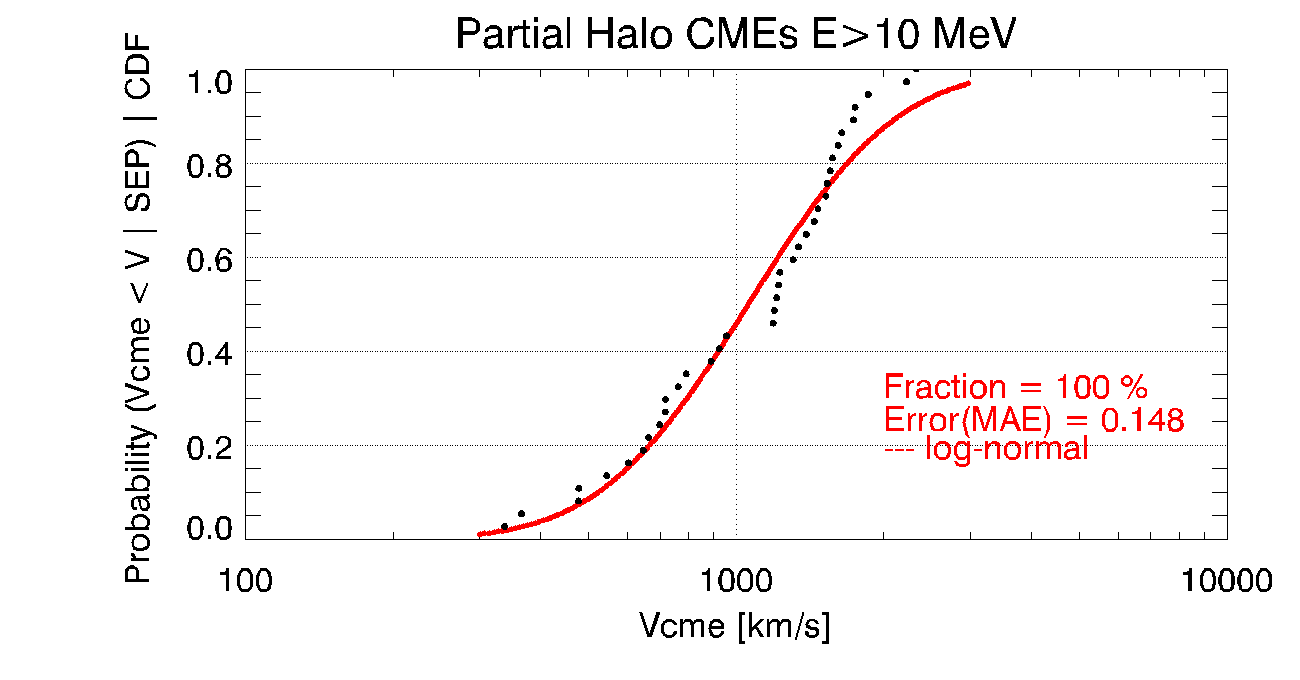}
\end{minipage}
\begin{minipage}[b]{0.48\linewidth}
\includegraphics[width=8cm,height=4.5cm]{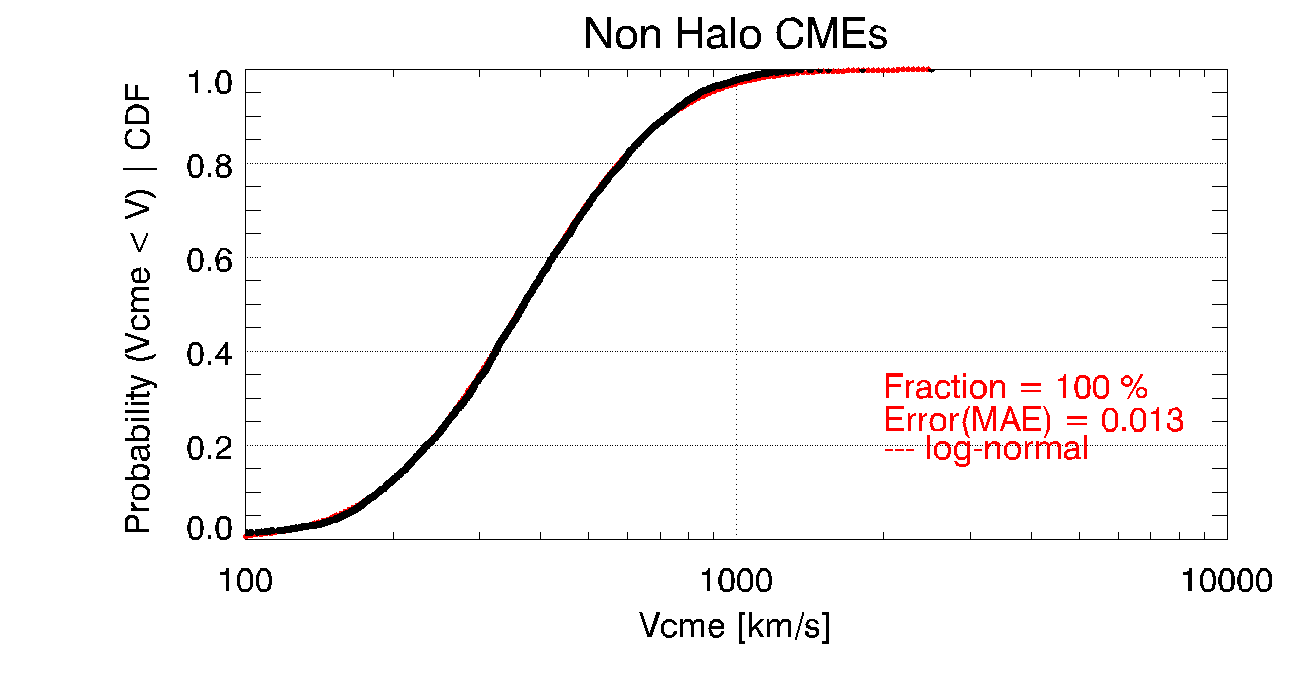}
\end{minipage}
\begin{minipage}[b]{0.45\linewidth}
\includegraphics[width=8cm,height=4.5cm]{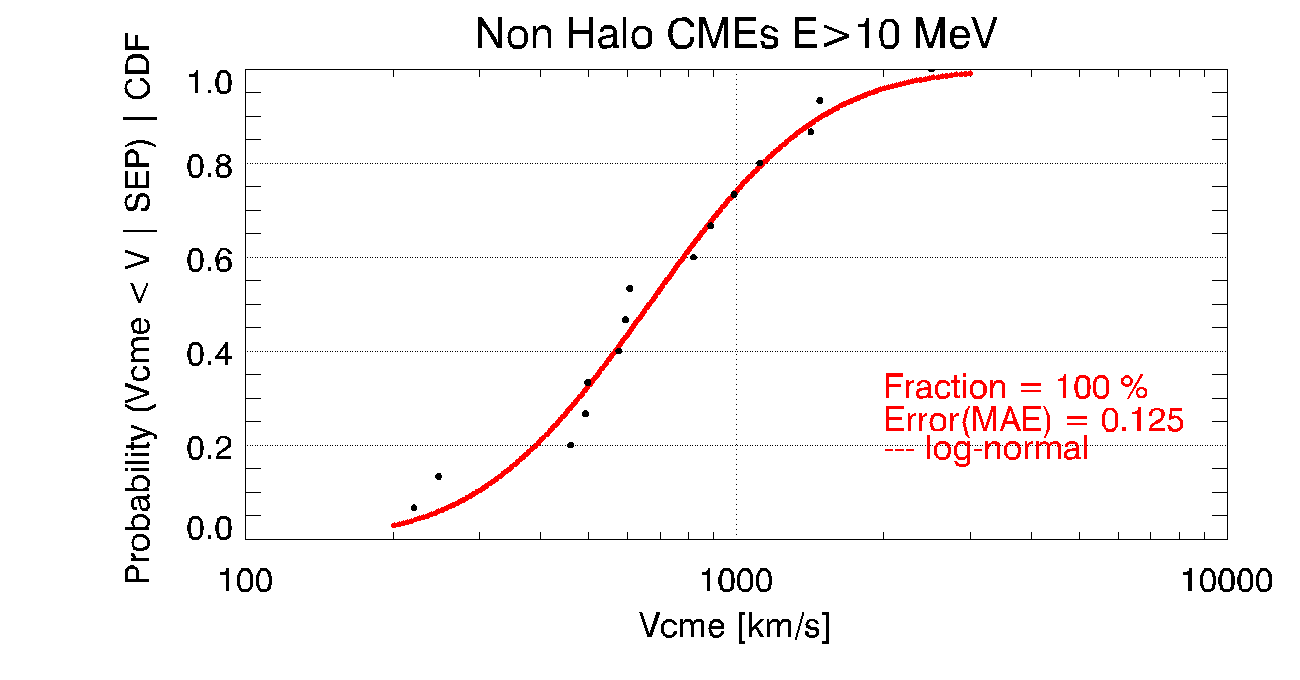}
\end{minipage}
\quad
\caption{The observational Cumulative Distribution Functions (CDFs) constructed by the database (i.e. all black points correspond to actual data). The red line depicts the log-normal fit to the data per case. Each plot further includes the fraction of the data explained by the fit (in [\%]) [i.e. a 100\% fraction means that all data points have been used in the corresponding fit], the mean absolute error (MAE) and the fit that was used (i.e. log-normal). From top to bottom the column on the left hand side corresponds to the CDFs per CME width (Halo, Partial Halo, Non-Halo) for all CMEs in each sample. The column on the right hand side corresponds to the CDFs in case that the CME was associated with an SEP at an energy of E$>$10 MeV, at each width bin. Similar fits (and plots) have been constructed for all other SEP integral energies of interest but are not shown.}
\label{fig:CDFs_fit}
\end{figure}

\begin{figure}
   \centering
   \includegraphics[width=12cm]{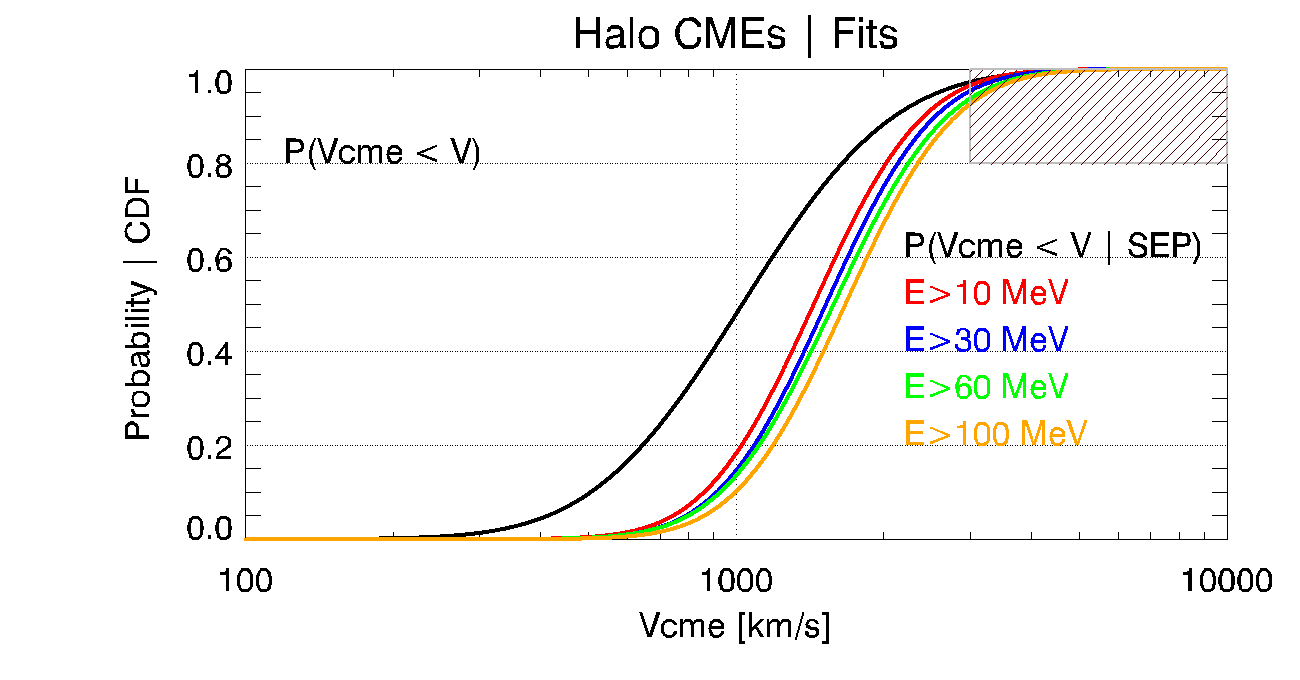}
   \includegraphics[width=12cm]{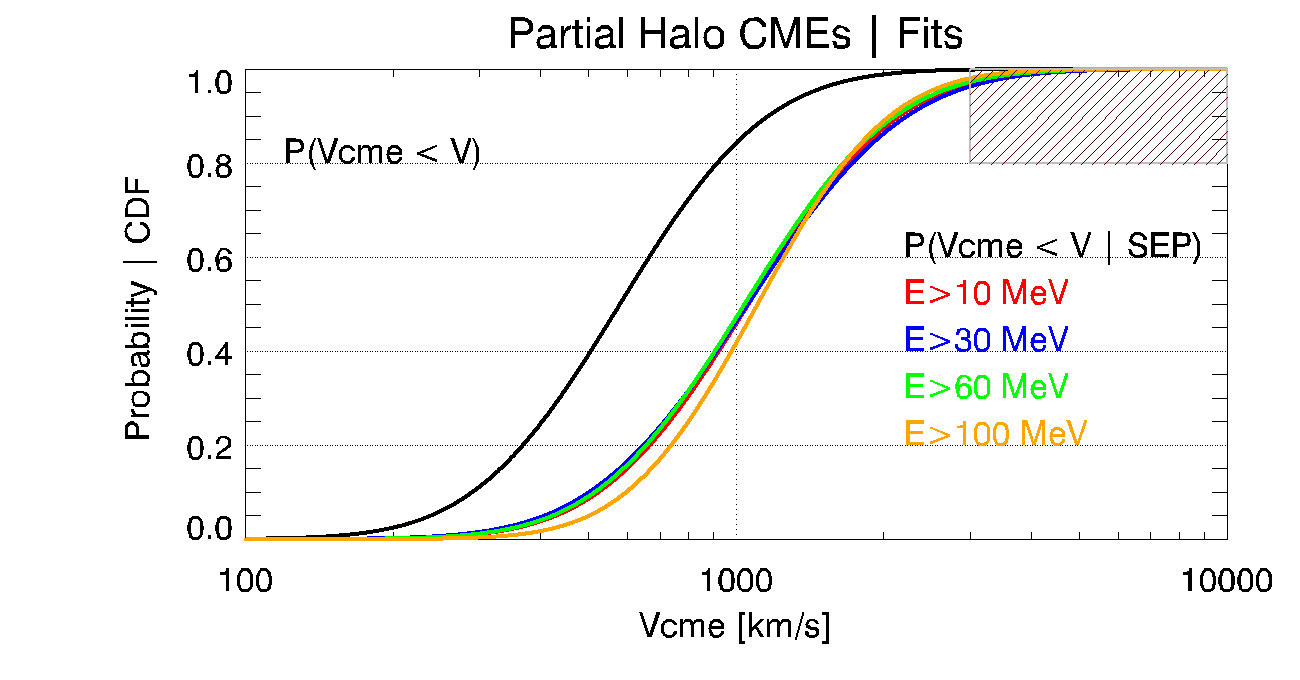}
   \includegraphics[width=12cm]{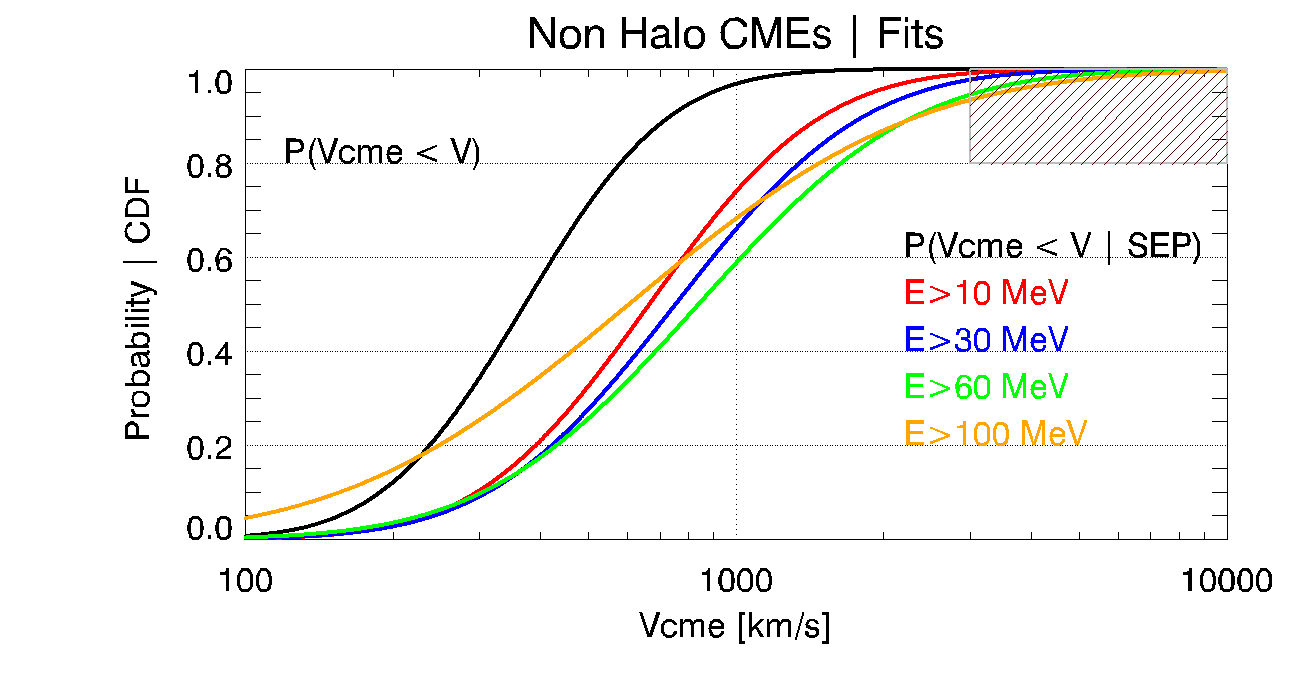}
      \caption{The Cumulative Distribution Functions (CDFs) for the case of Halo CMEs (top panel); Partial Halo CMEs (middle panel) and Non-Halo CMEs (bottom panel). The black color represents all CMEs in the respective sample, whereas CMEs associated with SEP events for different integral energies are colour coded. The gray border hatched rectangle area provides the limit for the $V_{CME}$.}
      \label{fig:cdfs_2}
   \end{figure}

\begin{figure}[ht]
\centering
\begin{minipage}[b]{0.48\linewidth}
\includegraphics[width=8cm,height=4.5cm]{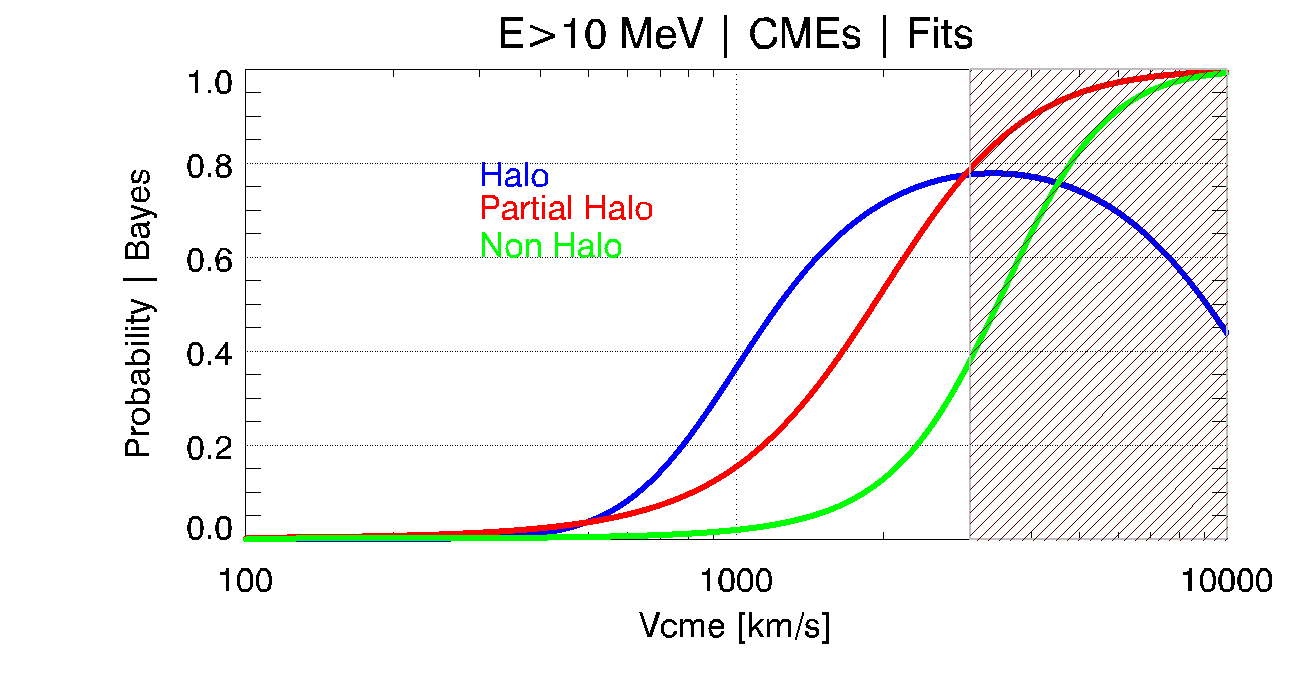}
\end{minipage}
\begin{minipage}[b]{0.48\linewidth}
\includegraphics[width=8cm,height=4.5cm]{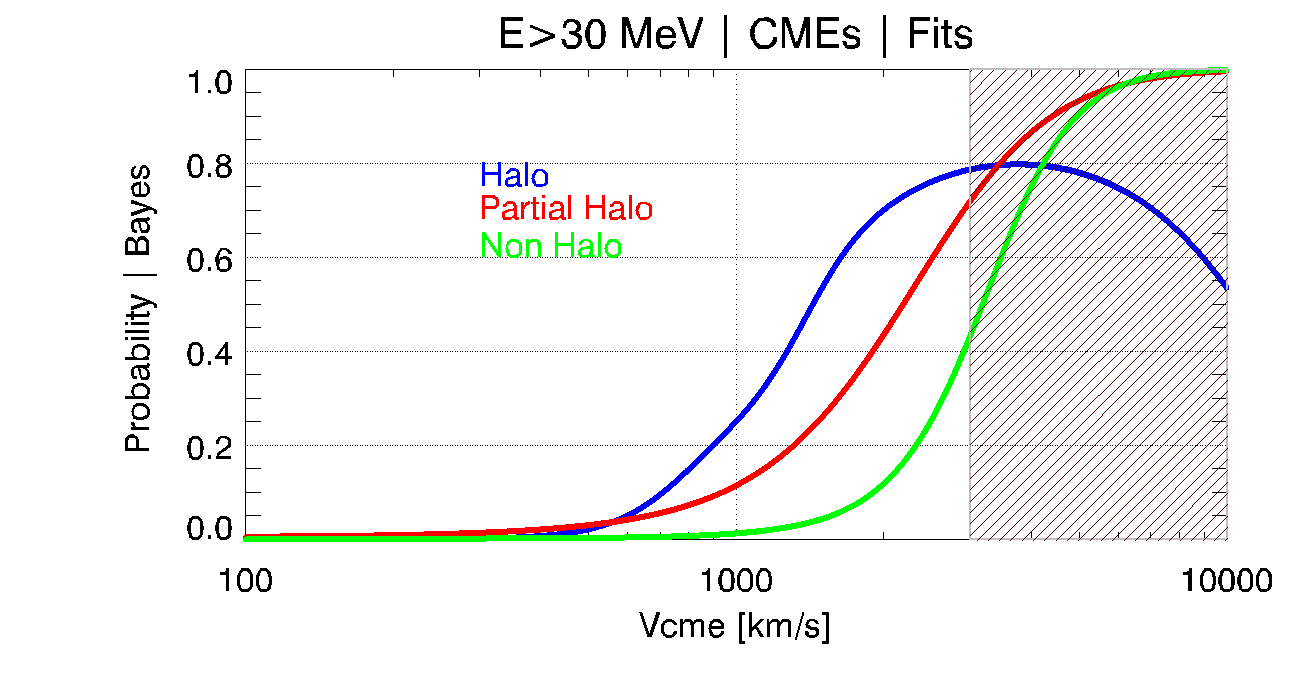}
\end{minipage}
\begin{minipage}[b]{0.48\linewidth}
\includegraphics[width=8cm,height=4.5cm]{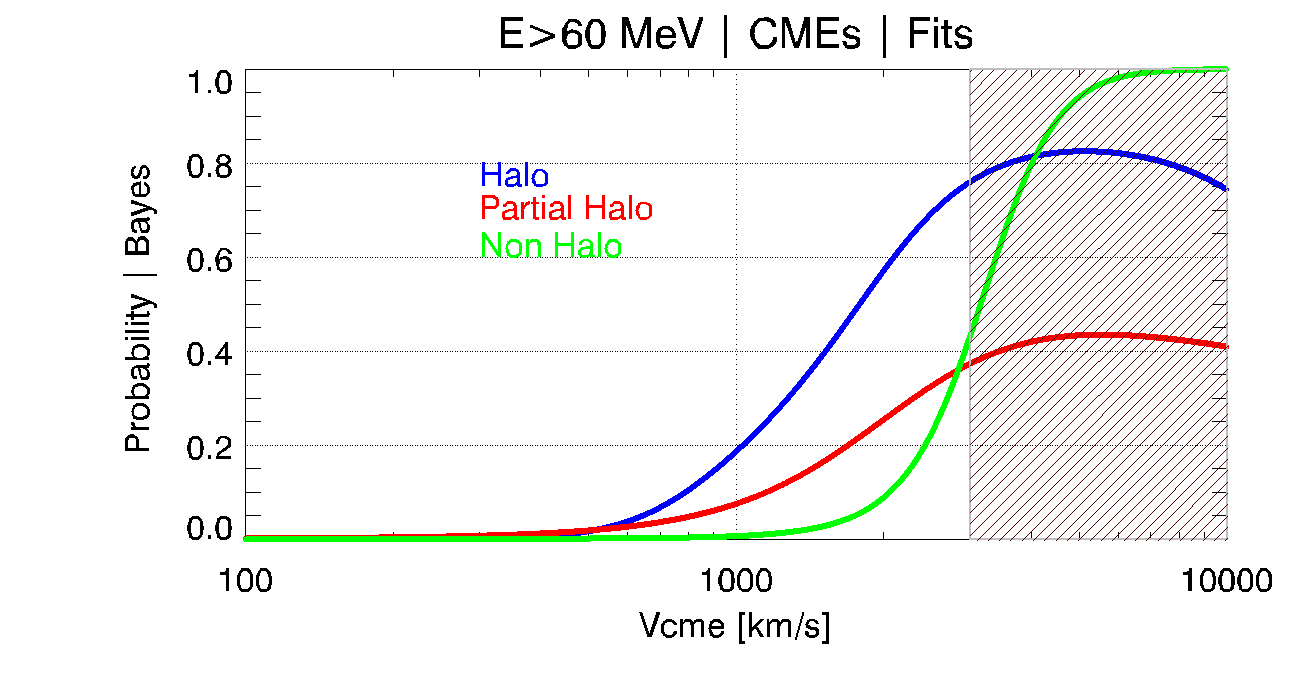}
\end{minipage}
\begin{minipage}[b]{0.48\linewidth}
\includegraphics[width=8cm,height=4.5cm]{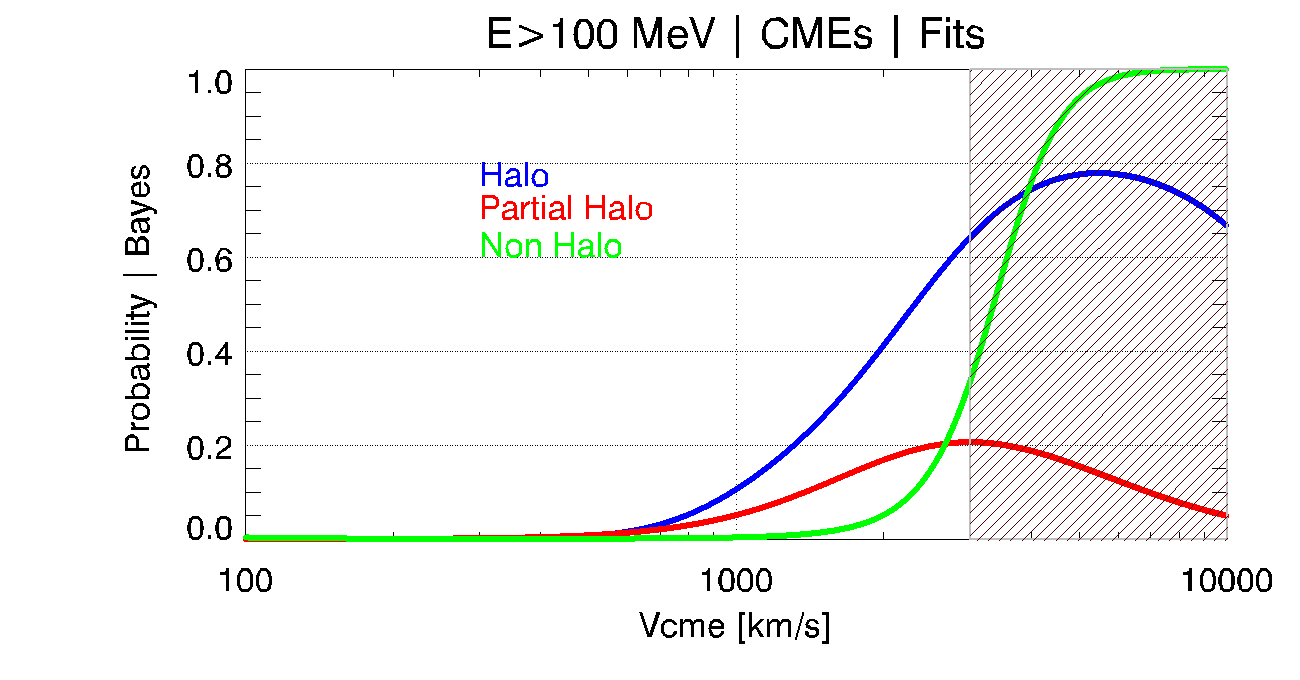}
\end{minipage}
\quad
\caption{The curves for the estimation of the probability of SEP occurrence when both the speed ($V_{CME}$, X-axis) and the width (Halo, Partial Halo and Non-Halo) of a CME are known. Each panel corresponds to an integral energy (i.e. E$>$10-; $>$30-; $>$60- and $>$100 MeV). Within each panel three fits are presented, one per CME width - colour coded as: Halo (blue), Partial Halo (red), Non Halo (green) CMEs. The gray border hatched rectangle area provides the limit for the $V_{CME}$. See text for details.}
\label{fig:bayes-fits}
\end{figure}

\begin{figure}[ht]
\centering
\begin{minipage}[b]{\linewidth}
\includegraphics[width=10cm]{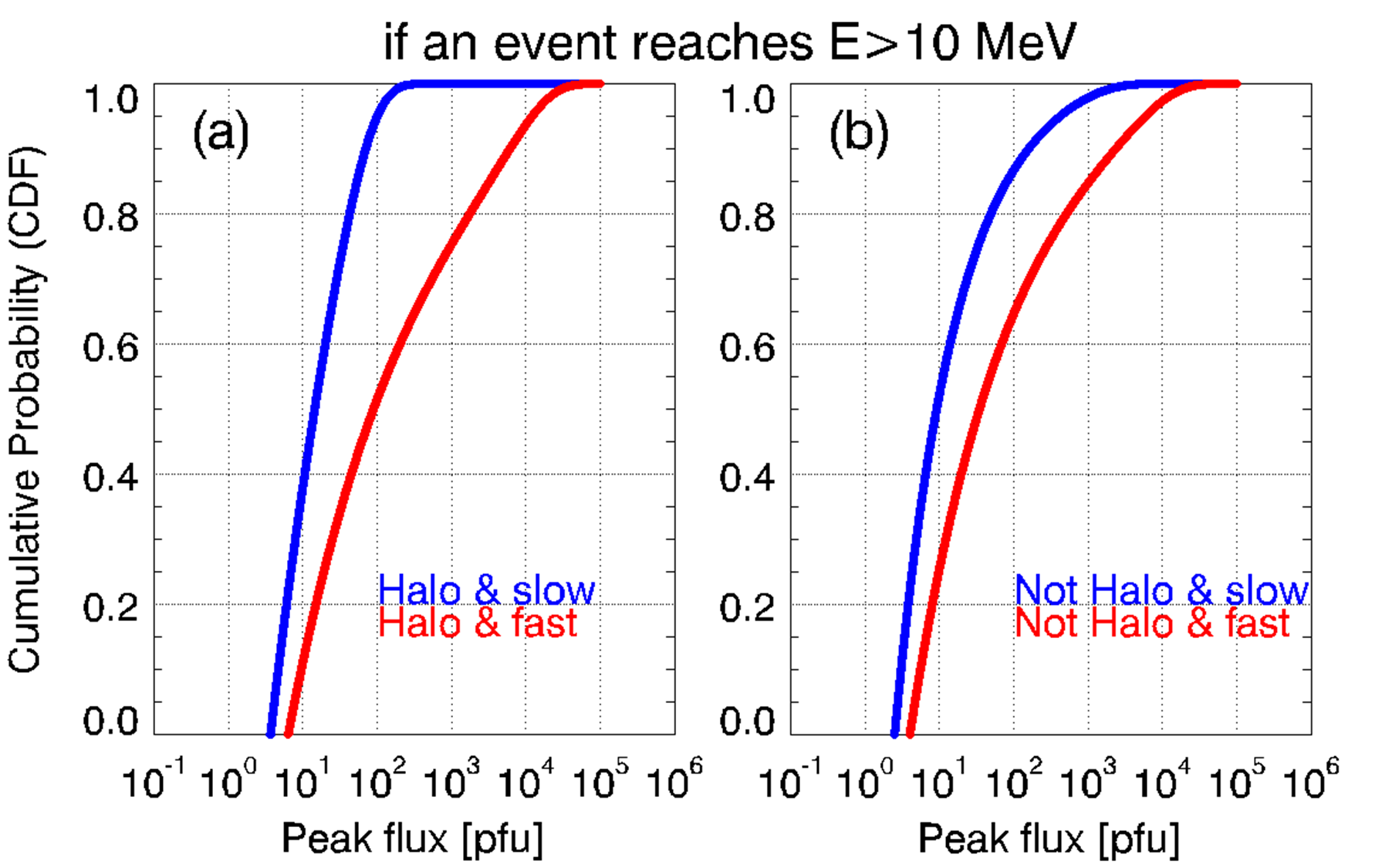}
\end{minipage}
\begin{minipage}[b]{\linewidth}
\includegraphics[width=10cm]{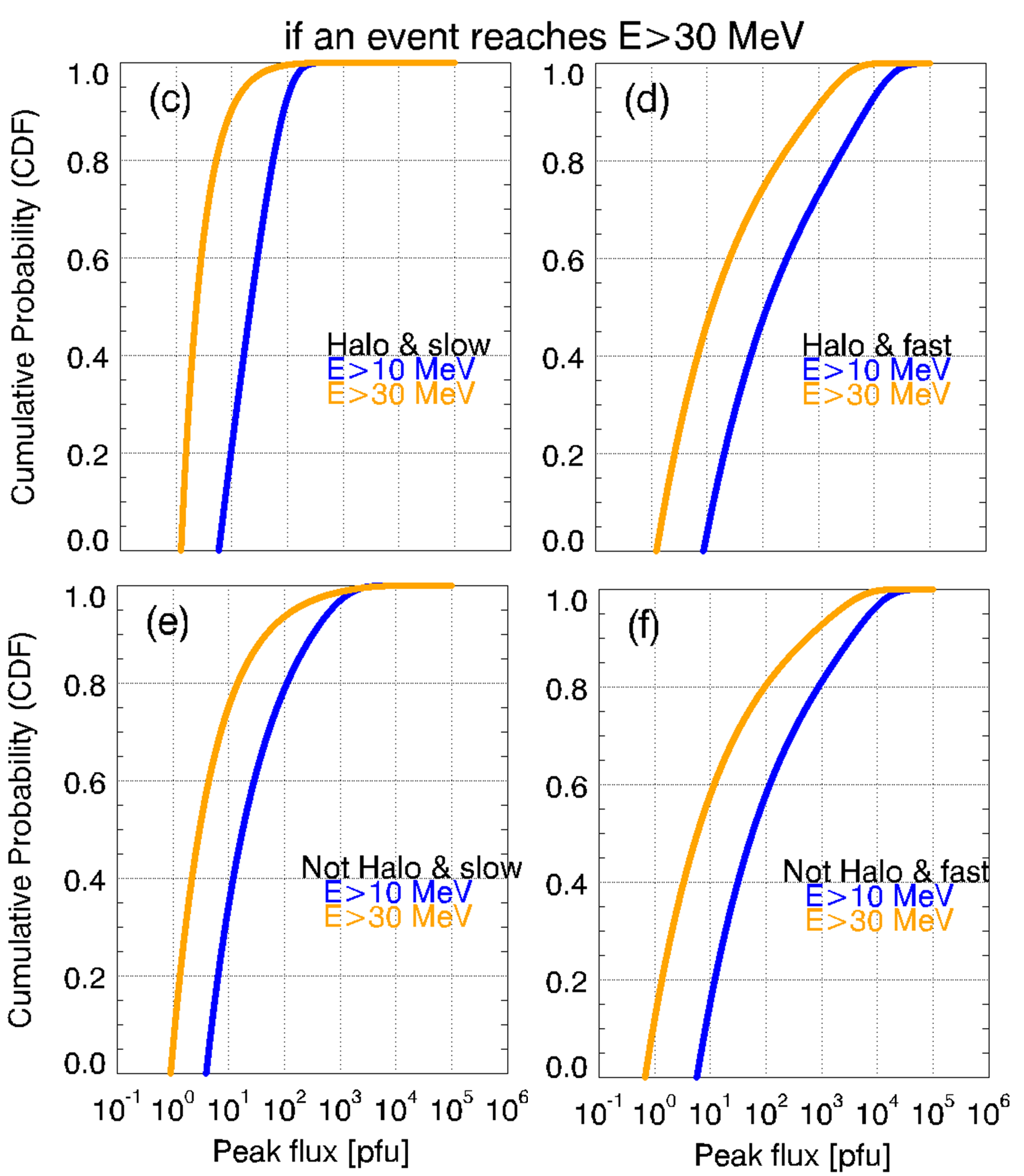}
\end{minipage}
\begin{minipage}[b]{\linewidth}
\includegraphics[width=10cm]{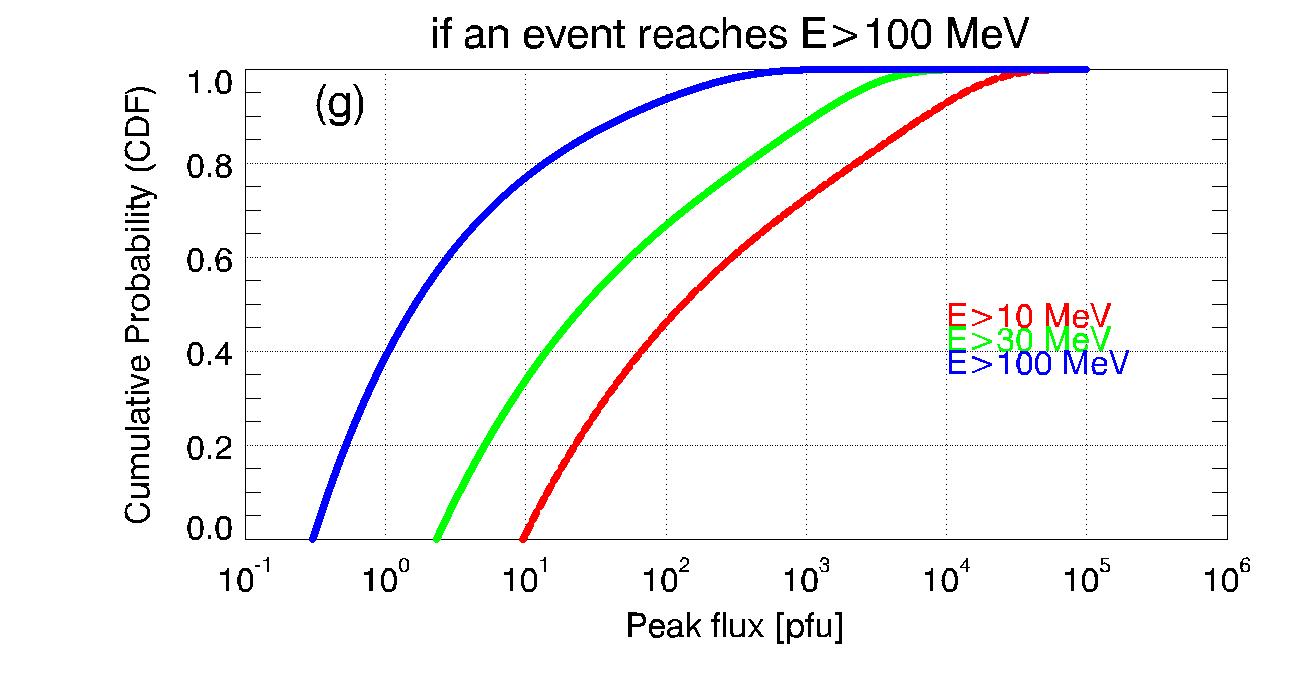}
\end{minipage}
\quad
\caption{The CDFs constructed by the database (i.e. all points correspond to actual data) for each integral energy of choice, for the estimation of the peak flux (Peak flux [pfu], X-axis), when both the speed and the width of a CME are known. Lines depict the exponential cut-off fits to the data points per case (from Equation \ref{eq:11}). Panels: (a) and (b) correspond to a filtering integral energy at E$>$10MeV; (c)-(f) to E$>$30 MeV and (g) to E$>$100 MeV). See text and Table \ref{tab:sf_peak} for details on the bins and the obtained parameters of the fit per case.}
\label{fig:PPF}
\end{figure}

\begin{figure}[ht]
\centering
\begin{minipage}[b]{0.48\linewidth}
\includegraphics[width=8cm,height=4.5cm]{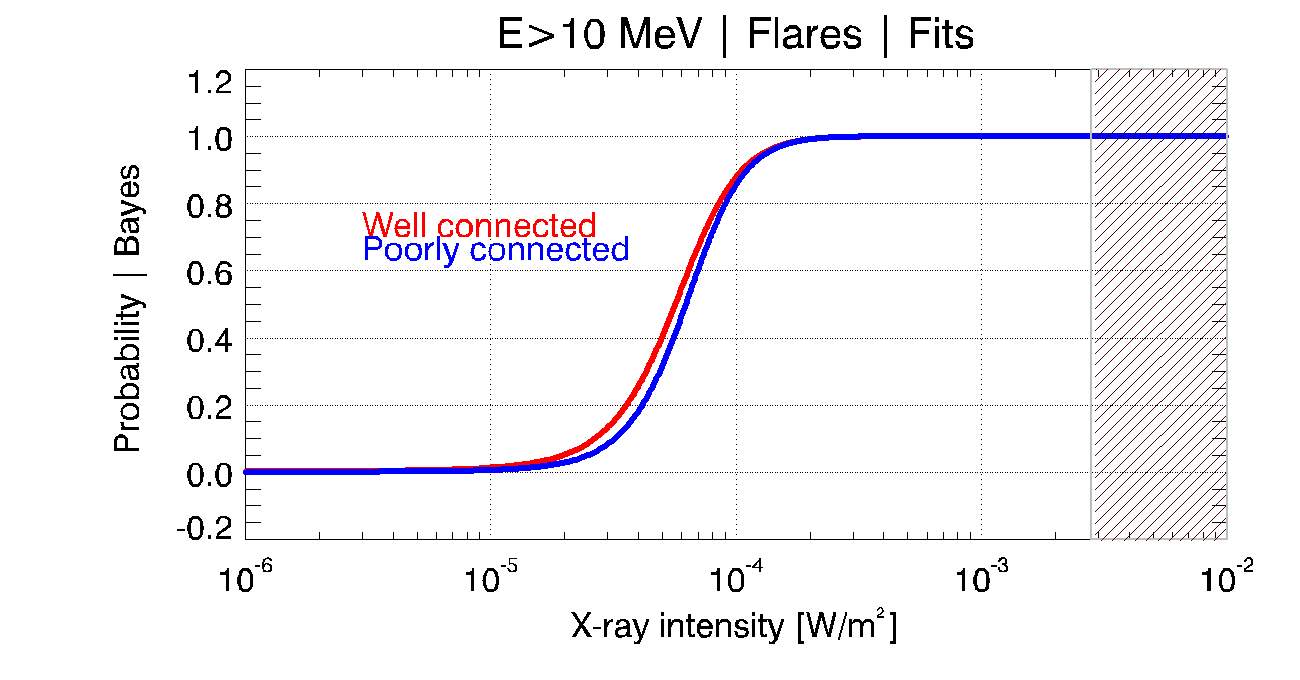}
\end{minipage}
\begin{minipage}[b]{0.45\linewidth}
\includegraphics[width=8cm,height=4.5cm]{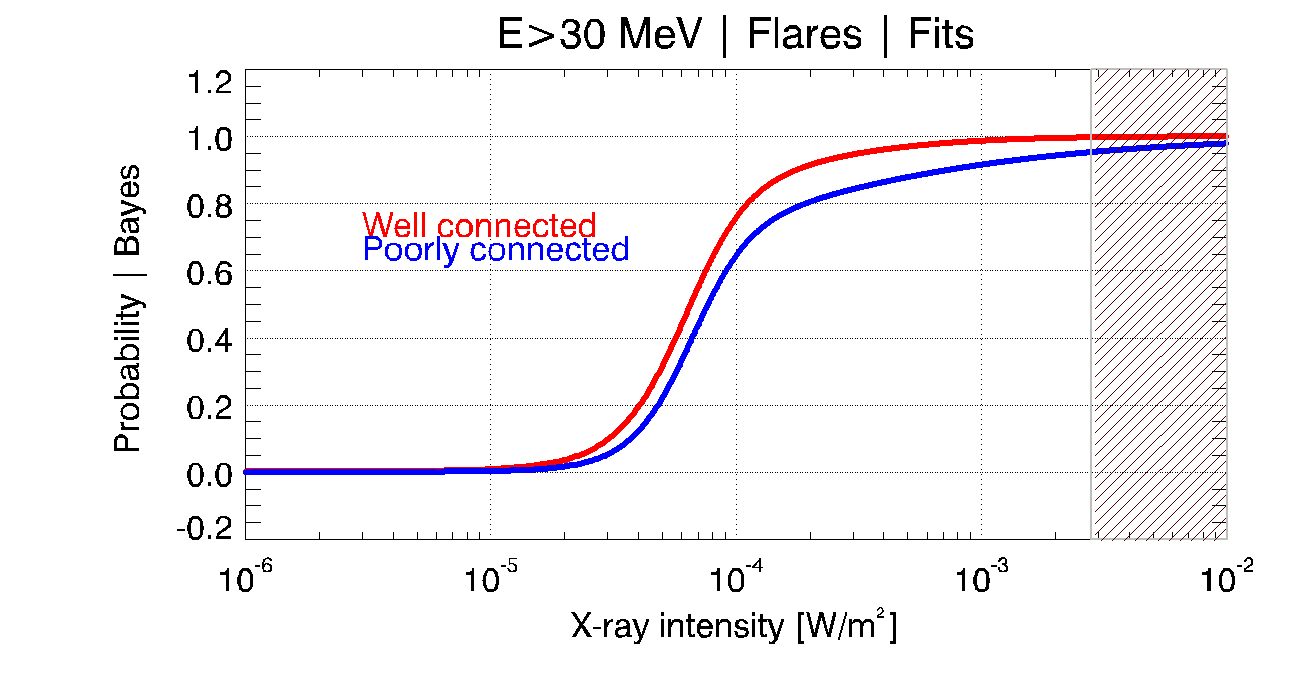}
\end{minipage}
\begin{minipage}[b]{0.48\linewidth}
\includegraphics[width=8cm,height=4.5cm]{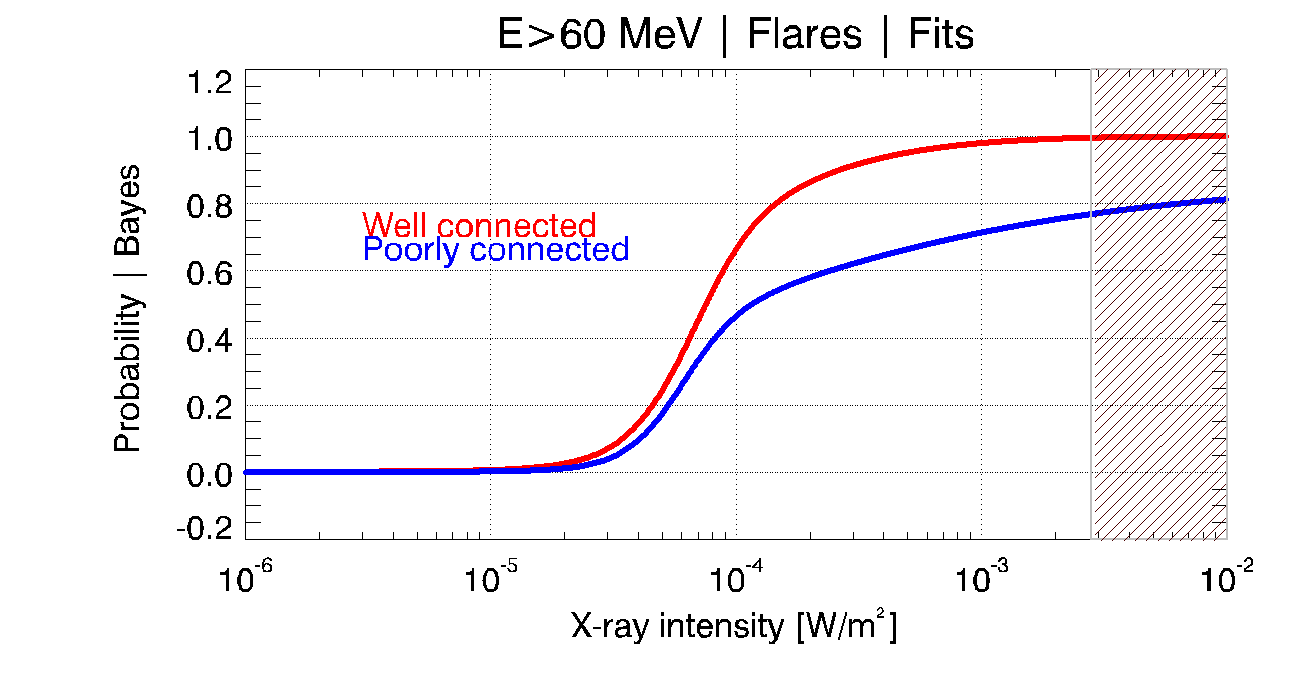}
\end{minipage}
\begin{minipage}[b]{0.45\linewidth}
\includegraphics[width=8cm,height=4.5cm]{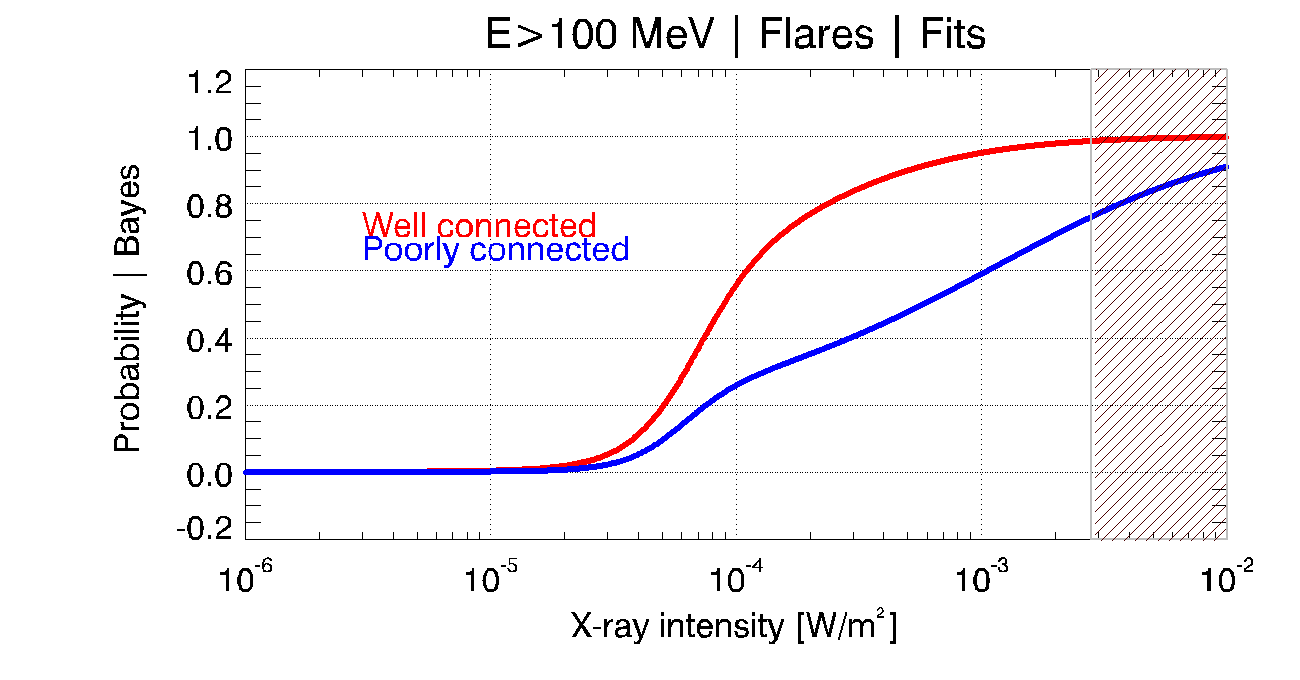}
\end{minipage}
\quad
\caption{The curves for the estimation of the probability of SEP occurrence when both the SXR magnitude and the longitude of a solar flare are known. Each panel corresponds to an integral energy (i.e. E$>$10-;$>$30-;$>$60- and $>$100 MeV). Within each panel two fits are presented, one per longitudinal bin - colour coded as: well connected (red), and poorly connected (blue) solar flares. See text for details. The gray border hatched rectangle area provides the limit for the $F_{SXR}$. This is similar to Figure \ref{fig:bayes-fits}.}
\label{fig:pf}
\end{figure}

\begin{figure}[ht]
\centering
\begin{minipage}[b]{\linewidth}
\includegraphics[width=16cm,height=8cm]{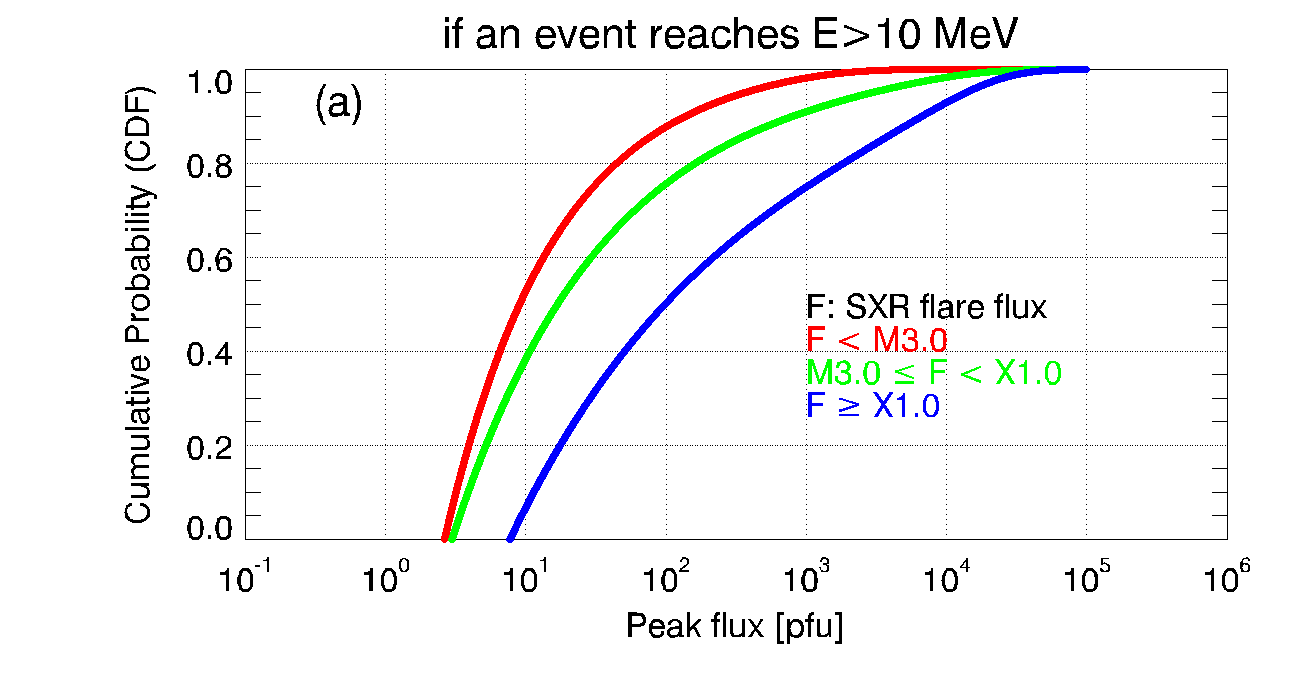}
\end{minipage}
\begin{minipage}[b]{0.48\linewidth}
\includegraphics[width=8cm,height=4.5cm]{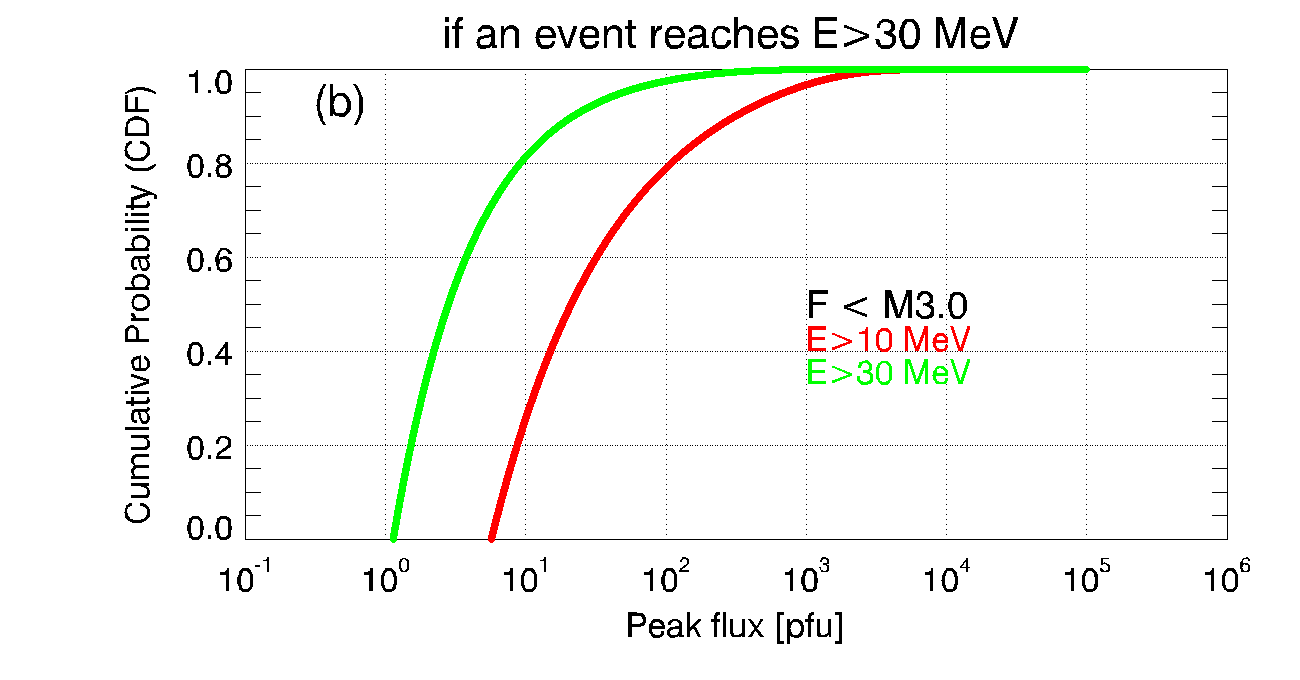}
\end{minipage}
\begin{minipage}[b]{0.48\linewidth}
\includegraphics[width=8cm,height=4.5cm]{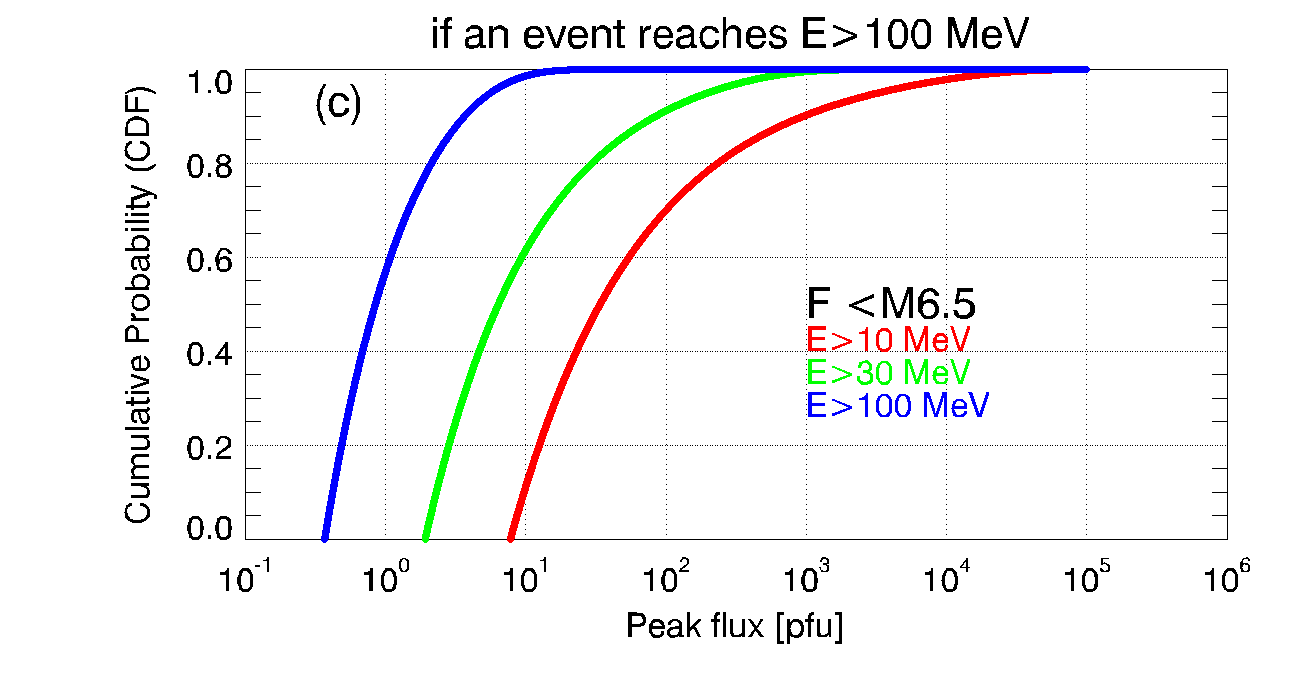}
\end{minipage}
\begin{minipage}[b]{0.48\linewidth}
\includegraphics[width=8cm,height=4.5cm]{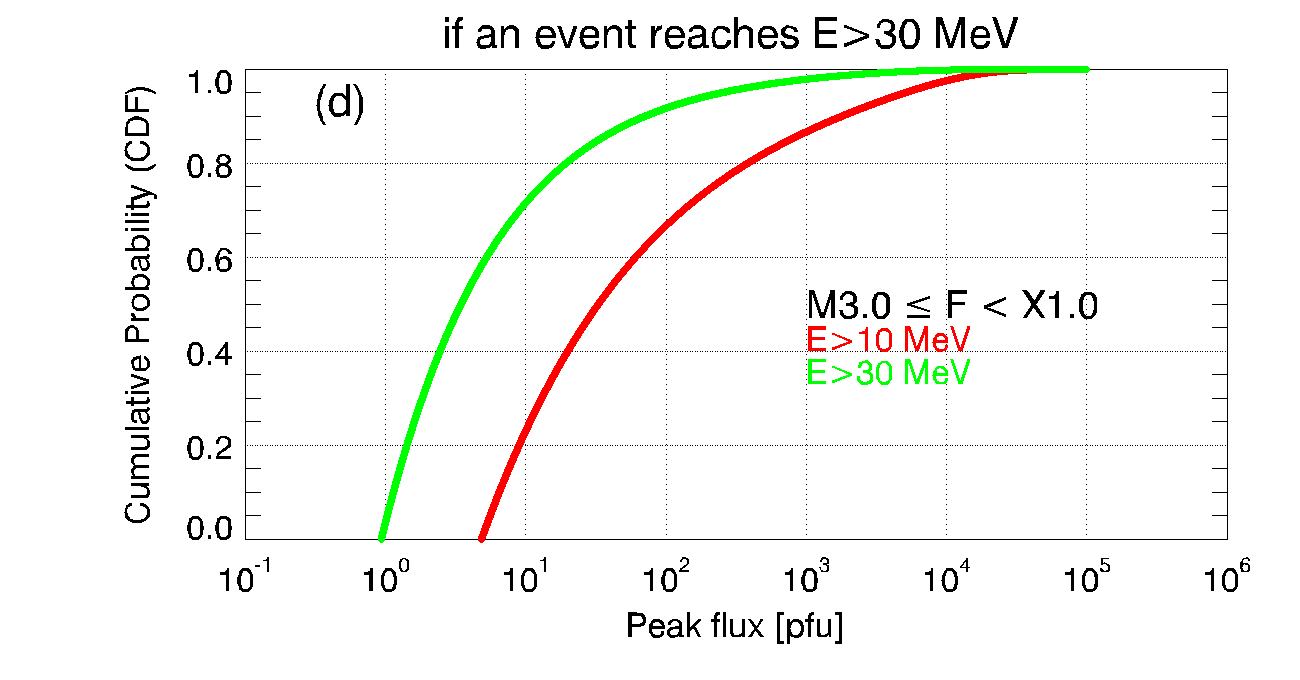}
\end{minipage}
\begin{minipage}[b]{0.48\linewidth}
\includegraphics[width=8cm,height=4.5cm]{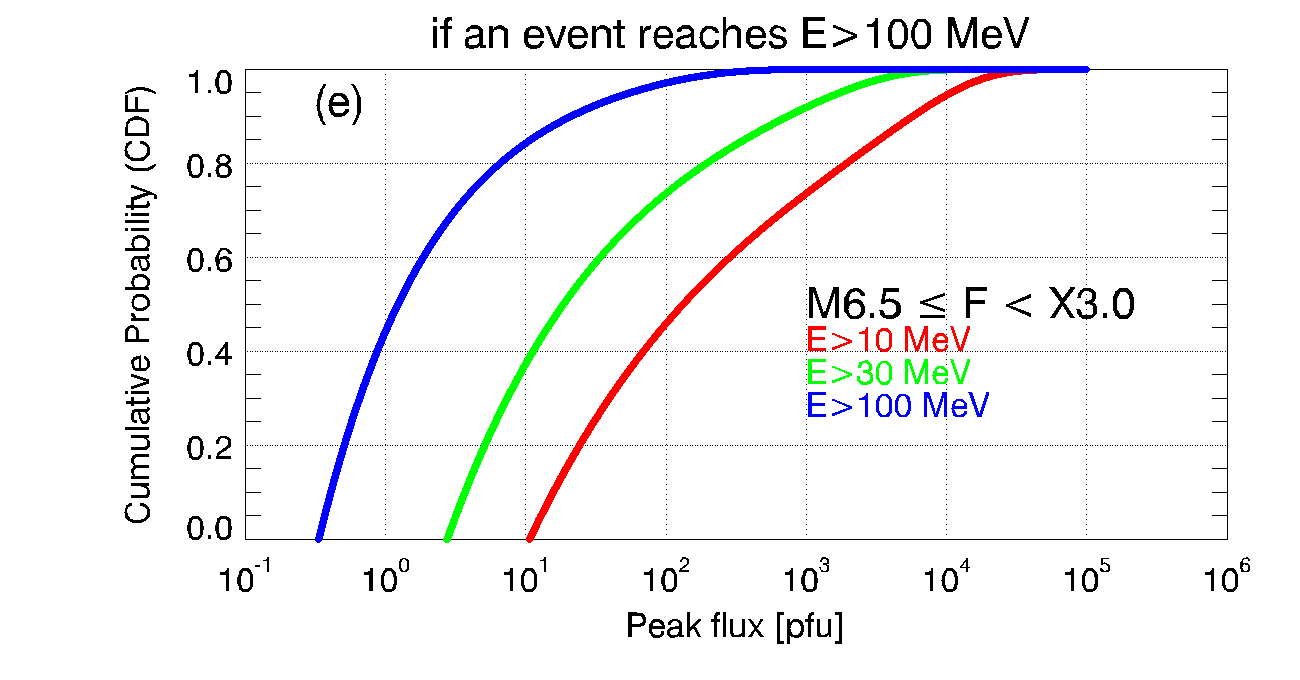}
\end{minipage}
\begin{minipage}[b]{0.48\linewidth}
\includegraphics[width=8cm,height=4.5cm]{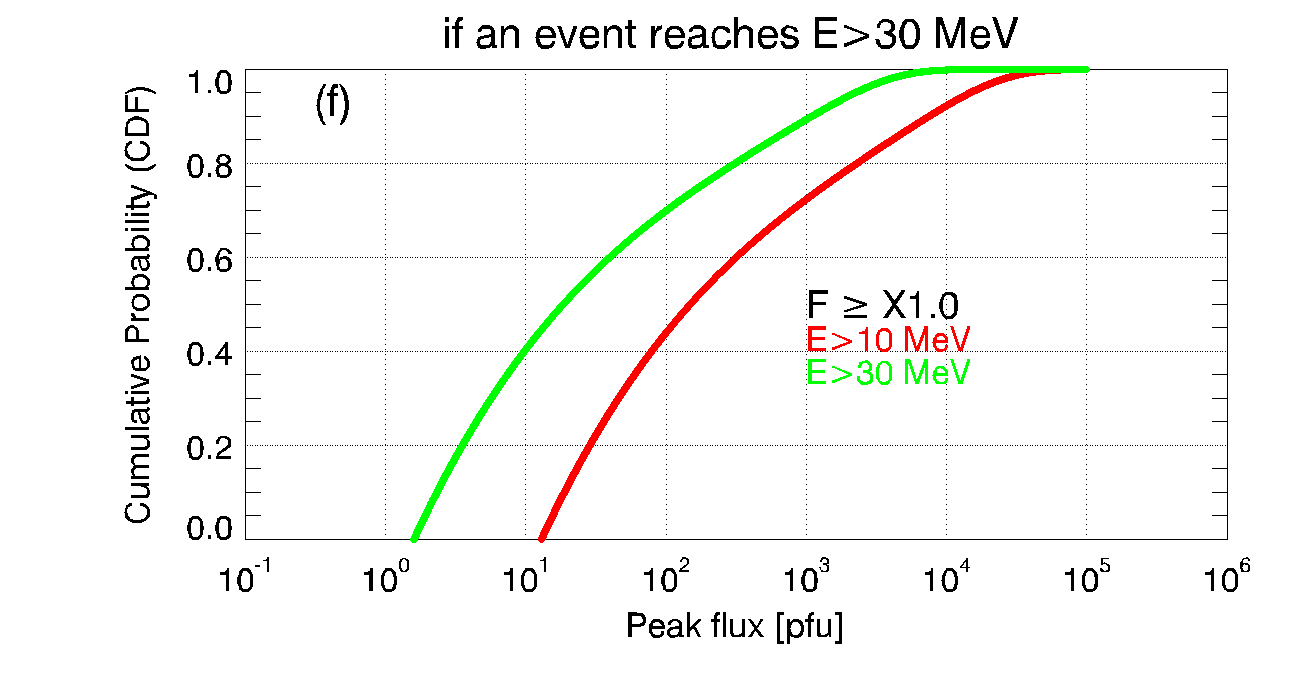}
\end{minipage}
\begin{minipage}[b]{0.48\linewidth}
\includegraphics[width=8cm,height=4.5cm]{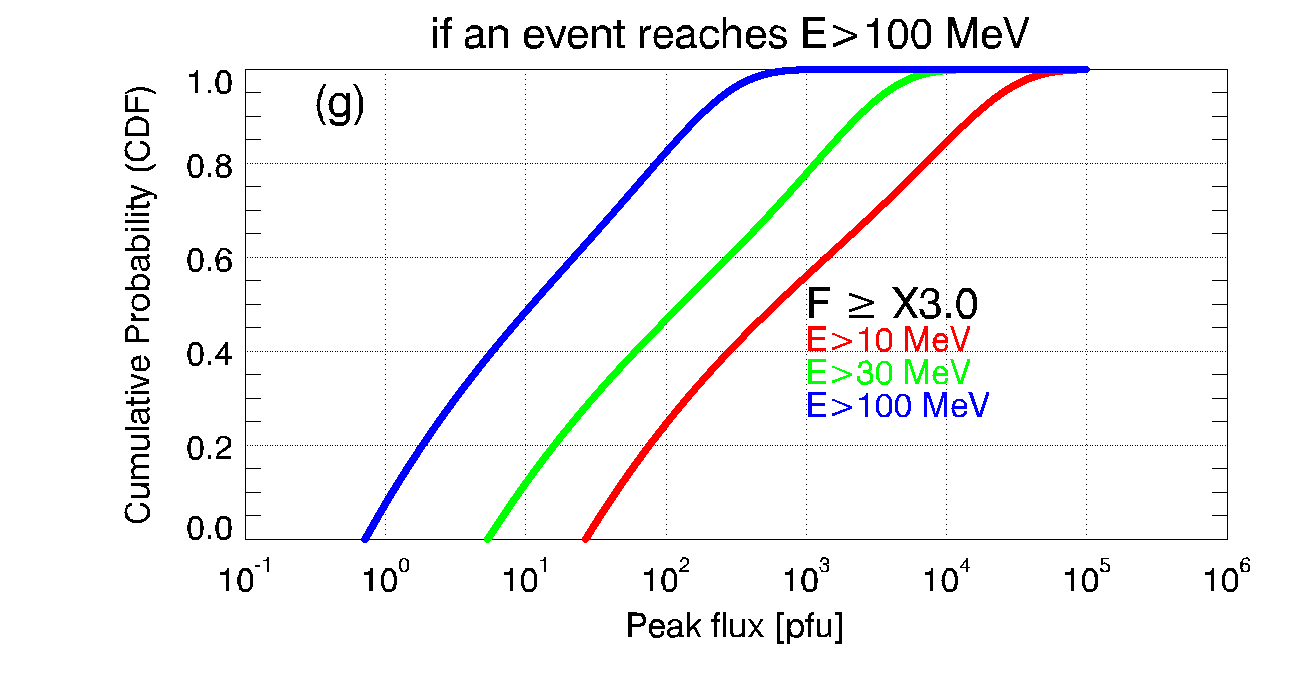}
\end{minipage}
\quad
\caption{The CDFs constructed by the database for each integral energy of choice, for the estimation of the peak flux (Peak flux [pfu], X-axis), when the SXR solar flare magnitude is known. Lines depict the exponential cut-off fits to the data points per case. Panel (a) corresponds to a filtering of the integral energy at E$>$10 MeV. The left hand column (panels (b), (d), (f)) corresponds to a filtering integral energy at E$>$30 MeV and the right hand column (panels (c), (d), (g)) to a filtering integral energy at E$>$100 MeV. Black color denotes the SXR flare flux bin in each panel. See text and Table \ref{tab:sf_peak} for details on the bins and the obtained parameters of the fit per case.}
\label{fig:PPF_sf}
\end{figure}

\begin{figure}[ht]
\centering
\includegraphics[width=12cm]{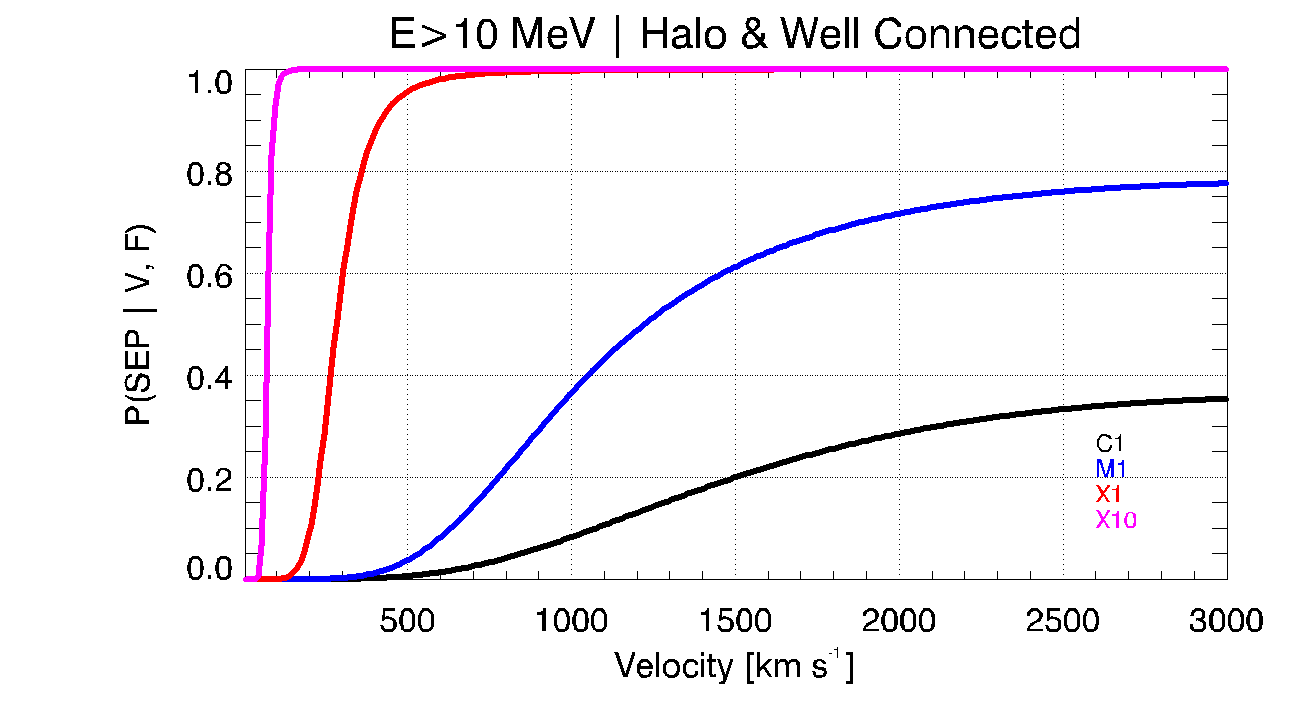}
\caption{The curves for the estimation of the probability of SEP occurrence when both a CME and a solar flare have taken place and thus the SXR magnitude (X-axis) and the longitude of a solar flare, as well as, the CME speed and AW are known. The figure presents the output of PROSPER for the case of E$>$10 MeV with Halo CMEs (AW = 360$^{o}$) and well connected flares (lon $\ge$20$^{o}$). Each fit corresponds to a solar flare class, color coded as: C1.0 (black line), M1.0 (blue line), X1.0 (red line) and X10 (magenta line). The X-axis shows the speed of the CME (in km s$^{-1}$).}
\label{fig:Pfcme}
\end{figure}

\begin{figure}[ht]
\centering
\begin{minipage}[b]{0.48\linewidth}
\includegraphics[width=8cm,height=4.5cm]{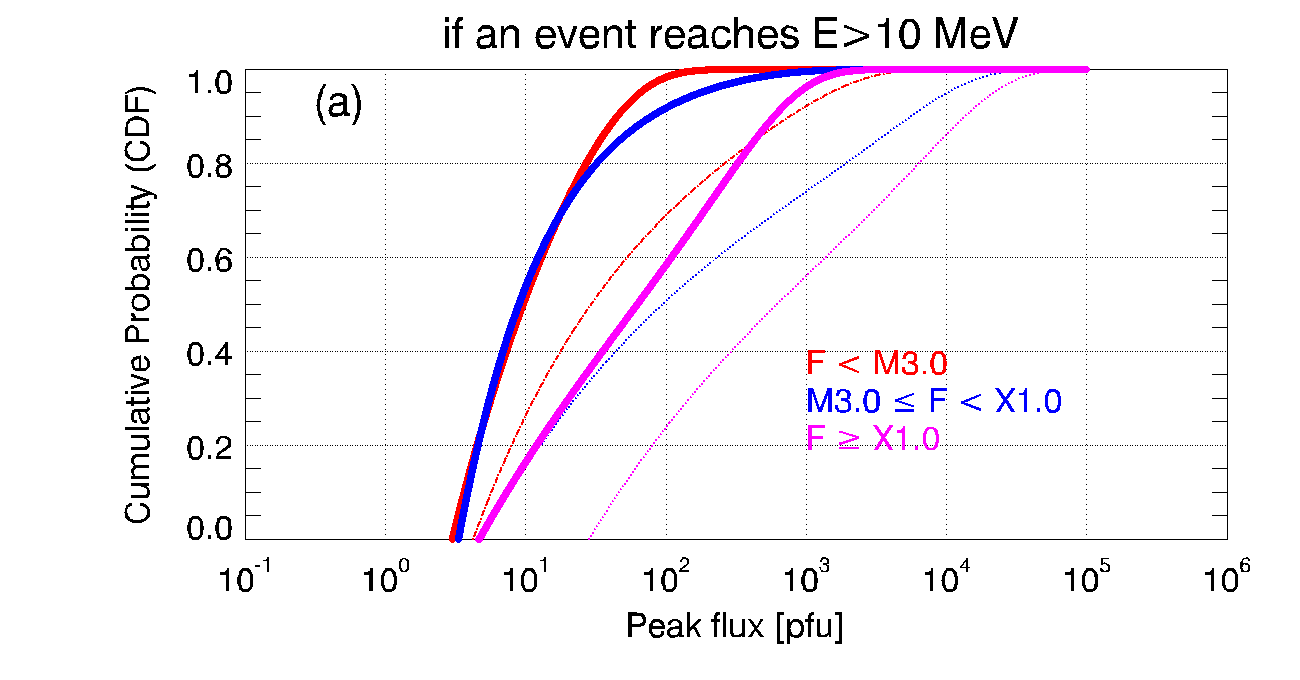}
\end{minipage}
\begin{minipage}[b]{0.48\linewidth}
\includegraphics[width=8cm,height=4.5cm]{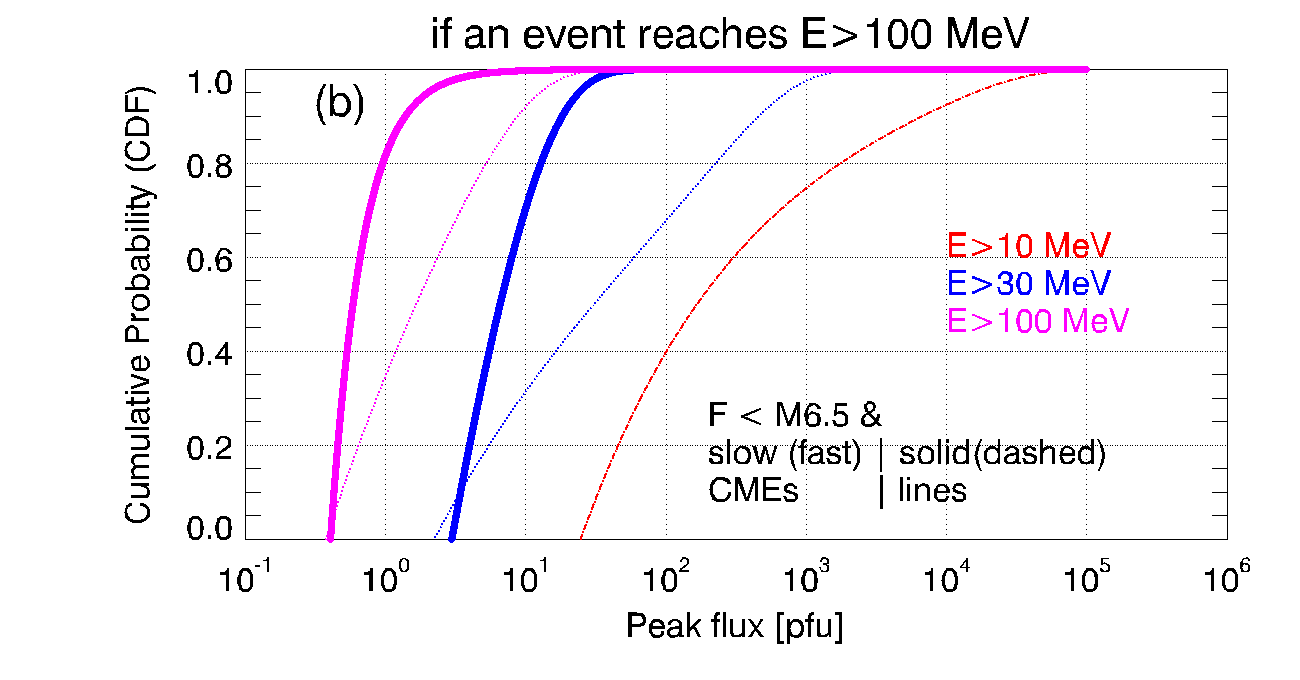}
\end{minipage}
\begin{minipage}[b]{0.48\linewidth}
\includegraphics[width=8cm,height=4.5cm]{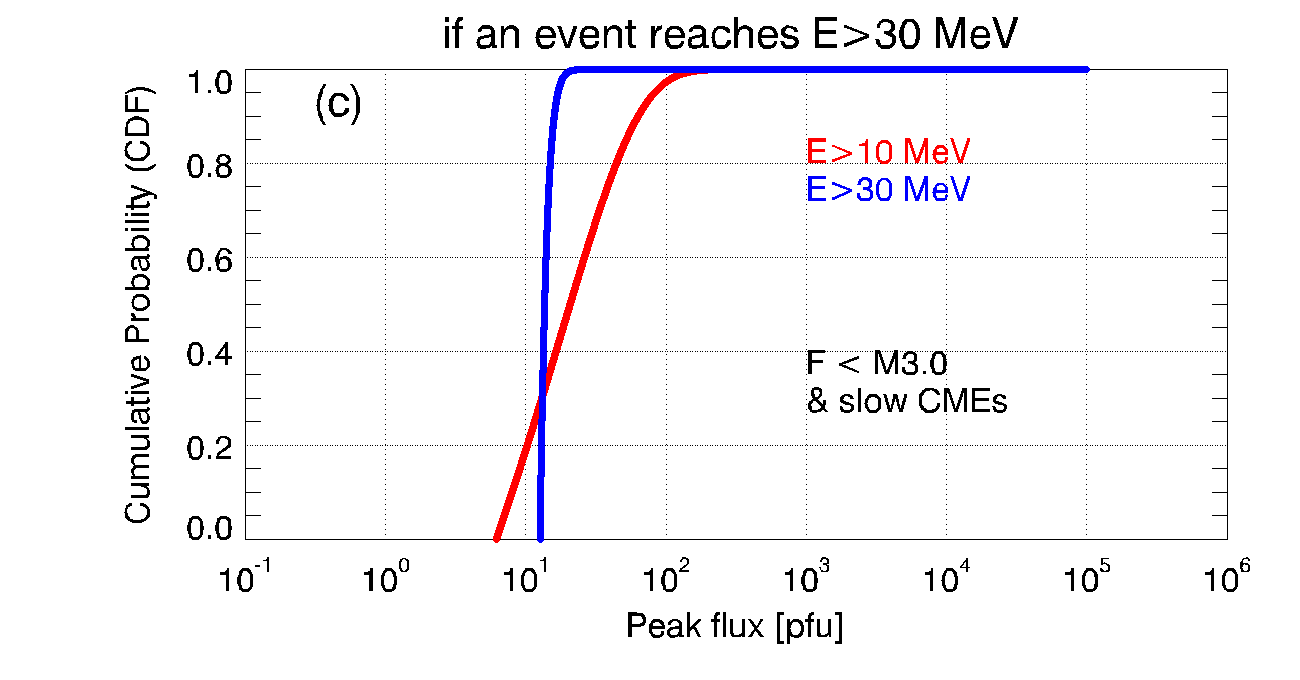}
\end{minipage}
\begin{minipage}[b]{0.48\linewidth}
\includegraphics[width=8cm,height=4.5cm]{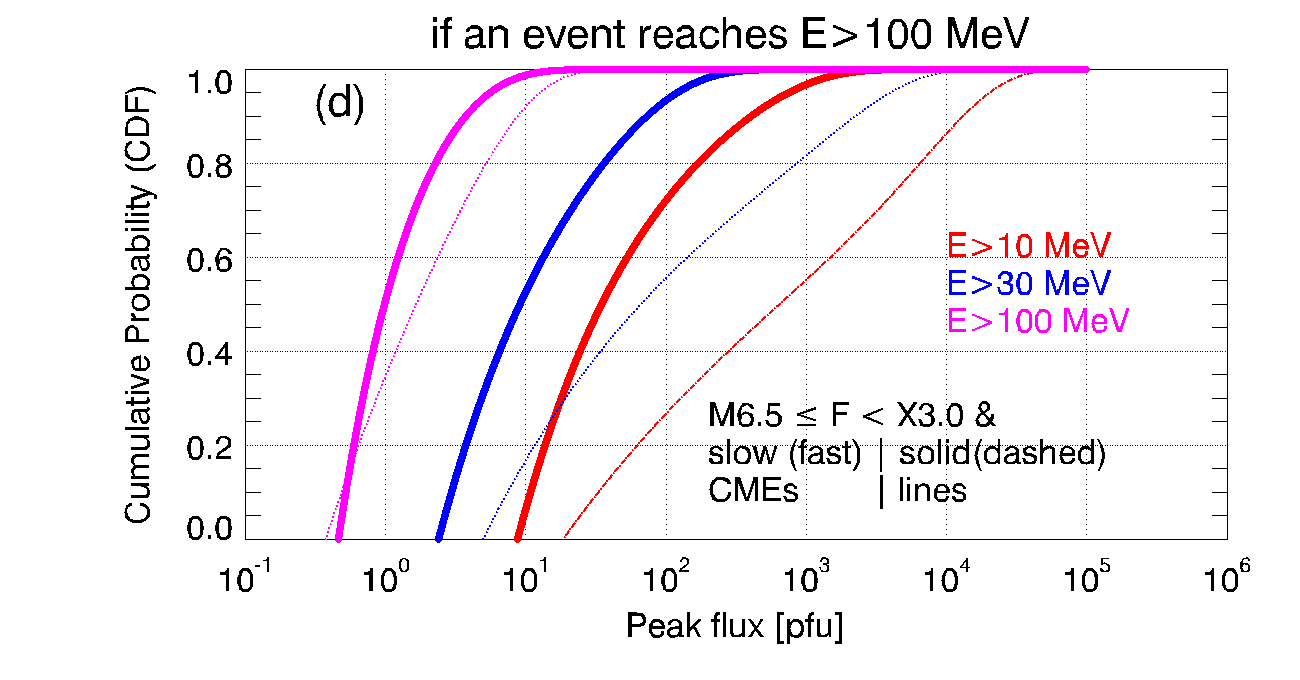}
\end{minipage}
\begin{minipage}[b]{0.48\linewidth}
\includegraphics[width=8cm,height=4.5cm]{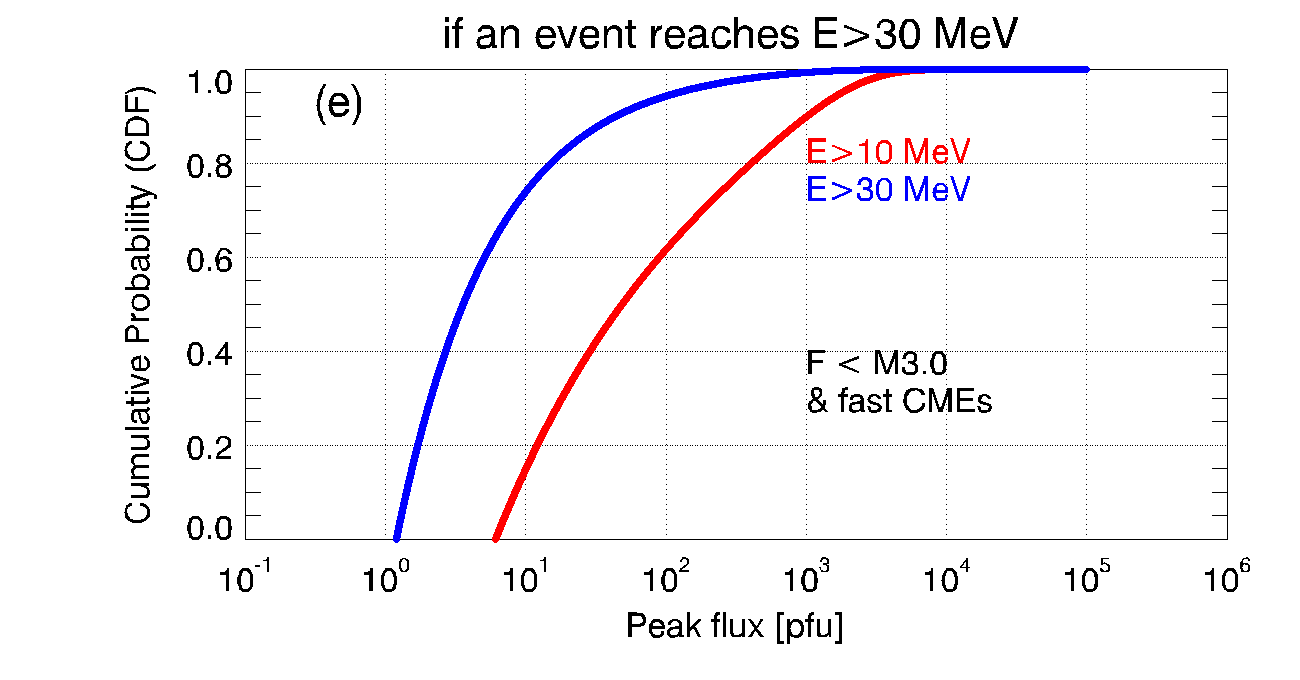}
\end{minipage}
\begin{minipage}[b]{0.48\linewidth}
\includegraphics[width=8cm,height=4.5cm]{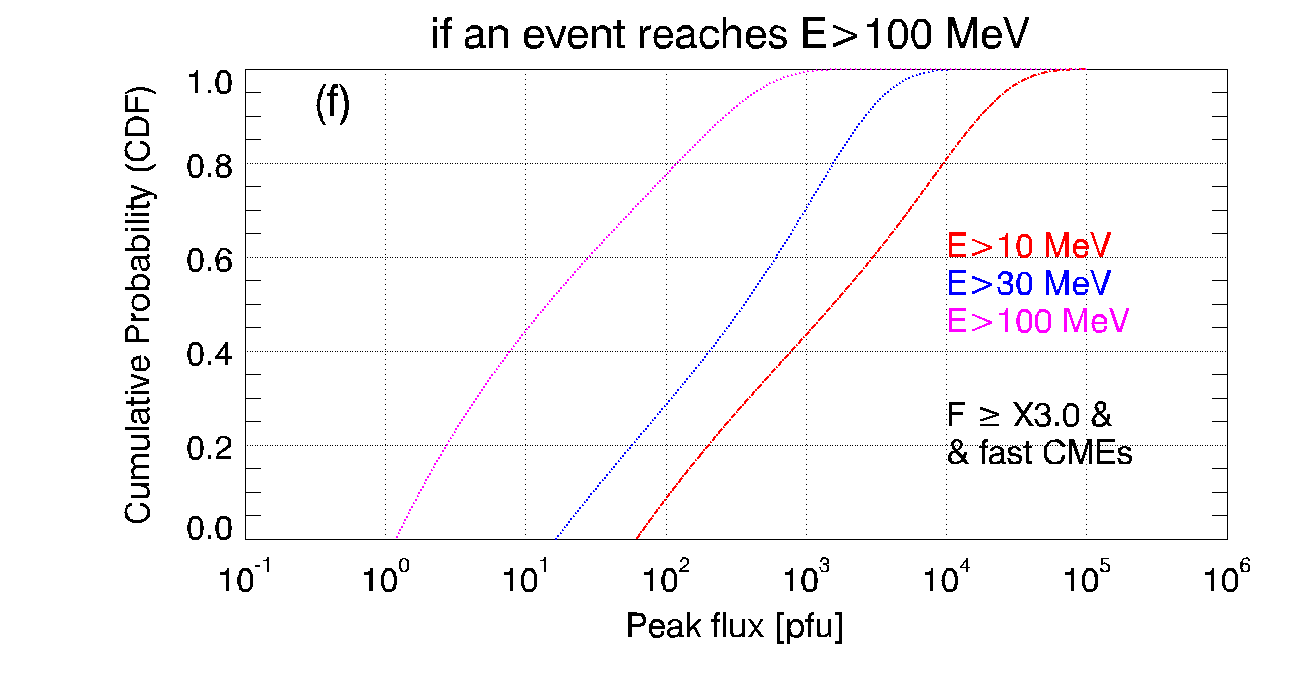}
\end{minipage}
\begin{minipage}[b]{0.48\linewidth}
\includegraphics[width=8cm,height=4.5cm]{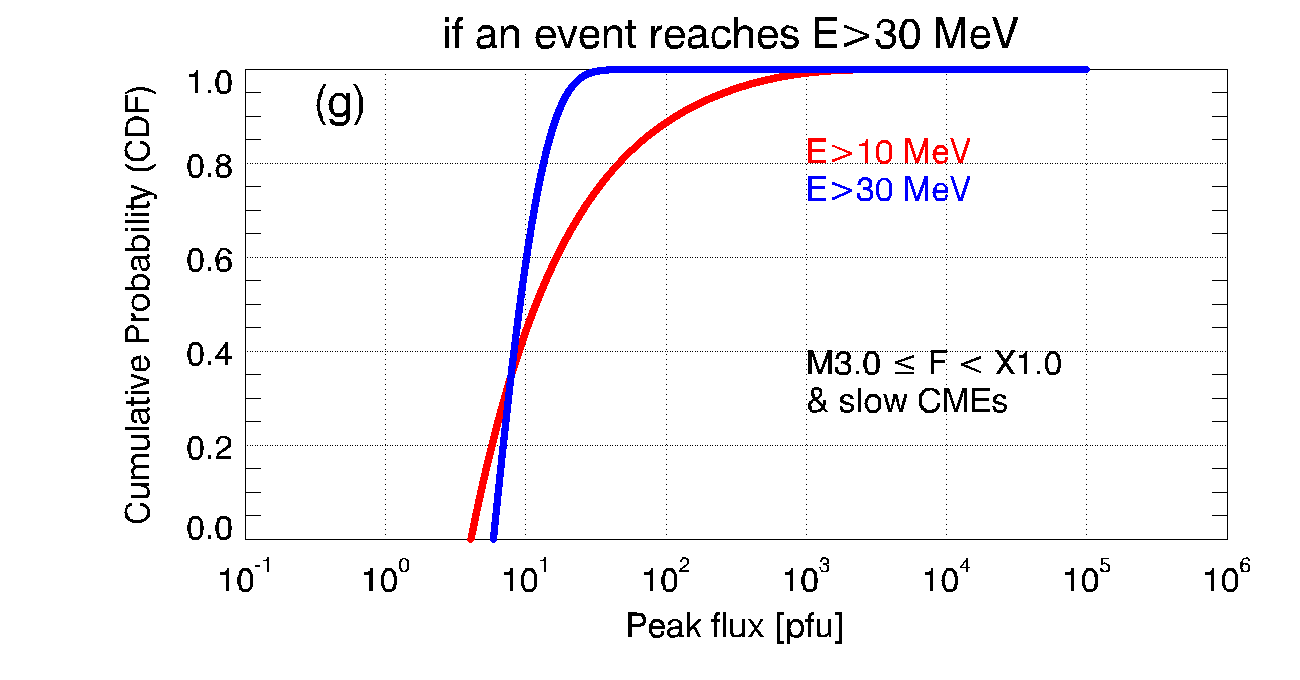}
\end{minipage}
\begin{minipage}[b]{0.48\linewidth}
\includegraphics[width=8cm,height=4.5cm]{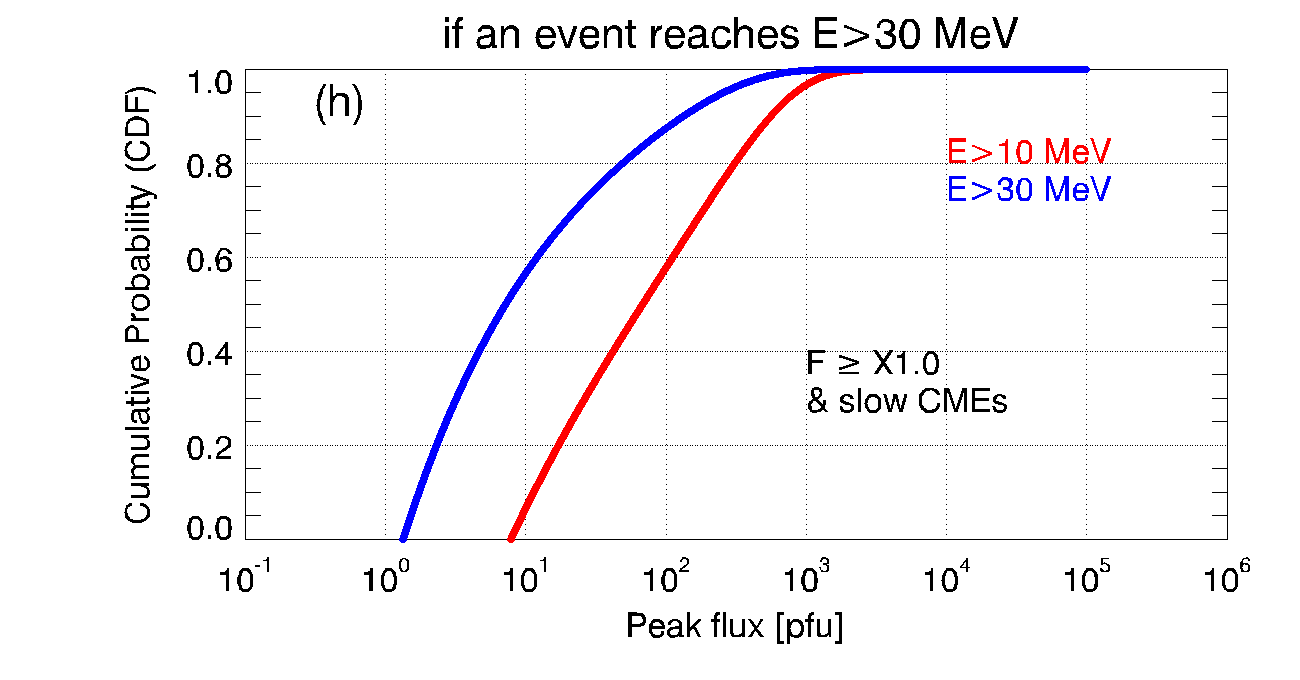}
\end{minipage}
\begin{minipage}[b]{0.48\linewidth}
\includegraphics[width=8cm,height=4.5cm]{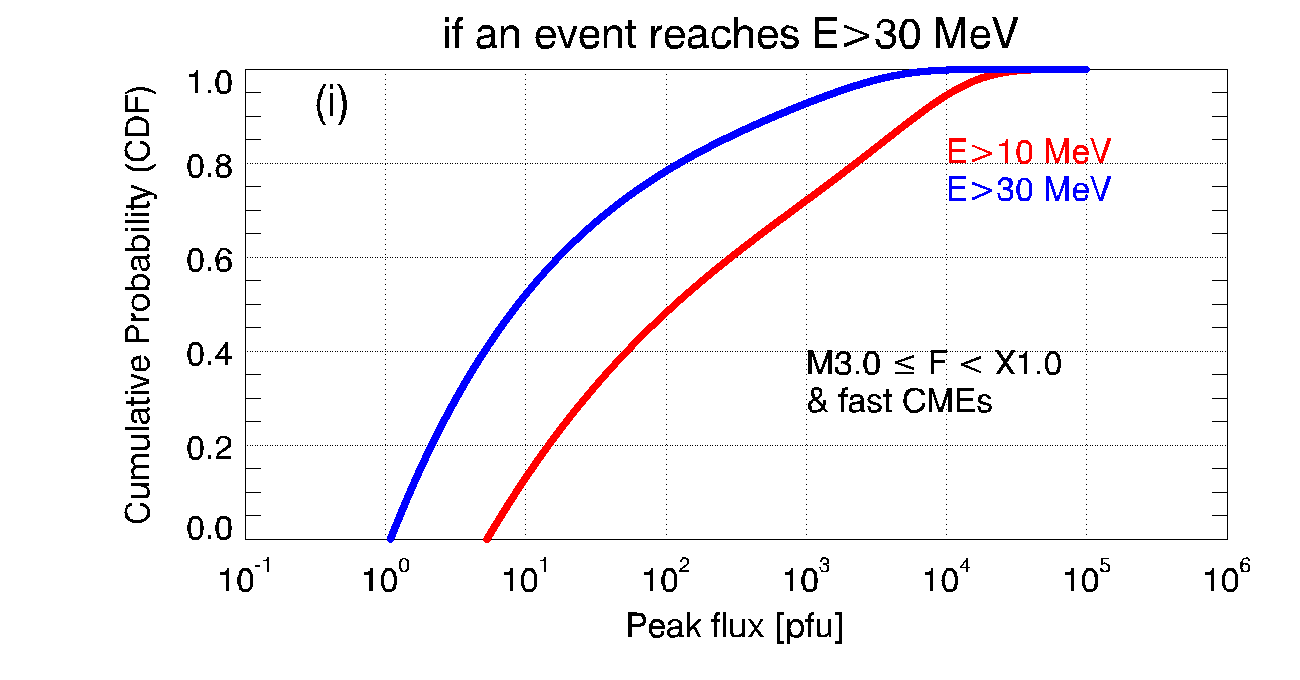}
\end{minipage}
\begin{minipage}[b]{0.48\linewidth}
\includegraphics[width=8cm,height=4.5cm]{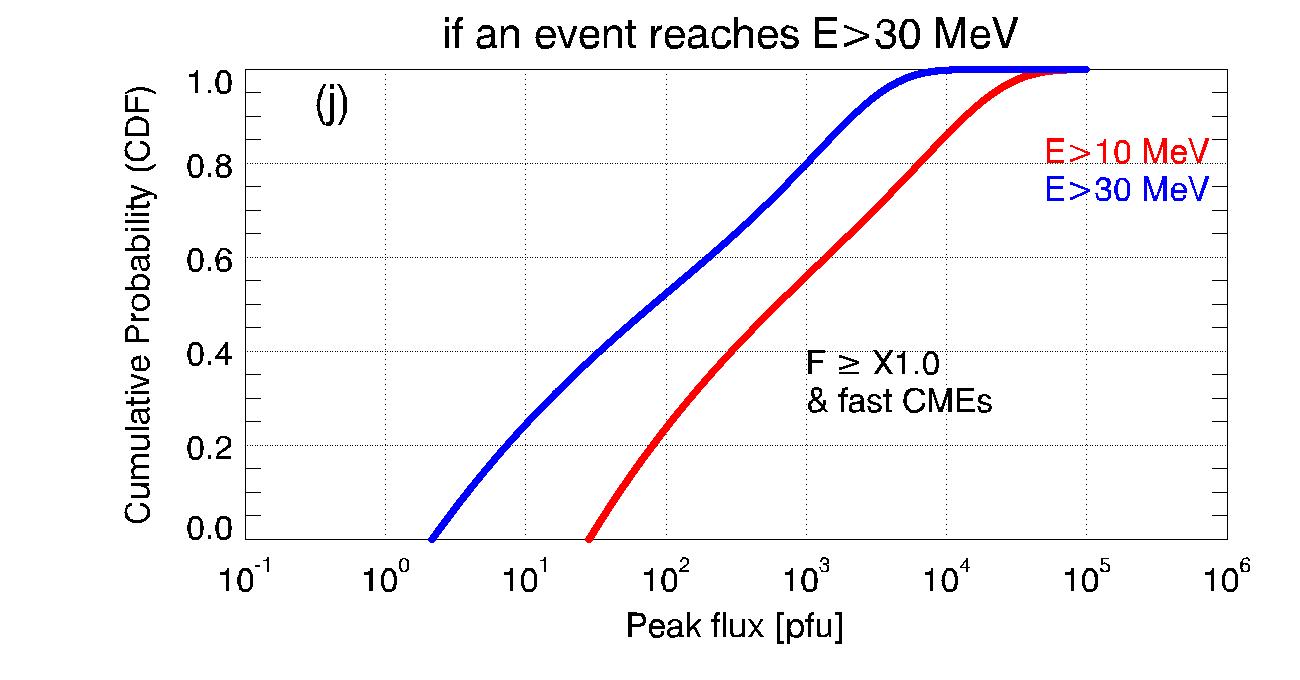}
\end{minipage}
\quad
\caption{The CDFs constructed by the database (i.e. all points correspond to actual data) for each integral energy of choice. See text and Table \ref{tab:sf_cme_peak} for details on the bins and the obtained parameters of the fit per case. Each panel further displays the selected bins and the displayed energies, if more than one are included in a panel.}
\label{fig:PPF_sf_cme}
\end{figure}

A new probabilistic model, PROSPER, that provides short-term forecasting (nowcasting) of SEP events (i.e. probability of occurrence and corresponding peak flux) has been presented in this work. In particular, PROSPER was implemented based on a straightforward application of the Bayes theorem using probability distribution functions constructed from a large database of solar flares, CMEs and SEPs. The output of the prediction is the probability of a solar event to lead to an SEP event, i.e. P($\rm SEP$). As a second step, the estimated peak proton flux of the SEP event is evaluated. PROSPER has three modes of operation based on the inputs received. There is one mode that utilizes solar flare data (magnitude and longitude), one that makes use of CME identifications (AW and speed) and a third one that makes use of the combined input of solar flares and CMEs (see Table \ref{tab:table1}). The basic implementations of PROSPER presented in this study are summarized as follows:

\begin{itemize}
    \item \textbf{Probability of SEP occurrence}. For the identification of $P(\rm SEP)$ in each case, log-normal fits (Equation \ref{eq:1}) to the data were utilized (see e.g. Figures \ref{fig:CDFs_fit} \& \ref{fig:cdfs_2}) and the resulting distributions were combined under the Bayes formula (Equation \ref{eq:9}). These Bayes fits were obtained for different integral energies spanning from E$>$10-; E$>$30- and E$>$100 MeV (see Figures \ref{fig:bayes-fits}, \ref{fig:pf} and \ref{fig:Pfcme}).  
    \item \textbf{Peak fluxes}. The modeling of the peak fluxes was based on exponential cut-off fits (Equation \ref{eq:12}) to cumulative distribution functions (CDFs) constructed by the recorded peak fluxes of the SEP events in our sample. In this case, first the data are filtered based on the highest expected integral energy that an SEP event will occur (e.g. applying a thresholding on the $P(\rm SEP)$ obtained in the previous step and thus utilizing the prediction of an event being expected to reach E$>$100 MeV, E$>$30 MeV or E$>$10 MeV). Detailed fits are presented in Figures \ref{fig:PPF}, \ref{fig:PPF_sf} and \ref{fig:PPF_sf_cme} as well as in Tables \ref{tab:cme_peak}, \ref{tab:sf_peak} and \ref{tab:sf_cme_peak}, respectively.   
 \end{itemize}

The ramifications of Bayes' rule are countless and have gained wide recognition in data driven, Space Weather related topics. Nonetheless, to our knowledge, PROSPER is the first SEP prediction model that allows the estimation of P($\rm SEP$) utilizing Bayes’ theorem \citep[see][and references there in]{2019SpWea..17.1166C}. PROSPER allows a direct estimation of P($\rm SEP$), based on a previous outcome having occurred in similar circumstances and in doing so, it takes into account all available data without any a priori bias. As a result, the benefit of such an approach is the natural and principled way of combining prior information with data, within a solid decision theoretical framework, providing exact inferences that are conditional on data \citep{hartigan2012bayes}.

Moreover, all PROSPER's modes of operation were validated based on detailed case studies of SEP events selected as part of the NASA CCMC SEP scoreboard challenge (see Table \ref{tab:valid}). Blind tests with archived parent solar data of these events were applied to PROSPER's modes in order to derive the probabilities of SEP detection per mode of operation and per SEP event. The obtained results were discussed in a comparative manner, showing that the magnetic connectivity (i.e. ``well connected" events), strong solar flares and fast CMEs lead to higher achieved $P(\rm SEP)$ values. The validation results of the PROSPER models are based on a sample of 10 SEP events and are summarized as follows: 
\begin{itemize}
    \item PROSPER's module that is based on solar flare input would spot $\sim$90\% (i.e. 8/9) of the events for which flare input was available.
    \item For the CME module the hit rate is 100\% (i.e. 10/10 events for which CME input was available were identified as SEP events).
    \item The solar flare \& CME mode of operation, also has a hit rate of 100\% (i.e. 9/9 events for which both flare and CME inputs were available).
\end{itemize}

Additionally, the largest differences in the achieved $P(\rm SEP)$ in lower energies (i.e. at an integral energy E$>$10 MeV) are due to the magnitude of the associated solar flare, $F_{SXR}$. For the higher energies (i.e. E$>$100 MeV) the differences in the obtained $P(\rm SEP)$ values become larger due to the additional separation of the fits, which is driven by their longitudinal difference. Finally, the solar flare \& CME module revealed that the better magnetic connection and the higher $V_{CME}$ seem to lead to a higher $P(\rm SEP)$.

Furthermore, we included scatter plots of observed versus predicted SEP peak fluxes to quantify the reliability of PROSPER's predictions. Although the SEP peak fluxes are difficult to infer based on the characteristics of their parent solar events, the usage of an upper (90\%) and lower (50\%) limit in PROSPER seemed to capture the expected peak flux of the SEP events, reasonably well. In particular, at E $>$ 10 MeV, for the solar flare module, the predicted peak flux, as a function of the observed peak flux shows a strong correlation especially in the lower limit (i.e. 50\% CL) with more than half of the the predicted values being close to the dichotomous line -- indicating a perfect prediction. Nonetheless, the CME module at the same energy seems not to capture the expected peak flux with granularity. However, the combined SF \& CME mode of operation shows that the predictions at both the lower and the upper limit seem to be very close to the dichotomous line, with a small tendency for over-forecasting in the 90\% limit predictions. Nonetheless, the predictions at 50\% for the SEP events achieving the largest peak flux at E$>$10 MeV seem to fall at the perfect dichotomous line. Additionally, in the case of E$>$100 MeV, the peak flux seems to be captured well by all three modes of operation, especially with the 90\% limit in the case of the SF and the flare \& CME modules the agreement is quite reasonable with most of the events predicted at 90\% being either on the dichotomous line or very close to it (i.e. within less then 1 order of magnitude).

The PROSPER model has been incorporated into the new operational Advanced Solar Particle Event Casting System (ASPECS) tool providing outputs in real-time through the web portal \url{http://phobos-srv.space.noa.gr/}. In addition, ASPECS offers the capability to interrogate PROSPER for historical cases via a run on demand functionality. ASPECS is the first realisation of ESA's SEP Advanced Warning System (SAWS) - a modular framework for forecasting solar energetic particle (SEP) events, their characteristics and profiles.

A point that still needs to be addressed in the validation is how the model reacts when solar events not associated with SEPs are introduced as inputs. This is a part of an ongoing follow-up study focusing on the performance of the ASPECS tool. On a preliminary basis, it can be commented that PROSPER achieves categorical scores in line with the expected range for data driven models. The range of scores are taken from Table 1 of \cite{anastasiadis2017predicting} (i.e. Probability of Detection -- POD $\sim$50-70\%, False Alarm Rate -- FAR $\sim$30-50\%). Such a range is realistic and inherently imposed due to the imbalanced dataset that is being used for SEP prediction efforts \cite[see the relevant detail discussion in][]{lavasa2021assessing}. In fact, recently, \cite{2021SpWea..1902794S} have shown that the major drawback in predicting the occurrence of SEPs, i.e. P($\rm SEP$), in the framework of statistical forecasting concepts, is the optimization of FAR which directly depends on the imbalance of the dataset used. That means that, the greater the imbalance, the greater the FAR is affected by the presence of false positives. Taking all of these into account, in our next study such a thorough validation step will be implemented.

It should be mentioned that, PROSPER's CME and flare \& CME modes of operation are subject to the CME input data used. In particular, while developing PROSPER, CME data from the CDAW CME catalogue have been used. It was pointed out by \cite{2015SoPh..290.1741R} that different CME catalogues provide different estimations of the same CME event, hence given that PROSPER is a data driven model, it is inherently affected by such differences. On top of the CDAW CME catalogue, other options include: (a) the Computer-Aided CME Tracking (CACTUS) catalog (\url{https://wwwbis.sidc.be/cactus/}), which is compiled using a specialized software package \citep{2004A&A...425.1097R}; (b) the SEEDS (Solar Eruptive Event Detection System) catalog (\url{http://spaceweather.gmu.edu/seeds/}) \citep{2008SoPh..248..485O}; (c) the CORIMP (coronal
image processing) method catalog (\url{http://alshamess.ifa.hawaii.edu/CORIMP/}) \citep{2012ApJ...752..145B} and (d) the Space Weather Database Of Notification, Knowledge, Information (DONKI) developed by CCMC (\url{https://ccmc.gsfc.nasa.gov/donki/}). Therefore, a natural next step of this study is to adapt PROSPER to each of these catalogues and quantify differences/changes following the study by \cite{2015SoPh..290.1741R}.

The usage of CME identifications (e.g.
angular width and speed) for the derivation of the SEP occurrence probabilities (P($\rm SEP$)) and expected peak proton flux have been explored, already, in a few operational efforts \citep[e.g.][]{dierckxsens2014relationship, papaioannou2018nowcasting, richardson2018prediction}. On top of that statistical studies have pointed out that P($\rm SEP$) increases as a function of solar flare magnitude ($F_{SXR})$, longitude and CME speed ($V_{CME}$) - especially when considering solar flare-CME couples situated on the west part of the visible solar
disk \citep[i.e.][]{dierckxsens2014relationship}. In addition, it was recently showed that $F_{SXR}$, $V_{CME}$ and SXR fluence have the largest potential for discriminating solar events that are associated with SEP events \citep{lavasa2021assessing}, while as many as targeted parameters need to be taken into account, in order to build a reliable and more accurate SEP predicting system \citep[see e.g.][]{papaioannou2018nowcasting}. PROSPER takes advantage of such findings and comes with three modes of operation. Nevertheless, its performance certainly benefits from the provision of reliable CME estimates or a proper proxy for the CME speed and width, which would be very much desirable to obtain in near-real time.


\begin{acknowledgements}
This work was supported through the ESA Contract No. 4000120480/NL/LF/hh ``Solar Energetic Particle (SEP) Advanced Warning System (SAWS)”. Athanasios Papaioannou and Angels Aran acknowledge the support from the project MDM-2014-0369 of ICCUB (Unidad de Excelencia ``María de Maeztu"). The CME Catalog used in this work is generated and maintained at the CDAW Data Center by NASA and The Catholic University of America in cooperation with the Naval Research Laboratory. SOHO is a project of international cooperation between ESA and NASA. Athanasios Papaioannou, Rami Vainio \& Anastasios Anastasiadis acknowledge the International Space Science Institute and the supported International Team 441: High EneRgy sOlar partICle Events Analysis (HEROIC, \url{http:// www.issibern.ch/teams/heroic/}). Finally, Athanasios Papaioannou acknowledges support from NASA/LWS project NNH19ZDA001N-LWS. The authors would like to thank Dr Manolis Georgoulis for stimulating discussions, Dr Katie Whitman for fruitful collaboration and exchange of ideas as concerns the validation of PROSPER and Mr George Vasalos for valuable technical assistance. Furthermore, the authors would like to thank the anonymous referees for a critical and constructive reading of the manuscript and for valuable comments that improved the contents of the paper.
 
\end{acknowledgements}


\bibliography{swsc}

\clearpage

  \clearpage

   \clearpage
  
   \clearpage
  
\clearpage

 \clearpage

\clearpage

 
  \clearpage

\clearpage

\clearpage

\clearpage

\clearpage

\clearpage

\end{document}